\documentclass[useAMS,usenatbib]{mn2e}

\usepackage{color}
\usepackage{colortbl}
\usepackage{multirow}
\usepackage{dcolumn}
\usepackage{amsmath}
\usepackage{amssymb}
\usepackage{graphicx}
\usepackage{epsfig}
\usepackage{changebar}
\usepackage{natbib}
\usepackage{multirow}
\usepackage{subfigure}
\usepackage{txfonts}
\usepackage{rotating}
\usepackage{threeparttable}
\usepackage{ulem}
\usepackage[backref,breaklinks,colorlinks,citecolor=blue]{hyperref}
\usepackage{caption}
\usepackage[all]{hypcap}

\usepackage{longtable,lscape}

\newcolumntype{d}[1]{D{.}{\cdot}{#1}}

\newcolumntype{.}{D{.}{.}{-1}}

\newcommand{\lsun}{\ensuremath{\mathrm{L}_\odot}}
\newcommand{\msun}{M$_\odot$}

\newcommand{\vlsr}{$v_{\rm{LSR}}$}

\newcommand{\kms}{km\,s$^{-1}$}

\newcommand{\hi}{H~{\sc i}}
\newcommand{\hii}{H~{\sc ii}}

\newcommand{\poi}{Poisson}

\newcommand{\Tant}{$T_A^*$}

\newcommand{\Tkin}{$T_{\mathrm{kin}}$}

\newcommand{\Tex}{$T_{\mathrm{ex}}$}

\newcommand{\rms}{r.m.s.}
\newcommand{\submm}{submillimetre}

\newcommand{\KS}{Kolmogorov-Smirnov}
\newcommand{\KSvalue}{$p$-$value$}

\newcommand{\Mal}{Malmquist}

\newcommand{\FW}{\texttt{FellWalker}}
\newcommand{\Kappa}{\texttt{Kappa}}

\newcommand{\arcs}{\mbox{\ensuremath{^{\prime\prime}}}}

\newcommand{\fieldnum}{66}
\newcommand{\clumpnum}{115}
\newcommand{\nonrmsclumpnum}{44}
\newcommand{\rmsclumpnum}{71}

\newcommand{\mum}{$\umu$m}

\title[Structure of massive star forming clumps]{The RMS Survey: Ammonia mapping of the environment of massive young stellar objects\thanks{The full version of Tables\,2, 3, 5 and 6 and Figs\,1, 5 and 6 are only available in electronic form at the CDS via anonymous ftp to cdsarc.u-strasbg.fr (130.79.125.5) or via http://cdsweb.u-strasbg.fr/cgi-bin/qcat?J/MNRAS/.}}
\author[J.\,S.\,Urquhart et al.]{J.\,S.\,Urquhart$^{1}$\thanks{E-mail:
jurquhart@mpifr-bonn.mpg.de (MPIfR)}, C.\,C.\,Figura$^{2}$, T.\,J.\,T.\,Moore$^{3}$, T.\,Csengeri$^{1}$, S.\,L.\,Lumsden${^4}$, T.\,Pillai$^{1}$,\newauthor  M.\,A.\,Thompson$^{5}$, D.\,J.\,Eden$^{6}$, L.\,K.\,Morgan$^{3,7}$ \\
$^{1}$Max-Planck-Institut f\"ur Radioastronomie, Auf dem H\"ugel
  69, D-53121 Bonn, Germany \\
$^{2}$Wartburg College, 100 Wartburg Blvd, Waverly, IA 50677, USA\\ 
$^{3}$Astrophysics Research Institute, Liverpool John Moores University, 146 Brownlow Hill, Liverpool, L3\,5RF, UK\\ 
$^{4}$School of Physics and Astrophysics, University of Leeds, Leeds, LS2\,9JT, UK \\
$^{5}$ Centre for Astrophysics Research, Science and Technology Research Institute, University of Hertfordshire, College Lane, Hatfield, AL10 9AB, UK \\
$^{6}$Observatoire astronomique de Strasbourg, Universite ́ de Strasbourg, CNRS, UMR 7550, 11 rue de l'Université, 67000, Strasbourg, France\\
$^{7}$Met Office, FitzRoy Road, Exeter, Devon, EX1\,3PB, UK\\
}

\begin{document}

\date{Accepted ??. Received ??; in original form ??}

\pagerange{\pageref{firstpage}--\pageref{lastpage}} \pubyear{2009}

\maketitle

\label{firstpage}

\begin{abstract}

We present the results of ammonia observations towards 66 massive star forming regions identified by the Red MSX source survey. We have used the Green Bank Telescope and the K-band focal plane array to map the ammonia (NH$_3$) (1,1) and (2,2) inversion emission at a resolution of 30\arcsec\ in 8\arcmin\ regions towards the positions of embedded massive star formation. We have identified a total of 115 distinct clumps, approximately two-thirds of which are associated with an embedded massive young stellar object or compact \hii\ region, while the others are classified as quiescent. There is a strong spatial correlation between the peak NH$_3$ emission and the presence of embedded objects. We derive the spatial distribution of the kinetic gas temperatures, line widths, and NH$_3$ column densities from these maps, and by combining these data with dust emission maps we estimate clump masses, H$_2$ column densities and ammonia abundances. The clumps have typical masses of $\sim$1000\,\msun\ and radii $\sim$0.5\,pc, line widths of $\sim$2\,\kms\ and kinetic temperatures of $\sim$16-20\,K. We find no significant difference between the sizes and masses of the star forming and quiescent subsamples; however, the distribution maps reveal the presence of temperature and line width gradients peaking towards the centre for the star forming clumps while the quiescent clumps show relatively uniform temperatures and line widths throughout. Virial analysis suggests that the vast majority of clumps are gravitationally bound and are likely to be in a state of global free fall in the absence of strong magnetic fields. The similarities between the properties of the two subsamples suggest that the quiescent clumps are also likely to form massive stars in the future, and therefore provide a excellent opportunity to study the initial conditions of massive pre-stellar and protostellar clumps.

\end{abstract}

\section{Introduction}
\label{sect:intro}

Although massive stars ($>$ 8\,\msun\ and 10$^3$\,\lsun) make up only a few per cent of the stellar population, they play a central role in many astrophysical processes. They have a  profound impact on their local environments through powerful outflows, strong stellar winds, copious amounts of optical/far-UV radiation, and chemical enrichment. The energy and processed material returned to the ISM play an important role in regulating star formation by changing the local chemistry and through the propagation of strong shocks in the surrounding molecular clouds. These feedback processes may be responsible for triggering subsequent generations of stars to form or disrupting conditions necessary for star formation in nearby clouds (\citealt{elmegreen1998}), ultimately governing the evolution of their host galaxy (\citealt{kennicutt2005}).

Despite their importance, our understanding of the initial conditions required and processes involved in the formation and early evolution of massive stars is still rather poor. There are a number of reasons for this: massive stars are rare and relatively few are located closer than a few kpc; they form almost exclusively in clusters, making it hard to distinguish between the properties of the cluster and individual members; and they evolve rapidly, reaching the main sequence while still deeply embedded in their natal environment, and consequently the earliest stages can only be probed at far-infrared, (sub)millimetre and radio wavelengths. Previous studies have been limited to high angular resolution observations of specific objects or larger low-resolution surveys which have tended to focus on bright far-infrared and/or radio sources that are already very luminous ($>$ 10$^3$\,\lsun; e.g., \citealt{wood1989, molinari1996, sridharan2002}). These studies have identified a large number of young massive stars; however, source confusion in complex regions resulted in these samples being biased away from the Galactic mid-plane where the majority of massive stars are located (scale height $\sim$30\,pc; \citealt{reed2000}). It is therefore unclear whether these samples are a good representation of the global population of young massive stars.

\setlength{\tabcolsep}{3pt}

\begin{table*}


\caption{Observed field parameters.}
\label{tbl:source_positions}

\begin{minipage}{\linewidth}
\begin{center}
\begin{tabular}{clcccc..}
\hline \hline
\multicolumn{1}{c}{Field}&\multicolumn{1}{c}{Field}&  \multicolumn{1}{c}{RA}&\multicolumn{1}{c}{Dec}&	\multicolumn{1}{c}{Map sensitivity}  & \multicolumn{1}{c}{Number}& \multicolumn{1}{c}{\vlsr}& \multicolumn{1}{c}{Distance}\\

\multicolumn{1}{c}{id}&\multicolumn{1}{c}{name}&  \multicolumn{1}{c}{(J2000)}&\multicolumn{1}{c}{(J2000)} &\multicolumn{1}{c}{(K \kms)} &\multicolumn{1}{c}{of clumps}&\multicolumn{1}{c}{(\kms)} &\multicolumn{1}{c}{(kpc)} \\

\hline
1	&	G010.300$-$00.143	&	18:08:54.88	&	$-$20:05:51.1	&	0.82	&	5	&	12.9	&	2.2	\\
2	&	G010.315$-$00.251	&	18:09:20.89	&	$-$20:08:10.1	&	0.66	&	3	&	27.7	&	4.0	\\
3	&	G010.472+00.033	&	18:08:36.89	&	$-$19:51:40.1	&	0.62	&	2	&	67.1	&	8.5	\\
4	&	G010.648$-$00.342	&	18:10:22.41	&	$-$19:53:19.5	&	1.98	&	3	&	-5.0	&	4.9	\\
5	&	G010.990$-$00.075	&	18:10:04.88	&	$-$19:27:37.4	&	0.54	&	1	&	29.7	&	3.7	\\
6	&	G011.112$-$00.395	&	18:11:31.25	&	$-$19:30:29.1	&	0.64	&	3	&	-1.0	&	4.9	\\
7	&	G011.501$-$01.482	&	18:16:21.87	&	$-$19:41:10.5	&	0.56	&	2	&	10.6	&	1.6	\\
8	&	G011.902$-$00.135	&	18:12:09.88	&	$-$18:41:25.4	&	0.67	&	3	&	36.4	&	12.8	\\
9	&	G011.924$-$00.613	&	18:13:58.89	&	$-$18:53:59.2	&	0.78	&	1	&	35.2	&	3.9	\\
10	&	G012.432$-$01.111	&	18:16:51.25	&	$-$18:41:29.4	&	0.36	&	2	&	\multicolumn{1}{c}{$\cdots$}	&	4.1	\\
11	&	G012.887+00.494	&	18:11:50.25	&	$-$17:31:27.2	&	0.54	&	3	&	33.5	&	2.4	\\
12	&	G013.197$-$00.122	&	18:14:43.63	&	$-$17:32:52.1	&	1.47	&	2	&	53.1	&	4.6	\\
13	&	G013.330$-$00.034	&	18:14:40.26	&	$-$17:23:18.1	&	0.42	&	2	&	54.7	&	4.7	\\
14	&	G013.656$-$00.595	&	18:17:23.25	&	$-$17:22:09.2	&	0.51	&	1	&	47.0	&	4.3	\\
15	&	G013.873+00.282	&	18:14:35.63	&	$-$16:45:39.0	&	0.53	&	1	&	48.5	&	4.4	\\
16	&	G014.330$-$00.639	&	18:18:53.25	&	$-$16:47:46.3	&	0.67	&	1	&	22.0	&	1.1	\\
17	&	G014.433$-$00.697	&	18:19:18.26	&	$-$16:43:57.4	&	0.60	&	1	&	16.9	&	1.1	\\
18	&	G014.608+00.019	&	18:17:01.10	&	$-$16:14:21.0	&	0.48	&	2	&	25.1	&	2.9	\\
19	&	G016.711+01.318	&	18:16:25.27	&	$-$13:46:18.3	&	0.48	&	1	&	20.0	&	2.2	\\
20	&	G016.804+00.817	&	18:18:25.26	&	$-$13:55:38.4	&	0.51	&	1	&	\multicolumn{1}{c}{$\cdots$}	&	2.1	\\
21	&	G016.927+00.961	&	18:18:08.28	&	$-$13:45:03.1	&	0.70	&	1	&	20.9	&	2.2	\\
22	&	G017.451+00.813	&	18:19:41.65	&	$-$13:21:34.0	&	0.41	&	1	&	21.3	&	2.2	\\
23	&	G017.636+00.156	&	18:22:26.27	&	$-$13:30:21.3	&	0.59	&	2	&	22.1	&	2.2	\\
24	&	G018.301$-$00.387	&	18:25:41.27	&	$-$13:10:20.4	&	0.58	&	1	&	32.3	&	3.0	\\
25	&	G018.461+00.001	&	18:24:35.27	&	$-$12:50:58.4	&	0.62	&	1	&	52.4	&	12.1	\\
26	&	G018.606$-$00.071	&	18:25:07.64	&	$-$12:45:19.1	&	0.61	&	3	&	46.0	&	3.7	\\
27	&	G018.662+00.030	&	18:24:52.00	&	$-$12:39:28.9	&	0.56	&	1	&	80.9	&	10.8	\\
28	&	G018.846$-$00.558	&	18:27:21.06	&	$-$12:46:11.3	&	0.87	&	2	&	65.2	&	4.5	\\
29	&	G019.078$-$00.285	&	18:26:48.28	&	$-$12:26:12.2	&	0.62	&	1	&	66.0	&	4.5	\\
30	&	G019.756$-$00.130	&	18:27:32.00	&	$-$11:45:54.9	&	0.53	&	2	&	60.2	&	4.3	\\
31	&	G019.885$-$00.535	&	18:29:14.64	&	$-$11:50:21.2	&	0.59	&	1	&	43.2	&	3.5	\\
32	&	G019.923$-$00.258	&	18:28:19.00	&	$-$11:40:36.8	&	0.63	&	1	&	64.7	&	4.5	\\
33	&	G020.747$-$00.074	&	18:29:12.64	&	$-$10:51:41.1	&	0.60	&	3	&	56.0	&	4.2	\\

\hline\\
\end{tabular}\\
\end{center}
\end{minipage}
\end{table*}

\setcounter{table}{0}

\begin{table*}

\caption{Continued.}

\begin{minipage}{\linewidth}
\begin{center}
\begin{tabular}{clcccc..}
\hline \hline
\multicolumn{1}{c}{Field}&\multicolumn{1}{c}{Field}&  \multicolumn{1}{c}{RA}&\multicolumn{1}{c}{Dec}&	\multicolumn{1}{c}{Map sensitivity}  & \multicolumn{1}{c}{Number}& \multicolumn{1}{c}{\vlsr}& \multicolumn{1}{c}{Distance}\\

\multicolumn{1}{c}{id}&\multicolumn{1}{c}{name}&  \multicolumn{1}{c}{(J2000)}&\multicolumn{1}{c}{(J2000)} &\multicolumn{1}{c}{(K \kms)} &\multicolumn{1}{c}{of clumps}&\multicolumn{1}{c}{(\kms)} &\multicolumn{1}{c}{(kpc)} \\

\hline
34	&	G021.373$-$00.241	&	18:30:59.63	&	$-$10:23:01.2	&	0.59	&	2	&	90.8	&	10.3	\\
35	&	G022.350+00.070	&	18:31:42.64	&	$-$09:22:26.0	&	0.62	&	1	&	84.2	&	5.2	\\
36	&	G022.414+00.315	&	18:30:56.99	&	$-$09:12:16.0	&	0.61	&	3	&	84.2	&	5.2	\\
37	&	G023.708+00.172	&	18:33:52.99	&	$-$08:07:21.0	&	0.60	&	3	&	113.0	&	9.1	\\
38	&	G024.183+00.120	&	18:34:57.01	&	$-$07:43:28.8	&	0.63	&	3	&	54.5	&	3.8	\\
39	&	G025.649+01.047	&	18:34:21.36	&	$-$05:59:46.6	&	0.67	&	2	&	42.2	&	3.1	\\
40	&	G025.716+00.046	&	18:38:03.36	&	$-$06:23:49.6	&	0.58	&	1	&	99.2	&	9.5	\\
41	&	G025.815$-$00.168	&	18:39:00.35	&	$-$06:24:28.7	&	0.60	&	1	&	93.7	&	5.0	\\
42	&	G027.269+00.147	&	18:40:33.35	&	$-$04:58:15.7	&	0.71	&	2	&	31.6	&	12.8	\\
43	&	G028.199$-$00.049	&	18:42:58.00	&	$-$04:14:00.9	&	0.58	&	2	&	98.2	&	5.9	\\
44	&	G028.293$-$00.377	&	18:44:18.35	&	$-$04:17:59.6	&	0.70	&	2	&	48.8	&	11.6	\\
45	&	G028.337+00.113	&	18:42:38.35	&	$-$04:02:10.7	&	0.84	&	1	&	81.0	&	5.0	\\
46	&	G029.596$-$00.615	&	18:47:32.34	&	$-$03:14:56.8	&	0.76	&	2	&	76.6	&	4.8	\\
47	&	G030.877+00.056	&	18:47:29.35	&	$-$01:48:09.6	&	0.56	&	2	&	74.5	&	4.9	\\
48	&	G031.271+00.061	&	18:48:11.35	&	$-$01:26:59.7	&	0.80	&	2	&	109.0	&	4.9	\\
49	&	G031.406+00.299	&	18:47:35.35	&	$-$01:13:15.6	&	0.56	&	1	&	97.6	&	4.9	\\
50	&	G032.052+00.068	&	18:49:35.35	&	$-$00:45:08.8	&	0.81	&	1	&	95.3	&	4.9	\\
51	&	G033.397$-$00.001	&	18:52:17.35	&	+00:24:48.2	&	0.55	&	1	&	103.9	&	7.1	\\
52	&	G033.913+00.109	&	18:52:50.35	&	+00:55:23.1	&	0.48	&	1	&	107.5	&	7.1	\\
53	&	G034.407+00.231	&	18:53:18.35	&	+01:25:06.2	&	0.68	&	1	&	57.9	&	3.8	\\
54	&	G035.196$-$00.744	&	18:58:12.99	&	+01:40:29.9	&	0.50	&	2	&	33.9	&	2.2	\\
55	&	G035.463+00.140	&	18:55:33.35	&	+02:18:58.3	&	0.53	&	1	&	74.1	&	8.8	\\
56	&	G037.554+00.201	&	18:59:09.99	&	+04:12:13.9	&	0.48	&	1	&	85.1	&	6.7	\\
57	&	G043.180$-$00.520	&	19:12:09.36	&	+08:52:11.3	&	0.49	&	2	&	57.5	&	8.2	\\
58	&	G043.306$-$00.213	&	19:11:17.36	&	+09:07:26.3	&	0.48	&	1	&	59.6	&	4.4	\\
59	&	G045.462+00.049	&	19:14:24.36	&	+11:09:20.1	&	0.56	&	1	&	62.1	&	6.7	\\
60	&	G048.989$-$00.301	&	19:22:26.36	&	+14:06:33.1	&	0.55	&	2	&	71.0	&	5.6	\\
61	&	G052.204+00.724	&	19:25:00.27	&	+17:25:34.8	&	0.54	&	3	&	0.5	&	\multicolumn{1}{c}{$\cdots$}	\\
62	&	G053.604+00.015	&	19:30:25.88	&	+18:19:06.5	&	0.52	&	2	&	25.3	&	1.9	\\
63	&	G058.468+00.437	&	19:38:57.00	&	+22:46:32.1	&	0.53	&	2	&	36.4	&	4.4	\\
64	&	G059.782+00.075	&	19:43:08.62	&	+23:44:18.9	&	0.50	&	1	&	22.3	&	2.2	\\
65	&	G075.766+00.358	&	20:21:37.30	&	+37:26:20.9	&	0.54	&	2	&	\multicolumn{1}{c}{$\cdots$}	&	1.4	\\
66	&	G078.977+00.363	&	20:31:09.48	&	+40:03:30.3	&	0.49	&	1	&	\multicolumn{1}{c}{$\cdots$}	&	1.4	\\
\hline\\
\end{tabular}\\
\end{center}
\end{minipage}
\end{table*}

\setlength{\tabcolsep}{6pt}

The Red MSX Source (RMS; \citealt{lumsden2013}) survey has used a combination of MSX and 2MASS point source catalogues to identify an unprecedented sample of candidate embedded massive young stars. A multi-wavelength campaign of follow-up observations (\citealt{urquhart_radio_south, mottram2007, urquhart_13co_north, urquhart_13co_south,urquhart_radio_north,urquhart2009_h2o, urquhart2011, Cooper2013}) has led to the identification of a combination of $\sim$1300 massive young stellar objects (MYSOs) and compact \hii\ regions (\citealt{lumsden2013}). This is the largest and most well-characterised sample yet compiled, and is an order of magnitude larger than previous catalogues. In a recent study (\citealt{urquhart2014b}), we investigated the bulk properties of these massive star forming regions using the \submm\  dust emission traced by the ATLASGAL survey (870\,\mum; \citealt{schuller2009}). We extracted clumps parameters by matching the positions of the RMS catalogue with the ATLASGAL Compact Source Catalogue (CSC; \citealt{contreras2013,urquhart2014c}). This study found strong correlations between the clump mass and the bolometric luminosity of the embedded source, and between clump mass and radius (partial-Spearman correlation coefficients $r_{\rm{AB,C}}=0.64$ and 0.85, respectively; see Section\,\ref{sect:corrlations} for definition); however, the available continuum data was not sufficient to investigate these correlations in detail.

In this paper we use a flux-limited sample of RMS sources to investigate the properties of their natal clumps and the role that environmental conditions play in the formation of massive stars. We have mapped 66 massive star forming regions in the lowest excitation ammonia inversion transitions (i.e., NH$_3$ (J,K) = (1,1) and (2,2)). These transitions are sensitive to cold ($\sim$10-40\,K; \citealt{ho1983,mangum1992}) and dense ($>$10$^4$\,cm$^{-3}$; \citealt{rohlfs2004}) gas and NH$_3$ does not deplete from the gas phase at high densities ($<$10$^6$\, cm$^{-3}$; \citealt{bergin1997}), which makes them an excellent probe of the dense gas properties of these clumps (\citealt{rydbeck1977,ho1983}). Furthermore, the ratio of the hyperfine components of the  NH$_3$ (1,1) can be used to estimate the optical depth and relative populations of different levels, which can be used to estimate the rotation and kinetic temperature of the gas. 

Ammonia is therefore one of the most useful high-density molecular gas tracers, and has been widely used to study the properties of massive star forming regions. These studies have sampled a range of evolutionary stages such as infrared dark clouds (IRDCs; \citealt{pillai2006, ragan2011, chira2013}), massive submillimetre clumps (\citealt{dunham2011, wienen2012}) and embedded mid-infrared bright sources (\citealt{urquhart2011}; hereafter Paper\,I). These have found that higher temperatures and larger line widths tend to be associated with more evolved protostars, which is generally attributed to increased feedback from the embedded objects.

We used these observations to trace the temperature and density structure of these massive star forming environments and probe the gas kinematics, which in turn provides an insight into the bulk motion of the gas and level of turbulence, and through virial analysis, an estimate of the global stability of these clumps. These data are combined with archival infrared and submillimetre data to investigate the  relationship between the embedded MYSOs and compact \hii\ regions and their natal clumps, and to evaluate the influence of the local environment on the structure and evolution of the clumps.

The combined dataset is extremely rich and will form the foundation for several studies. In this paper, the second in the series, we present the results of ammonia mapping observations and a statistical analysis of the mean clump parameters. Subsequent papers will focus on the relationship between the larger scale environment and the dense clumps, the density and temperature structure of the clumps, and detailed studies of more complicated regions. The structure of the paper is as follows: in Sect.\,\ref{sect:obs} we describe the observational set-up, data reduction and source extraction procedures, as well as the spectral line analysis. In Sect.\,\ref{sect:results} we present an overview of our results, derive physical properties for the clumps and compare their distributions with respect to their embedded protostellar content, and evaluate the impact of the external environment. We investigate the correlation between different derived properties in Sect.\,\ref{sect:discussion}. We summarise our results and present our conclusions in  Sect.\,\ref{sect:summary_conclusions}.

\section{Observations and data analysis}
\label{sect:obs}

\subsection{GBT K-band Mapping Observations}

\begin{figure*}\begin{center}
\includegraphics[width=0.48\textwidth, trim= 0 0 0 0]{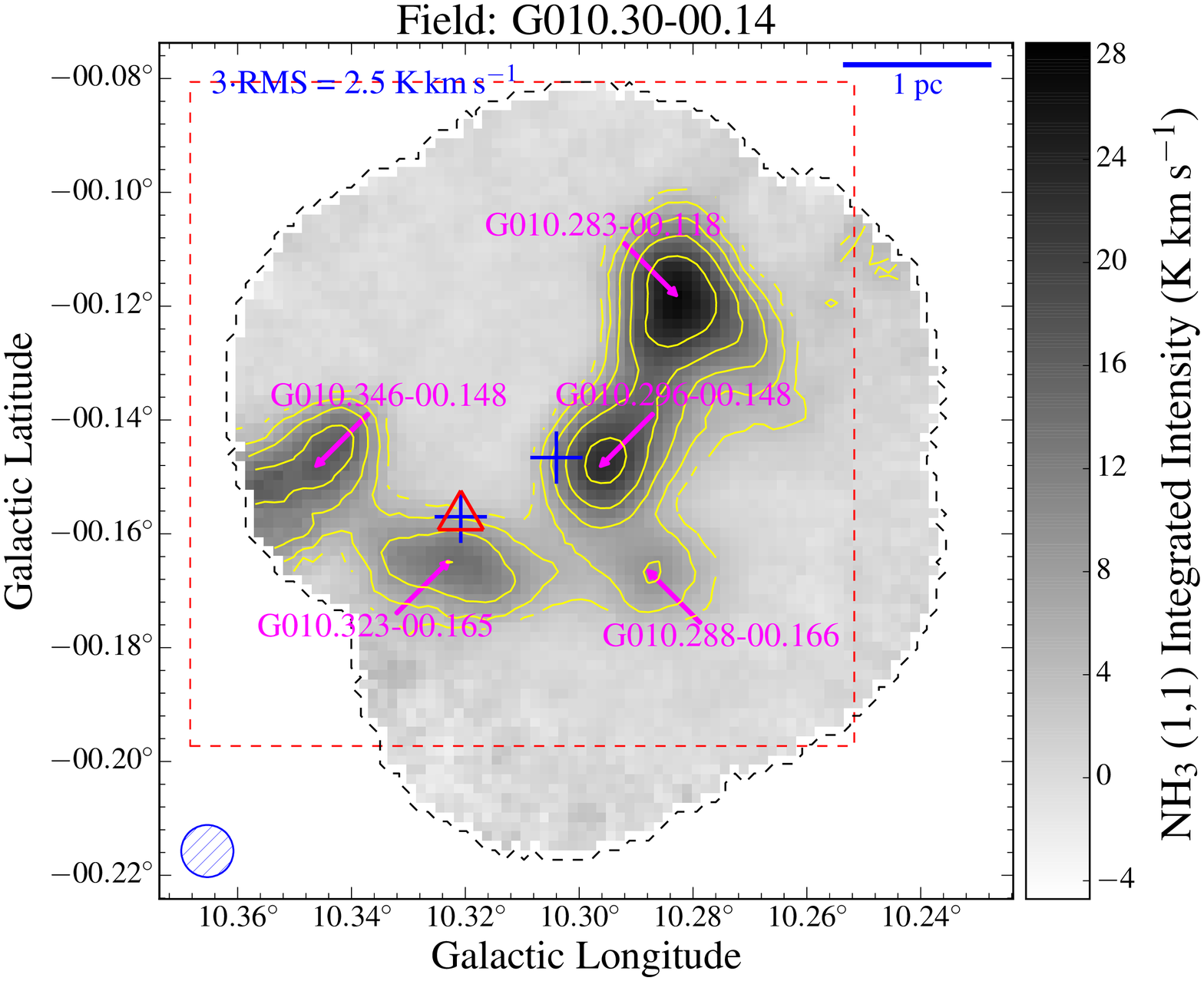}
\includegraphics[width=0.48\textwidth, trim= 0 0 0 0]{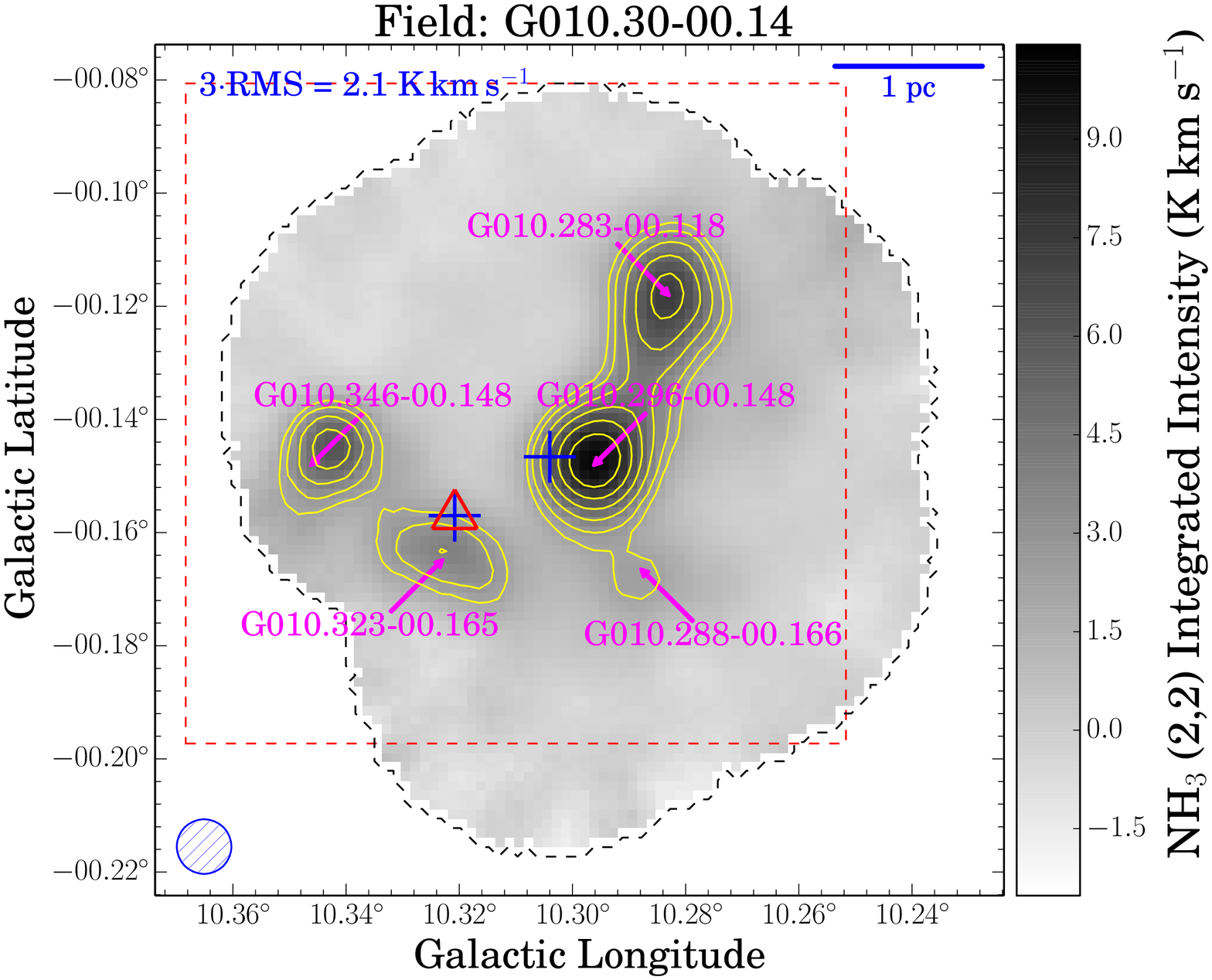}\\
\includegraphics[width=0.48\textwidth, trim= 0 0 0 0]{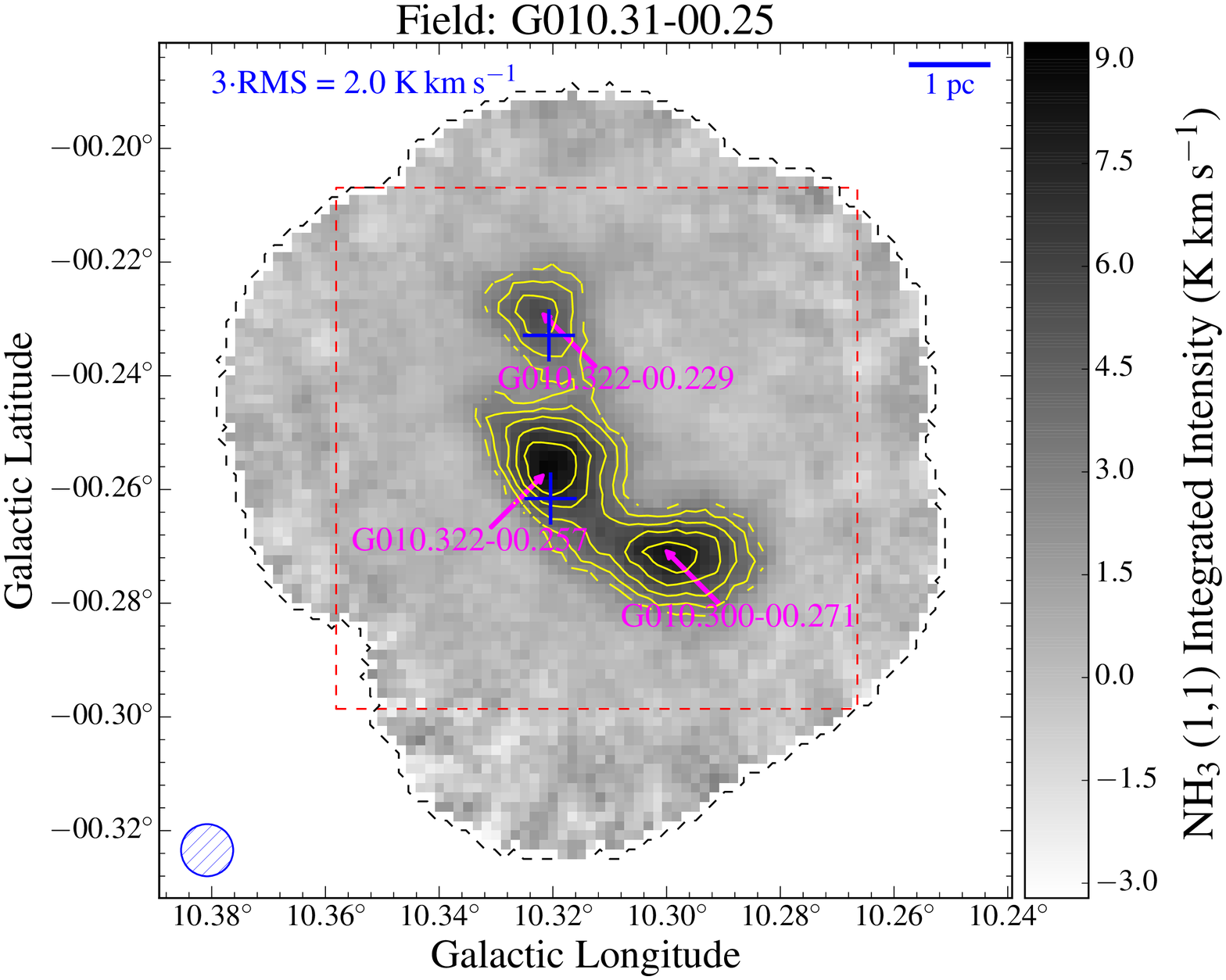}
\includegraphics[width=0.48\textwidth, trim= 0 0 0 0]{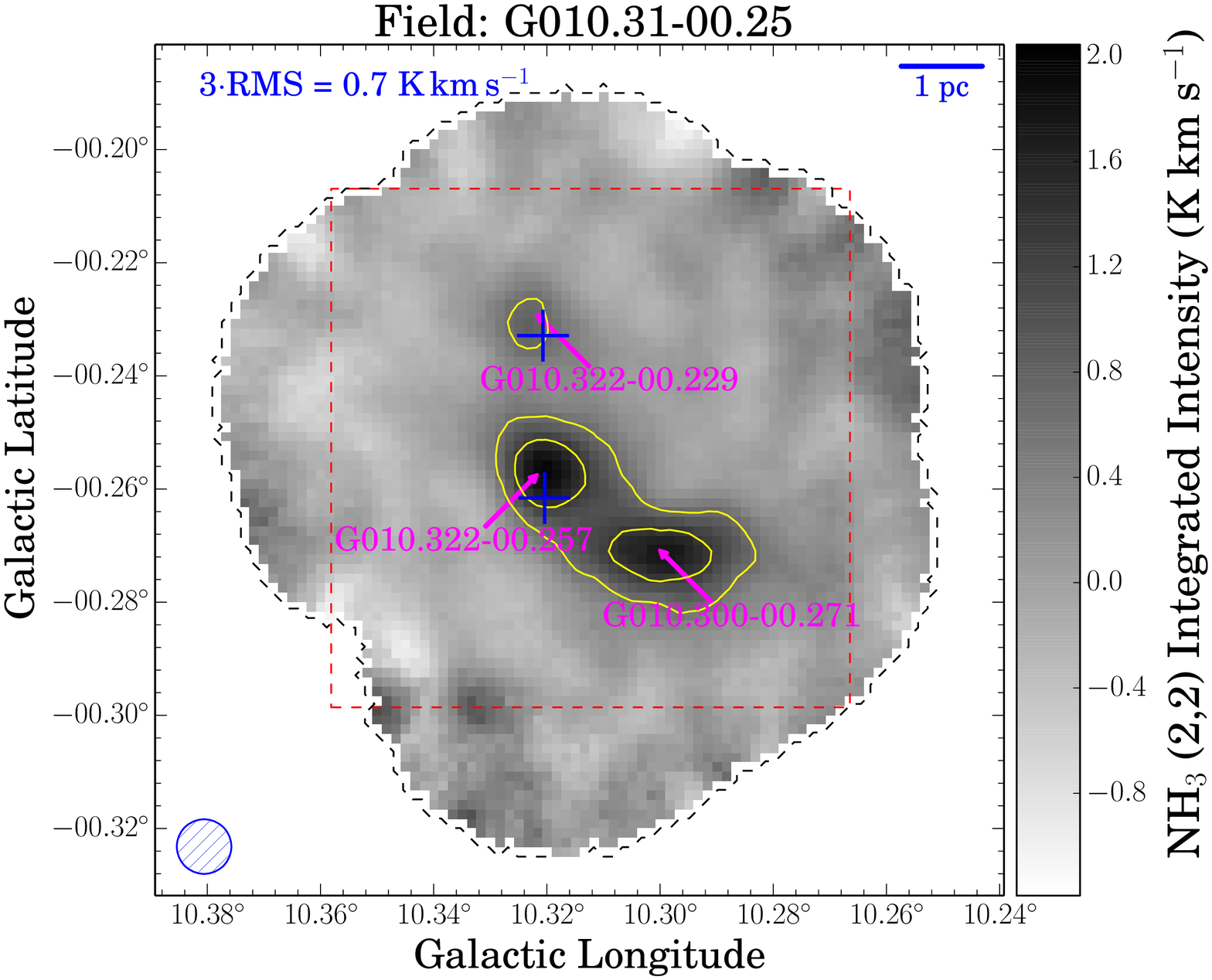}\\
\includegraphics[width=0.48\textwidth, trim= 0 0 0 0]{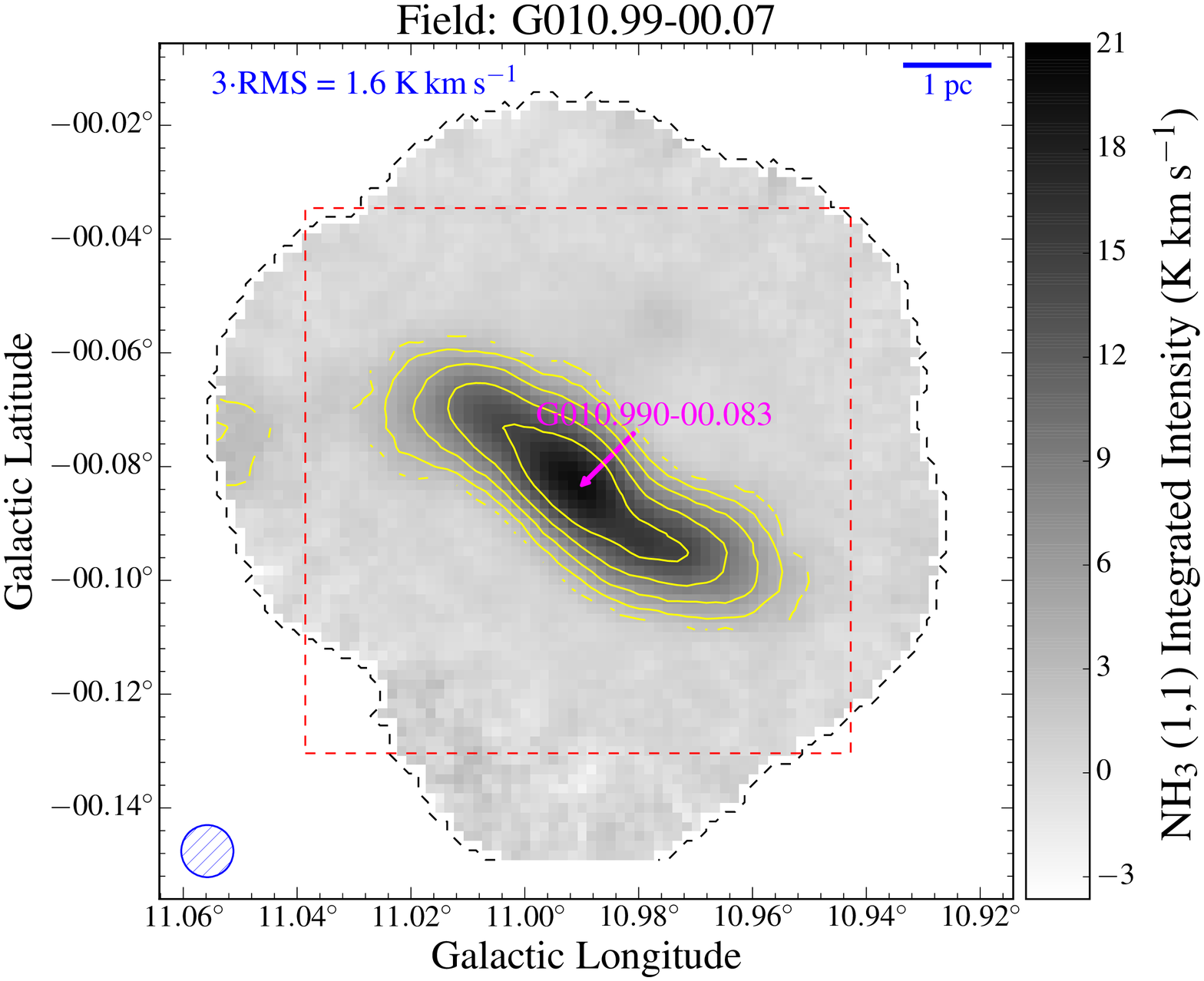}
\includegraphics[width=0.48\textwidth, trim= 0 0 0 0]{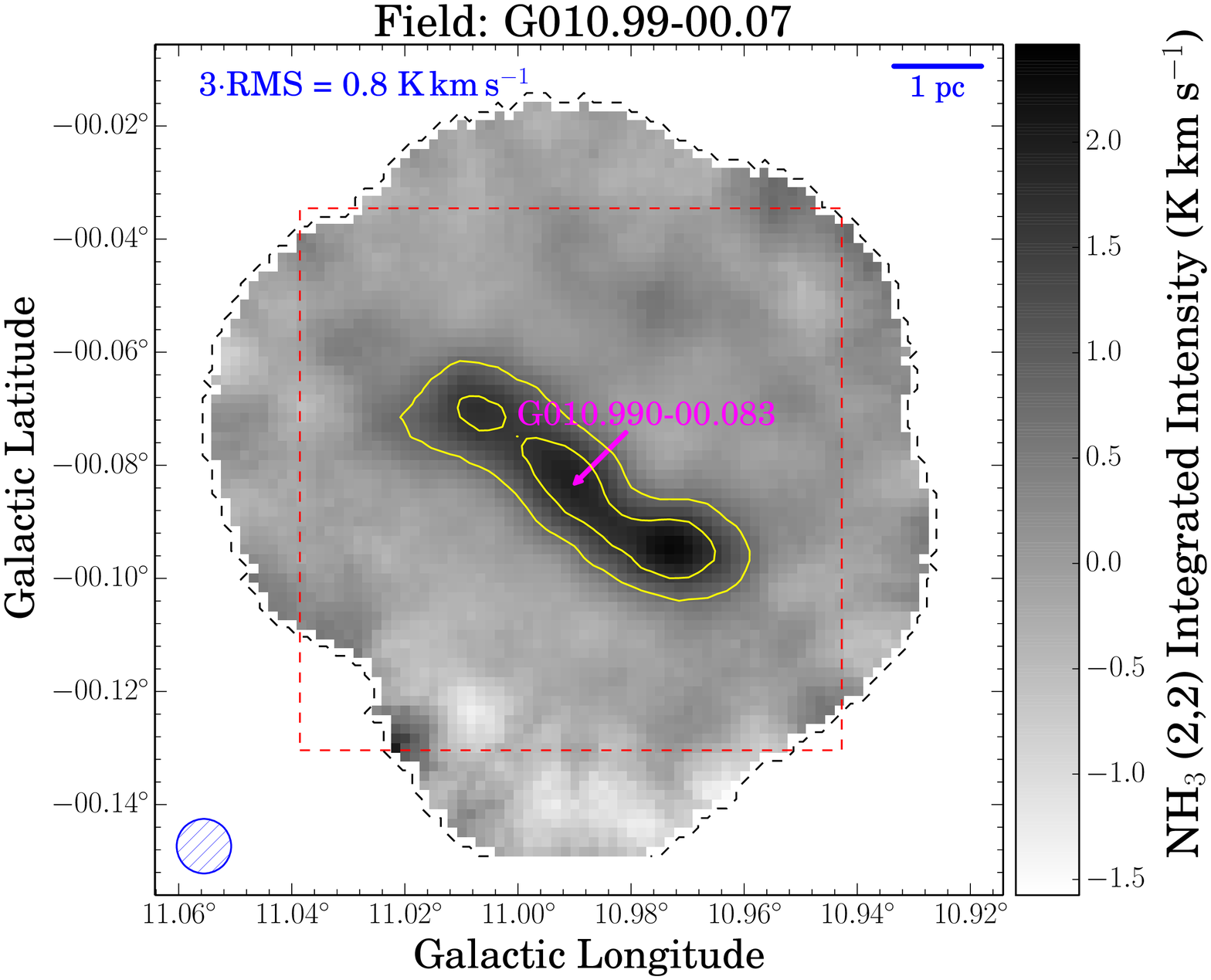}

\caption{\label{fig:example_emission_maps} Examples of the integrated NH$_3$ (1,1) and (2,2) emission maps obtained towards three fields are presented on the left and right panels, respectively. The NH$_3$  (1,1) emission is integrated over a velocity range of $\sim$50\,\kms\ in order to include the main line and four hyperfine components and is centred on the velocity of the central clump in each field. The NH$_3$  (1,1) emission is integrated over a velocity range of 3 times the standard deviation of the Gaussian fit to the peak profile. The greyscale and yellow contours show the distribution of the integrated emission. The positions of the MYSOs, \hii\ regions and MYSO/\hii\ regions are indicated by blue crosses, red triangles and purple squares, respectively. The dashed red lines  outline the regions we focus on in Figs.\,{\color{red}6} and {\color{red}9}. The contour levels start at 3$\sigma$ and increase in steps set by a dynamically determined power-law (see text for details). The 3$\sigma$ noise is given in the top left corner and a linear scale bar shown in the upper right corner provides an indication of the physical sizes of the clumps. The angular resolution of the GBT beam at this frequency is indicated by the blue hatched circle shown in the lower left corner of each map. }

\end{center}\end{figure*}

We have mapped 62 fields towards MYSOs and compact \hii\ regions identified by the RMS survey. A further 4 fields were included towards strong dust emission sources that are not associated with an RMS source to allow comparison between massive star forming (MSF) clumps and relatively quiescent clumps. These observations have been made using the K-Band Focal Plane Array (KFPA) on the National Radio Astronomy Observatory's\footnote{The National Radio Astronomy Observatory is a facility of the National Science Foundation operated under cooperative agreement by Associated Universities, Inc.} Green Bank Telescope (GBT). The observations were made in shared-risk time shortly after the KFPA was commissioned between March 2011 and February 2012  (Project Id.: GBT10C21).

The fields were selected using the results of a programme of targeted observations also made with the GBT towards $\sim$600 RMS sources (Paper\,I). This previous study detected NH$_3$ emission towards approximately 80\,per\,cent of the sources targeted. We have selected bright sources that have good detections in both ammonia transitions in order to provide the highest-sensitivity maps of the various parameters (e.g., \vlsr, line width, density and temperature) and trace their spatial distribution across these star forming regions. We chose to focus on sources located within the inner Galactic plane (i.e., $\ell < 60\degr$ and $|b| < 1\degr$) as this has been covered by a number of other surveys and ensures the availability of a wealth of complementary data. It was not always possible to meet these two criteria simultaneously and as a result a few fields were observed outside this region. 

The KFPA features seven 32\arcs\ beams in a hexagonal array with a central feed, with a nearest-neighbour spacing of 96\arcs. A 50\,MHz spectral bandpass was used to observe both of the ammonia (1,1) and (2,2) rotation inversion transitions simultaneously (at $\sim23.6945$ and $23.7226$\,GHz, respectively). The `Daisy' pattern was used to map a region $\sim$8\arcmin\ in diameter towards each target source. The petal-shaped scan trajectories produce fully-sampled maps within a radius of 3\farcm5 of map centre; beyond this point the integration time decreases, resulting in a decrease in signal-to-noise (SNR) ratios towards the map edges.  Maps were weighted by integration times during the data reduction in order to compensate for this decrease.

Sky subtraction was accomplished via off-source observations, and a noise diode was used to calibrate fluxes to the \Tant\ scale. The zenith atmospheric opacity was determined using weather models.\footnote{\url{http://www.gb.nrao.edu/~rmaddale/Weather.}} Time series data from the seven individual receivers in the array was processed through the GBT reduction pipeline.\footnote{ \url{https://safe.nrao.edu/wiki/pub/Kbandfpa/KfpaReduction/kfpaDataReduceGuide-11Dec01.pdf} see also \citet{morgan2014}.} Sky subtraction and calibration was performed by the pipeline, and spatial image cubes were produced. The weather conditions were stable for all observations, and the typical pointing corrections were found to be $\sim$4\arcsec. 

The reduced data cubes are approximately 8\arcmin\ in diameter and gridded  using 6\arcsec\ pixels. All maps were smoothed spatially with a 10\arcsec\ gaussian kernel, which results in a final image resolution of $\sim32\arcsec$, while the velocity axis was smoothed using a top-hat function to produce a resolution of $\sim$0.4\,\kms\,channel$^{-1}$. In a few cases where the observed fields overlapped the adjacent maps were mosaicked together using the \texttt{wcsmosaic} routine from the Starlink \Kappa\ suite to maximise the fully-sampled region.\footnote{\url{http://www.starlink.ac.uk/docs/sun95.htx/sun95.html}} In Fig.\,\ref{fig:example_emission_maps} we present a selection of integrated NH$_3$ (1,1) and (2,2) emission maps for 3 fields. The contour levels start at 3$\sigma$ and increase in steps set by a dynamically determined power-law of the form $D=3\times N^i+2$, where $D$ is the dynamic range of the \submm\ emission map (defined as the peak brightness divided by the local r.m.s. noise), $N$ is the number of contours used (6 in this case), and $i$ is the contour power-law index.  The lowest power-law index used was one, which results in linearly spaced contours starting at 3$\sigma$ and increasing in steps of 3$\sigma$ (see \citealt{thompson2006} for more details). The complete set of maps are provided in Fig.\,{\color{red}A1}. In Table\,\ref{tbl:source_positions} we give the names, centre coordinates, the noise in the integrated maps of the observed fields along with the distance and velocities of the target sources.

\subsection{Source extraction and structure analyses}
\label{sect:clump_structure}
\begin{table*}

\begin{center}\caption{The \FW\ source catalogue. The parameters given in this table have been obtained from the higher signal to noise NH$_3$ (1,1) integrated emission maps. The columns are as follows: (1) Field identification; (2) name derived from Galactic coordinates of the maximum intensity in the source; (3)-(4) right ascension  and declination in the J2000 coordinate system; (5)-(7) semi-major and semi-minor size and source position angle measured anti-clockwise from Galactic north; (8) aspect ratio; (9) effective radius of source; (10)-(12) peak and integrated flux densities and their associated uncertainties; (13) signal to noise ratio.}
\label{tbl:fw_parameters}

\begin{minipage}{\linewidth}
\begin{tabular}{clccrrr......}
  \hline \hline
   \multicolumn{1}{c}{Field} &\multicolumn{1}{c}{Clump}
  &  \multicolumn{1}{c}{RA} &  \multicolumn{1}{c}{Dec}&
  \multicolumn{1}{c}{$\sigma_{\rm{maj}}$} &  \multicolumn{1}{c}{$\sigma_{\rm{min}}$} &  \multicolumn{1}{c}{PA} &
  \multicolumn{1}{c}{Aspect} &  \multicolumn{1}{c}{$\theta_{\rm{R}}$} & \multicolumn{1}{c}{NH$_3$$_{\rm{peak}}$} & \multicolumn{1}{c}{$\Delta$ NH$_3$$_{\rm{peak}}$} & \multicolumn{1}{c}{NH$_3$$_{\rm{int}}$} & \multicolumn{1}{c}{SNR} \\
   
  \multicolumn{1}{c}{Id.} &\multicolumn{1}{c}{name}  &  \multicolumn{1}{c}{(J2000)} &
  \multicolumn{1}{c}{(J2000)} &  \multicolumn{1}{c}{($''$)}
  &  \multicolumn{1}{c}{($''$)} &  \multicolumn{1}{c}{($^{\circ}$)}&  \multicolumn{1}{c}{ratio}
  &  \multicolumn{1}{c}{($''$)}
  &\multicolumn{2}{c}{(K\,\kms\,beam$^{-1}$)} &  \multicolumn{1}{c}{(K\,\kms)}& \\
 
  \multicolumn{1}{c}{(1)} &  \multicolumn{1}{c}{(2)} &  \multicolumn{1}{c}{(3)} &  \multicolumn{1}{c}{(4)} &
  \multicolumn{1}{c}{(5)} &  \multicolumn{1}{c}{(6)} &  \multicolumn{1}{c}{(7)} &  \multicolumn{1}{c}{(8)} &
  \multicolumn{1}{c}{(9)} &  \multicolumn{1}{c}{(10)} &  \multicolumn{1}{c}{(11)}&  \multicolumn{1}{c}{(12)}&  \multicolumn{1}{c}{(13)} \\
  \hline

1	&	G010.283$-$00.118	&	18:08:54.88	&	$-$20:05:51.1	&	38	&	28	&	42	&	1.3	&	29.8	&	29.2	&	0.82	&	158.2	&	35.4	\\
1	&	G010.288$-$00.166	&	18:08:54.88	&	$-$20:05:51.1	&	22	&	16	&	106	&	1.4	&	12.0	&	11.3	&	0.82	&	27.1	&	13.7	\\
1	&	G010.296$-$00.148	&	18:08:54.88	&	$-$20:05:51.1	&	26	&	20	&	134	&	1.3	&	18.3	&	25.3	&	0.82	&	90.0	&	30.7	\\
1	&	G010.323$-$00.165	&	18:08:54.88	&	$-$20:05:51.1	&	34	&	21	&	79	&	1.6	&	21.9	&	15.4	&	0.82	&	66.6	&	18.7	\\
1	&	G010.346$-$00.148	&	18:08:54.88	&	$-$20:05:51.1	&	21	&	18	&	86	&	1.2	&	13.9	&	19.6	&	0.82	&	54.5	&	23.8	\\
2	&	G010.300$-$00.271	&	18:09:20.89	&	$-$20:08:10.1	&	29	&	19	&	89	&	1.6	&	18.2	&	9.4	&	0.66	&	34.3	&	14.3	\\
2	&	G010.322$-$00.229	&	18:09:20.89	&	$-$20:08:10.1	&	22	&	22	&	94	&	1.0	&	17.6	&	6.4	&	0.66	&	17.8	&	9.8	\\
2	&	G010.322$-$00.257	&	18:09:20.89	&	$-$20:08:10.1	&	26	&	22	&	44	&	1.2	&	19.2	&	9.8	&	0.66	&	34.5	&	15.0	\\
3	&	G010.440+00.003	&	18:08:36.89	&	$-$19:51:40.1	&	17	&	15	&	140	&	1.1	&	7.6	&	4.1	&	0.62	&	7.0	&	6.6	\\
3	&	G010.474+00.028	&	18:08:36.89	&	$-$19:51:40.1	&	31	&	23	&	80	&	1.3	&	22.7	&	61.4	&	0.62	&	259.9	&	98.9	\\
  \hline
\end{tabular}\\
\end{minipage}
Notes: Only a small portion of the data is provided here, the full table is only  available in electronic form at the CDS via anonymous ftp to cdsarc.u-strasbg.fr (130.79.125.5) or via http://cdsweb.u-strasbg.fr/cgi-bin/qcat?J/A+A/.
\end{center}
\end{table*}

\setlength{\tabcolsep}{6pt}


The emission maps reveal a mixture of isolated clumps with rather simple elliptical distributions and a smaller number of more irregular multi-peaked morphologies that are likely to consist of two or more distinct clumps. We have used the \FW\ source extraction algorithm (\citealt{berry2015}) to identify clumps and determine their properties in a consistent manner.\footnote{\FW\ is part of the Starlink-CUPID  software suite and more details can be found here: \url{http://docs.jach.hawaii.edu/star/sun255.htx/sun255.html}.} This algorithm has been applied to the integrated NH$_3$ (1,1) emission maps as they have the highest sensitivity and will most accurately trace the full extent of the clumps. 

We set a 3$\sigma$\ detection threshold and required that all sources identified consist of more than 30 pixels in order to reject sources with fewer pixels than the beam integral as these are likely to be spurious detections. We also excluded clumps located towards the edges of the fields as these tend to have lower SNRs, and it is likely that the emission is not entirely captured in the maps leading to larger uncertainties. Multiple clumps are found in 35 of the 66 observed fields (eleven fields have 3 clumps, and one has 5 clumps; the number of clumps identified in each field is given in Table\,\ref{tbl:source_positions}).  In the majority of cases, clumps found in the same field have similar radial velocities (typically differing by less than a few \kms) and are therefore likely to be associated with the same giant molecular cloud (GMC) complex and can be assumed to be at a similar distance. There are only two fields where the clumps are at significantly different velocities (G024.18+00.12 and G028.29$-$00.38). We will discuss these two fields in more detail in Sect.\,\ref{sect:distance}. In fields where two or more clumps are identified, we find that the RMS source is nearly always associated with the brightest and most prominent of the clumps identified.  

In total, \clumpnum\ clumps have been detected and these are identified by labels on the emission maps presented in Fig.\,\ref{fig:example_emission_maps} and their parameters are given in Table\,\ref{tbl:fw_parameters}. The source names are based on the Galactic coordinates of the peak flux position, which are given in Cols.\,2 and 3. The source sizes describe an ellipse with semi-major and semi-minor axis lengths and position angle; these are determined from the standard deviation of the pixel co-ordinate values about the centroid position weighted by the pixel values (see CUPID manual for more details). 

\begin{figure}
\begin{center}

\includegraphics[width=0.45\textwidth, trim= 0 0 0 0]{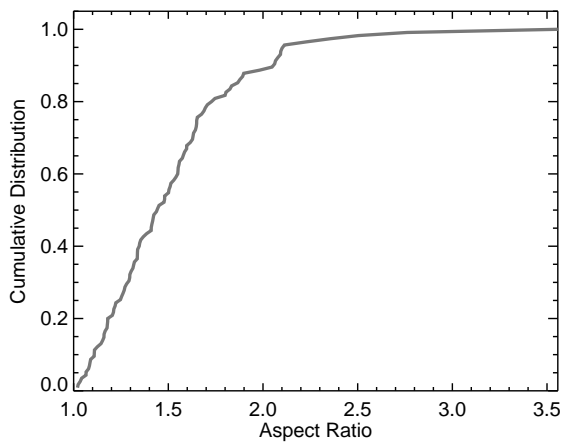}
\caption{\label{fig:aspect_ratio} Cumulative distribution function of the aspect ratio of all of the clumps (i.e., $\sigma_{\rm{maj}}/\sigma_{\rm{min}}$). } 

\end{center}
\end{figure}

The ratio of the semi-major and semi-minor axes is used to estimate the aspect ratio of each source, which in turn is used to classify each clump into one of two groups: spheroidal and filamentary structures. In Fig.\,\ref{fig:aspect_ratio} we present the cumulative distribution of the aspect ratio for all of the clumps. This plot shows a break at an aspect ratio of approximately 1.8 and we used this value to distinguish between spheroids and filaments, with the former having smaller aspect ratios. This results in 94 spheroidal and 21 filamentary structures. The fraction of filaments identified is significantly lower than has been found for several other high resolution ammonia surveys (i.e., almost ubiquitous by \citealt{lu2014} at 3-40\arcs\ and over 50\,per\,cent by \citealt{ragan2011} at 4 and 8\arcs\ resolution). It is therefore expected that many of the spheroidally structured clumps will turn out to be filamentary at higher resolution, but the sources classified as filamentary are unlikely to be affected and so this may prove a useful distinction. We investigated the possibility that sources that were identified as filamentary were more likely to be closer and thus more easily resolved.  There was, however, no statistically significant difference between filamentary and spheroidal subsample distances, indicating that filaments can have a large range of size scales.

\begin{figure}
\begin{center}

\includegraphics[width=0.45\textwidth, trim= 0 0 0 0]{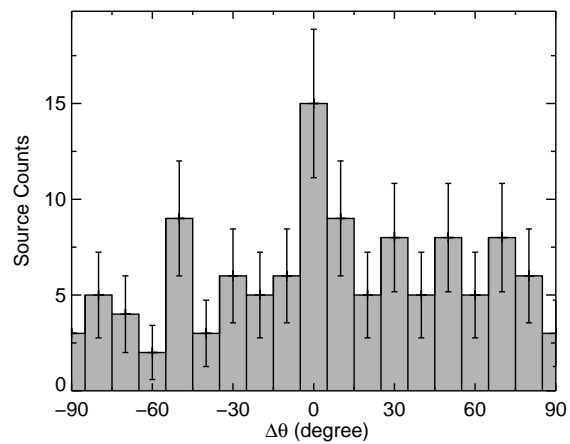}
\caption{\label{fig:pa_hist} Difference between the clump position angles and the Galactic mid-plane. The bin size is 10\degr\ and the errors are estimated using \poi\ statistics.} 

\end{center}
\end{figure}

The clump position angle is given as anti-clockwise from Galactic north. In Fig.\,\ref{fig:pa_hist} we show the difference between the clump position angles and the  Galactic mid-plane. It is clear from this plot that the clumps are preferentially aligned along the Galactic mid-plane, suggesting that Galactic rotation, magnetic field, or Galactic shear may be influencing their structure, although the last of these is less likely (\citealt{dib2012}). We find no correlation between the clump orientation and the angular separation from the Galactic mid-plane (i.e., $|b|$). \citet{li2014} present evidence that the B-field in cores is aligned with the local large-scale field in the diffuse ISM and that the latter is in turn aligned with the Galactic plane. The orientation of the elongation of the cores is then aligned either parallel or perpendicular to the direction of the B-field. It therefore seems likely that the preferred orientation of the clumps is due to the influence of large scale magnetic fields, or that perhaps both are influenced by Galactic kinematics.

The source sizes are only determined from pixels above the detection threshold and are therefore likely to underestimate the true source sizes. This can some times result in sizes that are smaller than the FWHM of the telescope beam as the moment method of determining sizes truncates the low-significance emission in the clump's outer envelope (cf \citealt{rosolowsky2010} sect. 7.4). We estimate the angular radius ($\theta_{\rm{R}}$) from the geometric mean of the deconvolved major and minor axes, multiplied by a factor $\eta$ that relates the \rms\ size of the emission distribution of the source to its angular radius (Eqn.\,6 of \citealt{rosolowsky2010}):

\begin{equation}
\label{radius}
\theta_{\rm{R}}= \eta \left[(\sigma_{\rm{maj}}^2-\sigma_{\rm{bm}}^2)(\sigma_{\rm{min}}^2-\sigma_{\rm{bm}}^2)\right]^{1/4},
\end{equation}

\noindent where $\sigma_{bm}$ is the \rms\ size of the beam (i.e.,
$\sigma_{bm}=\theta_{\mathrm{FWHM}}/\sqrt{8\ln 2} \simeq 13''$). Following \citet{rosolowsky2010} we adopt a value for $\eta$ of 2.4 to estimate the  effective radius of each source. We are able to estimate the effective radius for all but eight clumps: these eight have at least one of their axes smaller than the beamwidth; however, all of these are relatively weak (SNR $\sim$8) and it is likely that the observations do not trace the full extent of their extended envelopes.

\begin{figure}
\begin{center}

\includegraphics[width=0.45\textwidth, trim= 0 0 0 0]{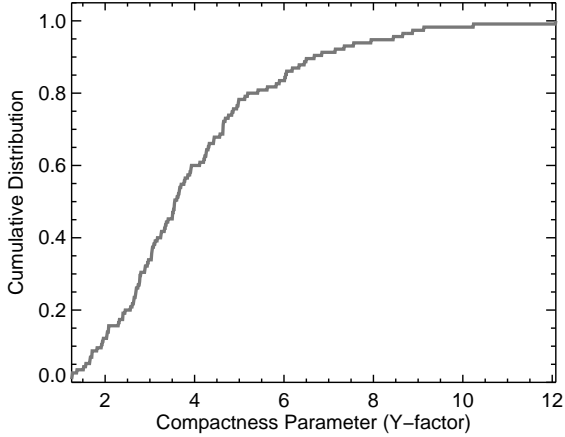}
\caption{\label{fig:yfactor_hist} Cumulative distribution function of the $Y$-factor of the clumps (i.e., $S_\nu({\rm{int}})/S_\nu({\rm{peak}})$). } 

\end{center}
\end{figure}

The peak flux is directly obtained from \FW; however, the integrated emission is the sum of all pixels above the threshold and does not take account of the beam size. To obtain a value for the total emission we have divided the derived flux by the beam integral (i.e., $1.133 \times FWHM^2 \simeq 32.2$ pixels). In Fig.\,\ref{fig:yfactor_hist} we present the cumulative distribution of the $Y$-factors: this is the ratio of the integrated and peak fluxes and gives an estimate of how centrally concentrated the structures are. The mean and median values for this sample of clumps are $4.04\pm0.19$ and 3.56, respectively: these are smaller than the values derived from the dust emission ($5.80\pm0.11$ and 4.66, \citealt{urquhart2014b}), and are likely a result of the larger beam used for these observations. However, this does indicate that these clumps are centrally condensed.

\subsection{Ammonia Spectral Line Analysis}
\label{sect:line_analysis}

\begin{figure}
\begin{center}

\includegraphics[width=0.45\textwidth, trim= 0 0 0 0]{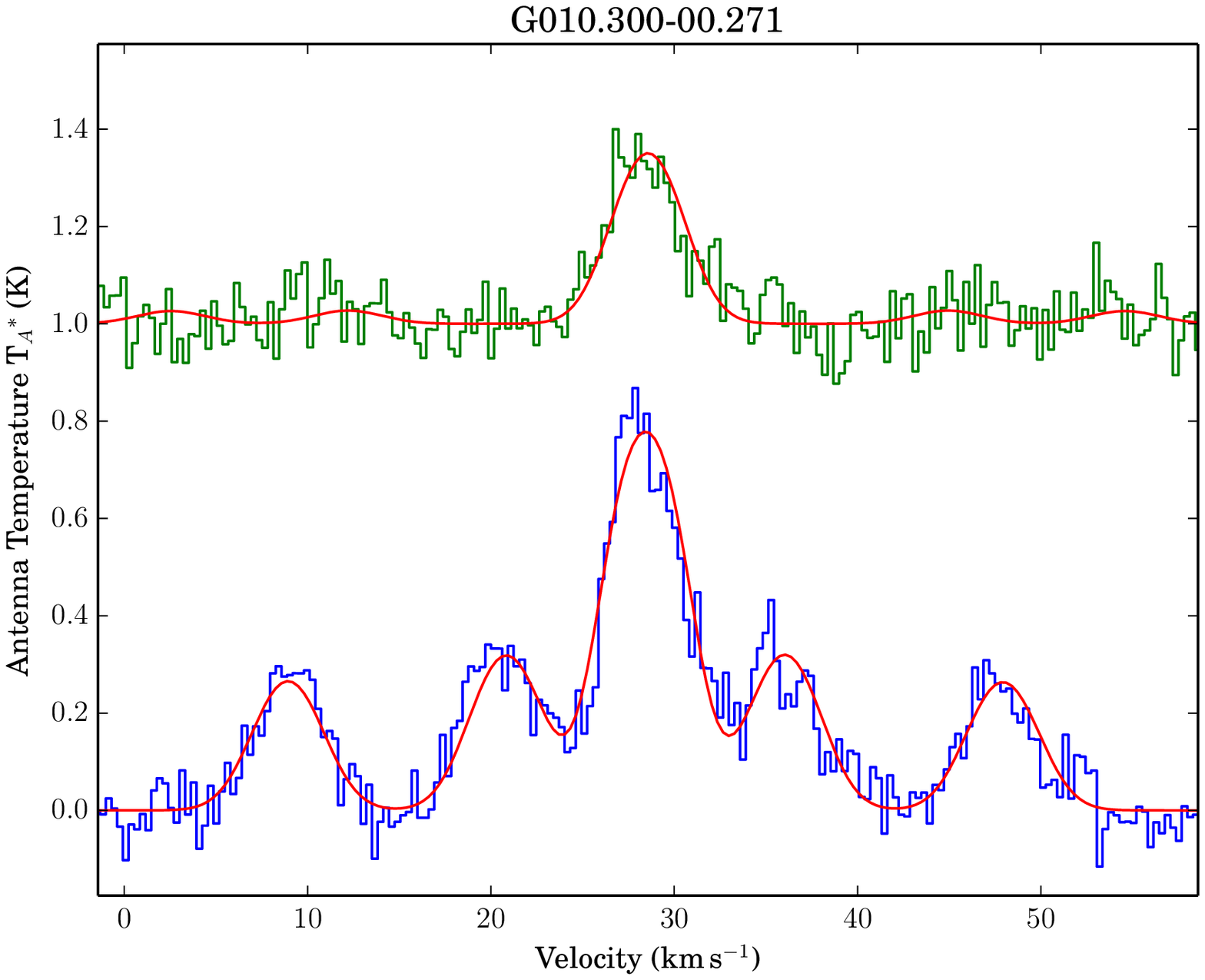}
\includegraphics[width=0.45\textwidth, trim= 0 0 0 0]{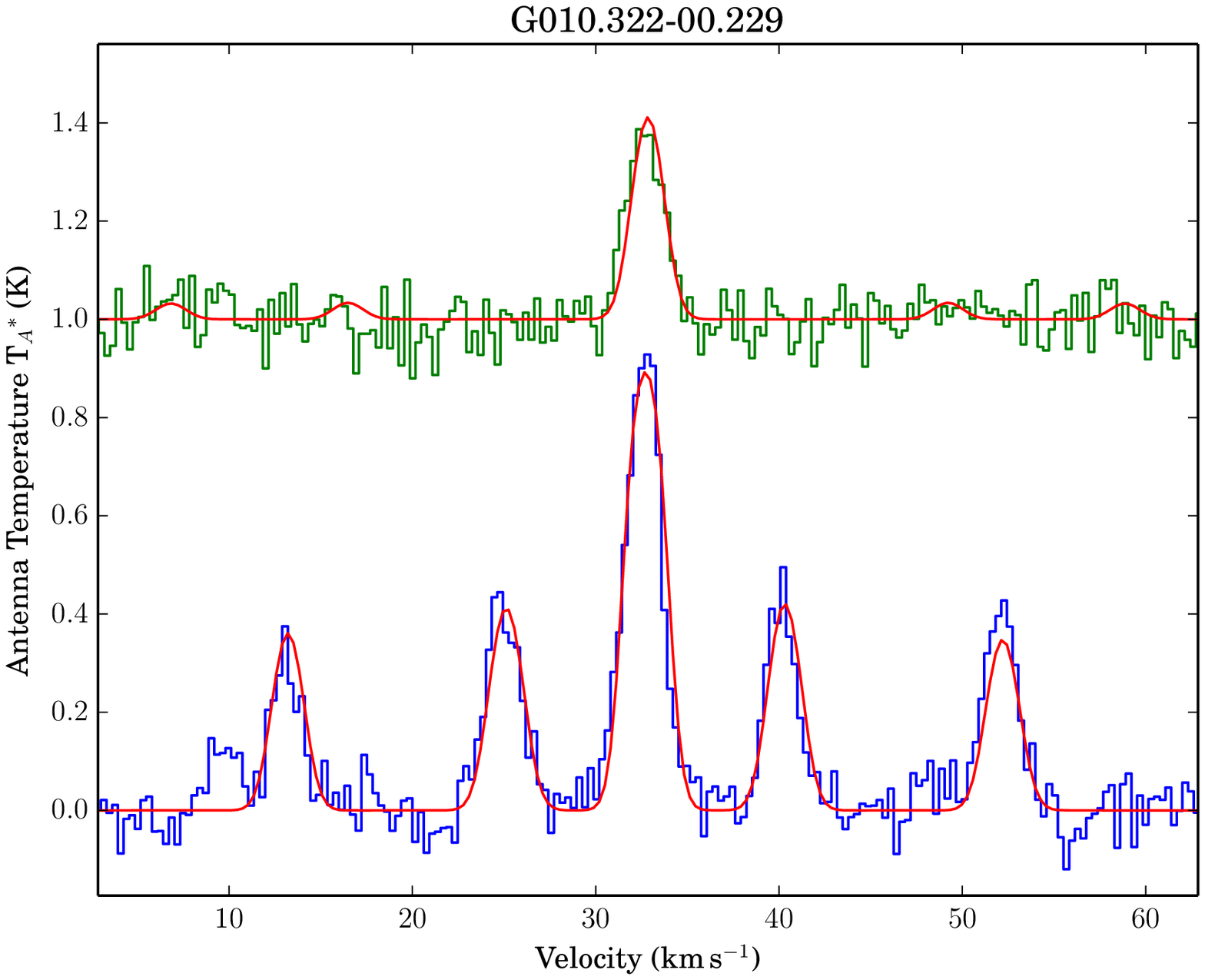}
\includegraphics[width=0.45\textwidth, trim= 0 0 0 0]{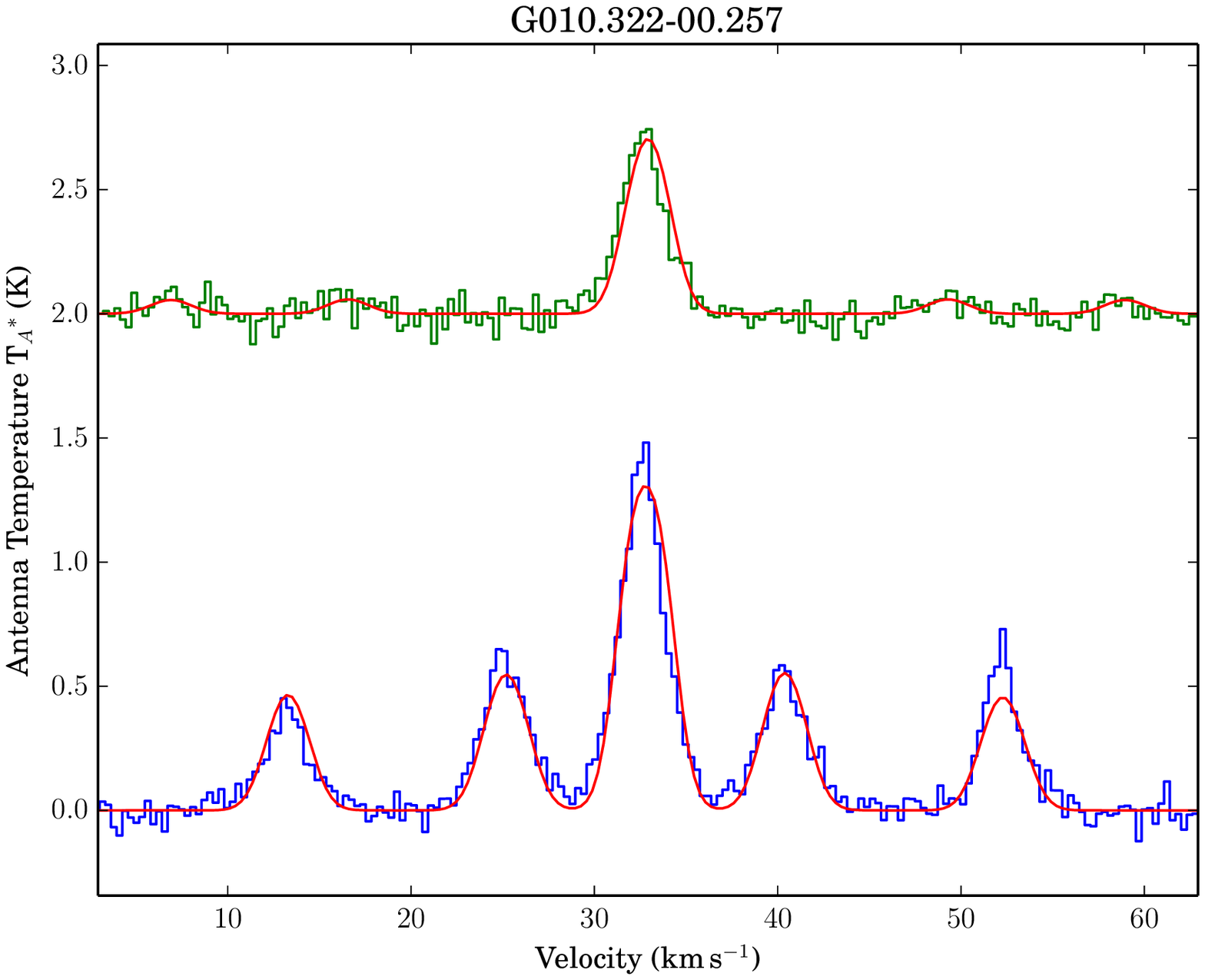}

\caption{\label{fig:example_spectra} Examples of the NH$_3$ (1,1) and (2,2) spectra taken towards the position of the peak emission of the three clumps identified in the G010.315$-$00.251 field. The NH$_3$ (1,1) and (2,2) lines are shown as blue and green histograms, respectively, and the fits to these transitions are shown in red.} 

\end{center}
\end{figure}

We fit the NH$_3$ (1,1) and (2,2) spectra simultaneously using \texttt{pyspeckit}\footnote{\url{http://pyspeckit.bitbucket.org/html/sphinx/index.html}}, which is a Python implementation of the method outlined by \citet{rosolowsky2008}.  This method utilises a model whose free parameters include the kinetic (\Tkin) and excitation temperatures (\Tex), the optical depth ($\tau$), $FWHM$ line width ($\Delta v$), the radial velocity of the clump (\vlsr) and NH$_3$ column density to fit the 18 individual (1,1) and 21 (2,2) magnetic hyperfine transitions. The satellite lines of the weaker (2,2) transition are not detectable in a large fraction of the pointings: simultaneous fitting of the two inversion transitions allows an adequate solution to be obtained with a sufficiently detectable (2,2) main line. This method has been used in many recent studies (cf. \citealt{dunham2011,battersby2014a}).

When using this algorithm we are implicitly assuming that the emission from the two transitions arises from the same spatial volume of gas that fills the beam (e.g., see figure\,11 of Paper\,I), that the excitation conditions are similar for all of the hyperfine components. Furthermore, the model assumes that the kinetic temperature is much less than $T_0 = 41.5$\,K, the temperature associated with the energy difference between the (2,2) and (1,1) inversion transitions, implying that only these first two states are significantly populated. To avoid contaminating our results with many low signal-to-noise pixels and anomalous values we restricted the fitting to pixels with a SNR $>3\sigma$.

\begin{figure*}
\begin{center}

\includegraphics[width=0.33\textwidth, trim= 0 0 0 0]{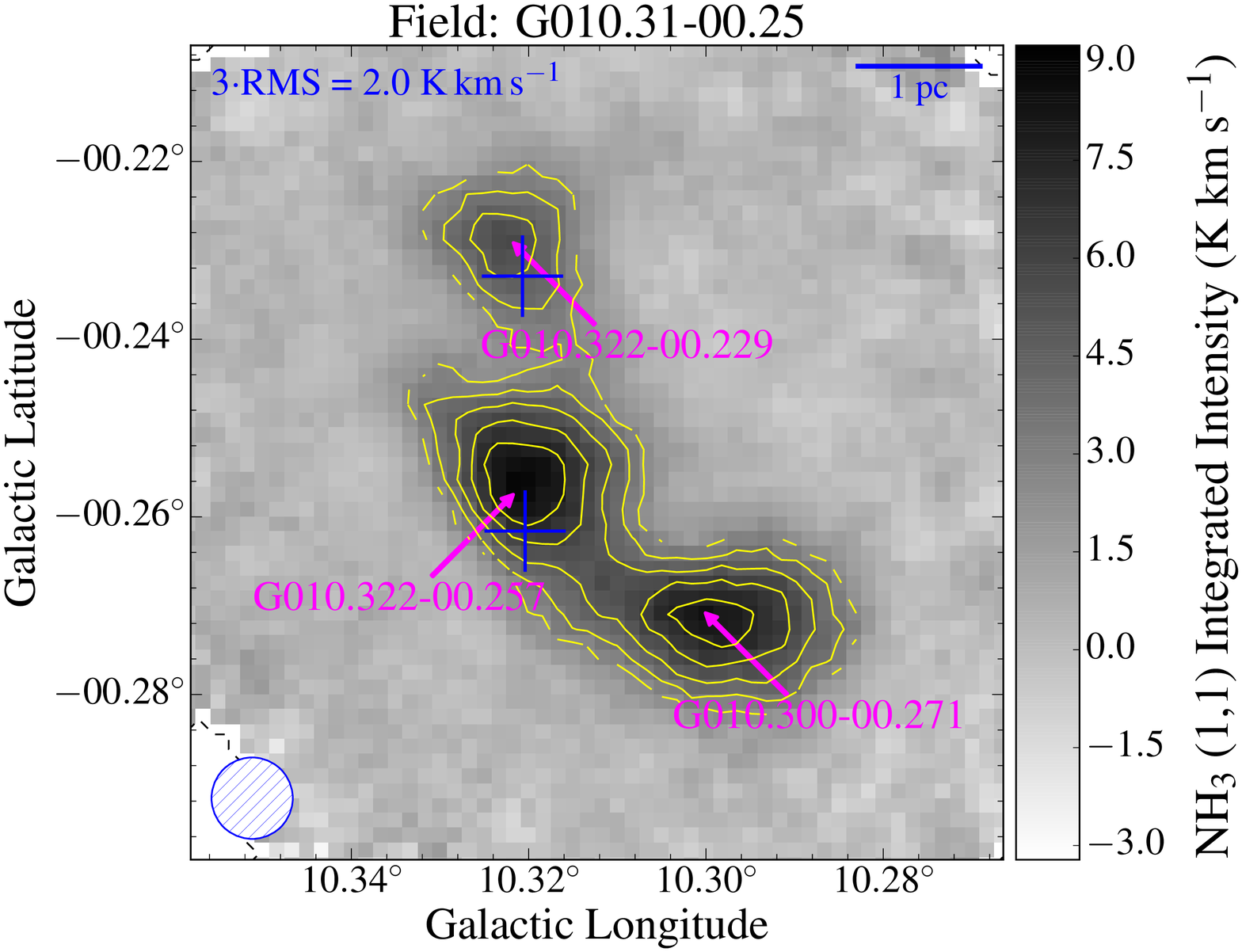}
\includegraphics[width=0.33\textwidth, trim= 0 0 0 0]{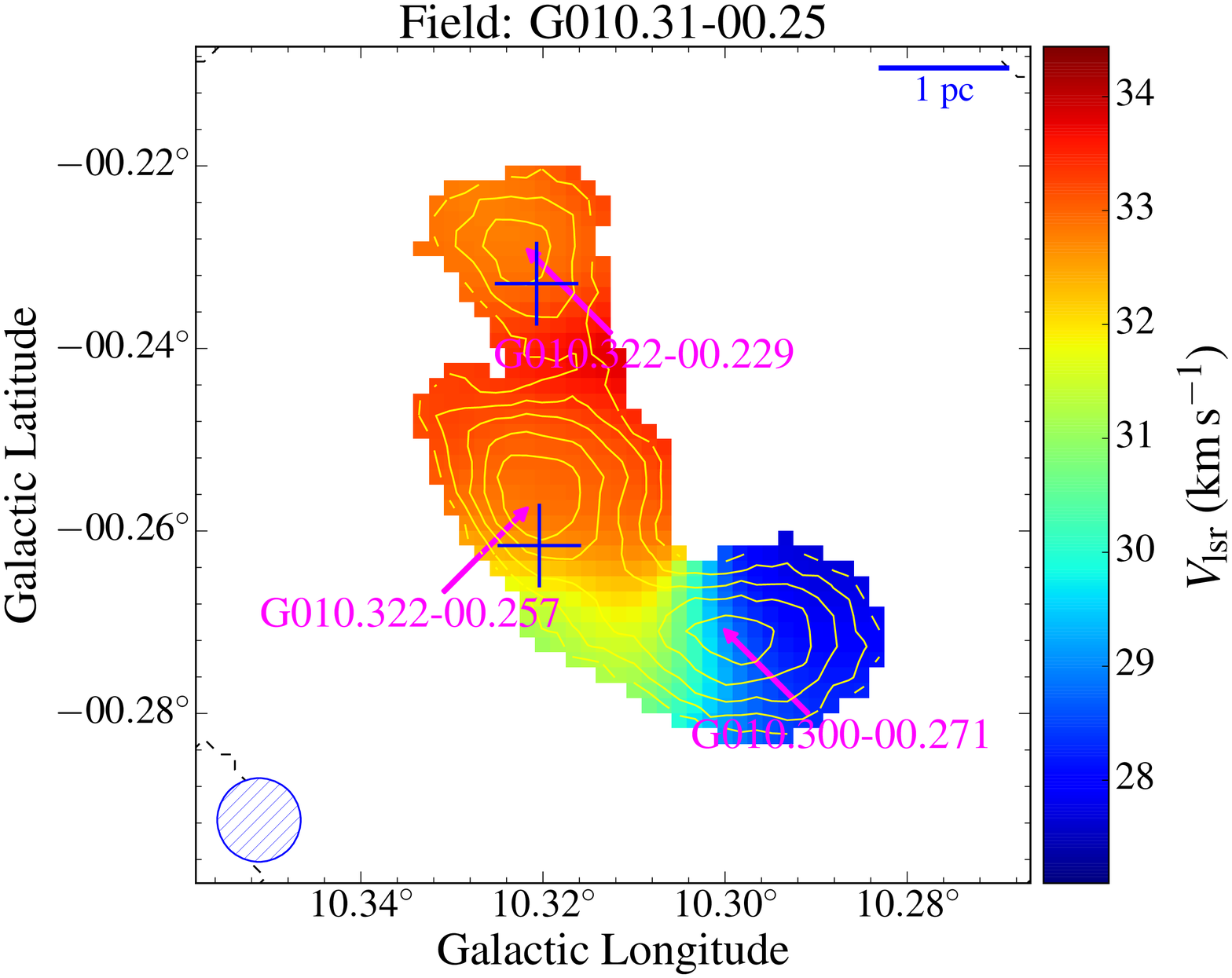}
\includegraphics[width=0.33\textwidth, trim= 0 0 0 0]{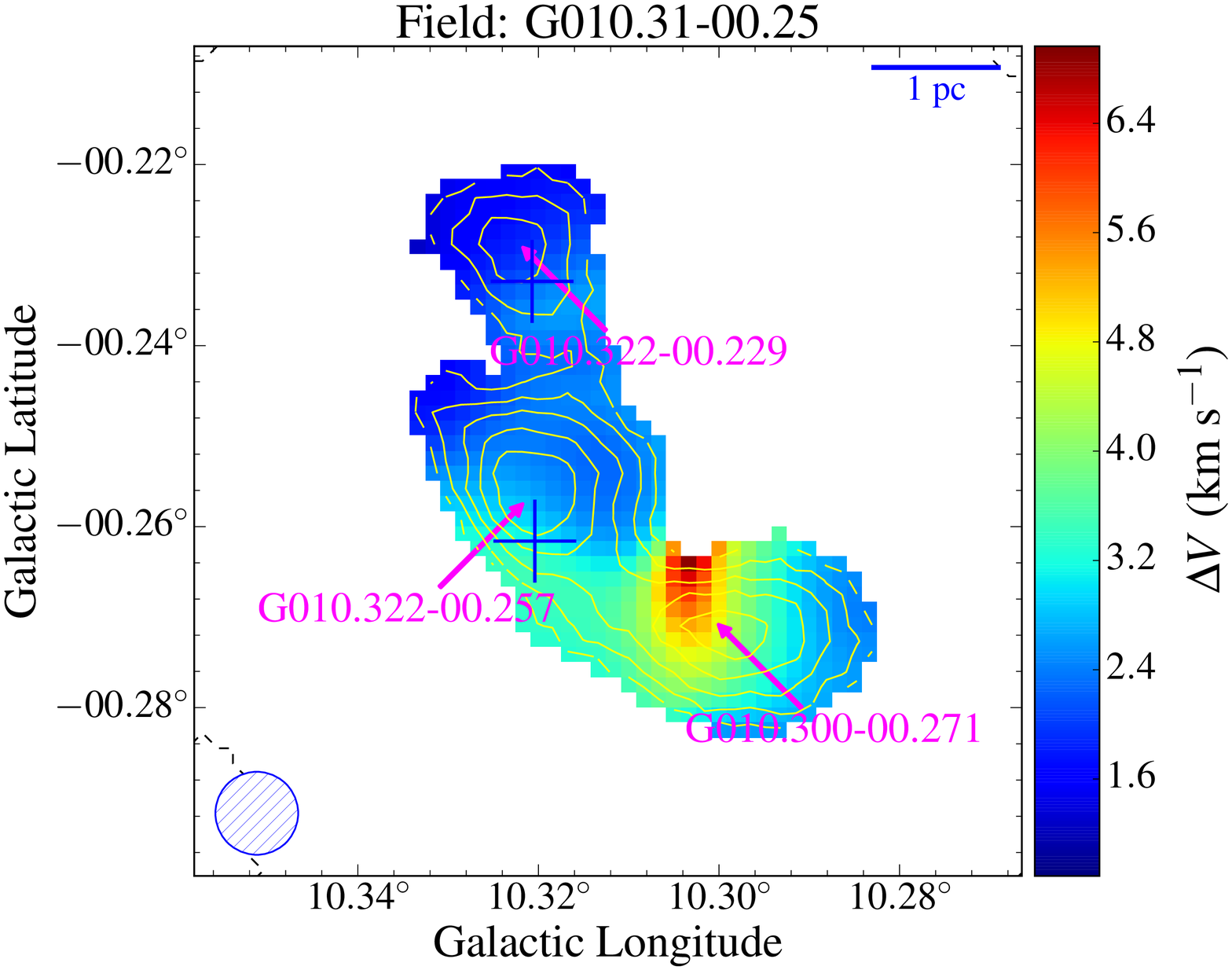}\\
\includegraphics[width=0.33\textwidth, trim= 0 0 0 0]{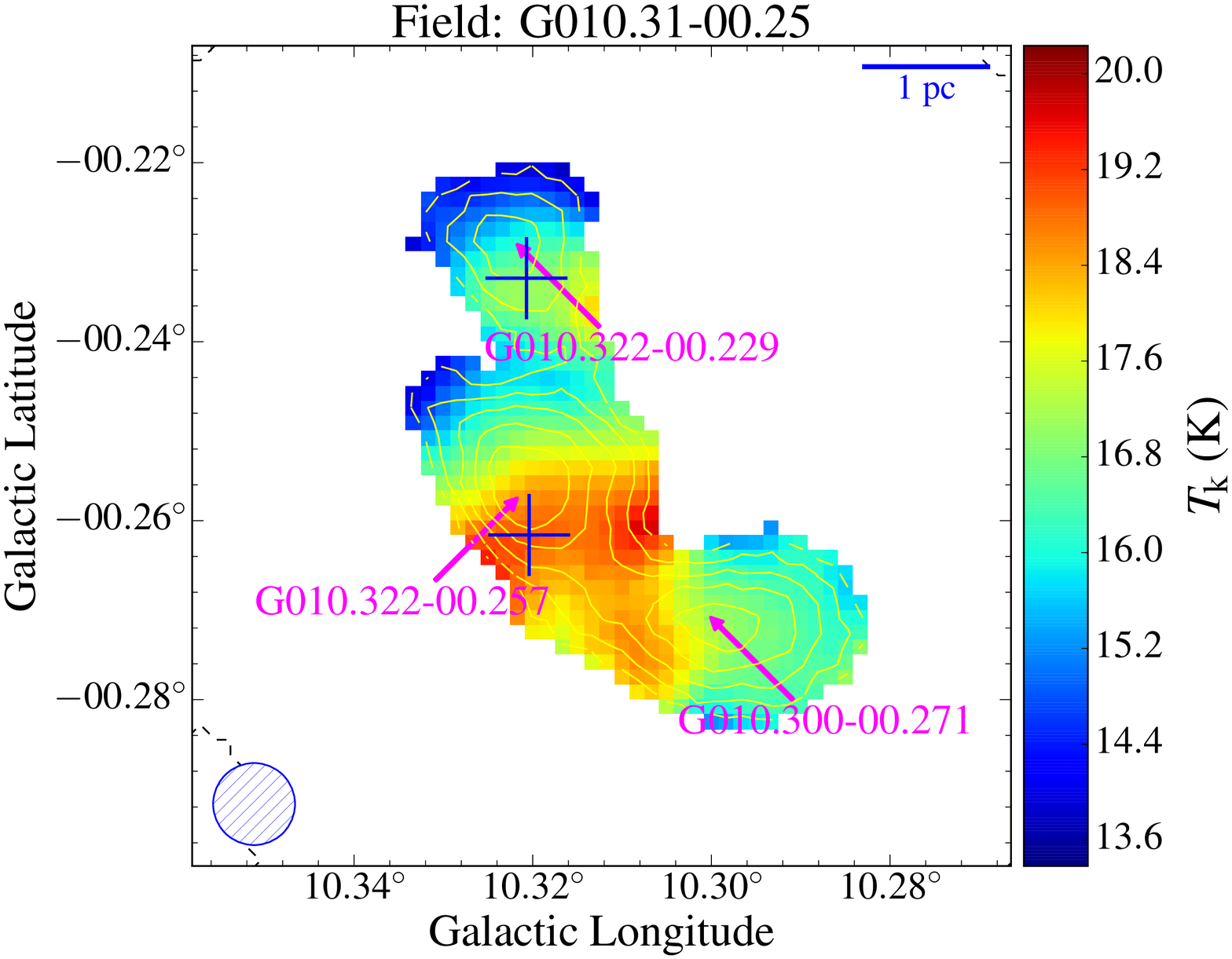}
\includegraphics[width=0.33\textwidth, trim= 0 0 0 0]{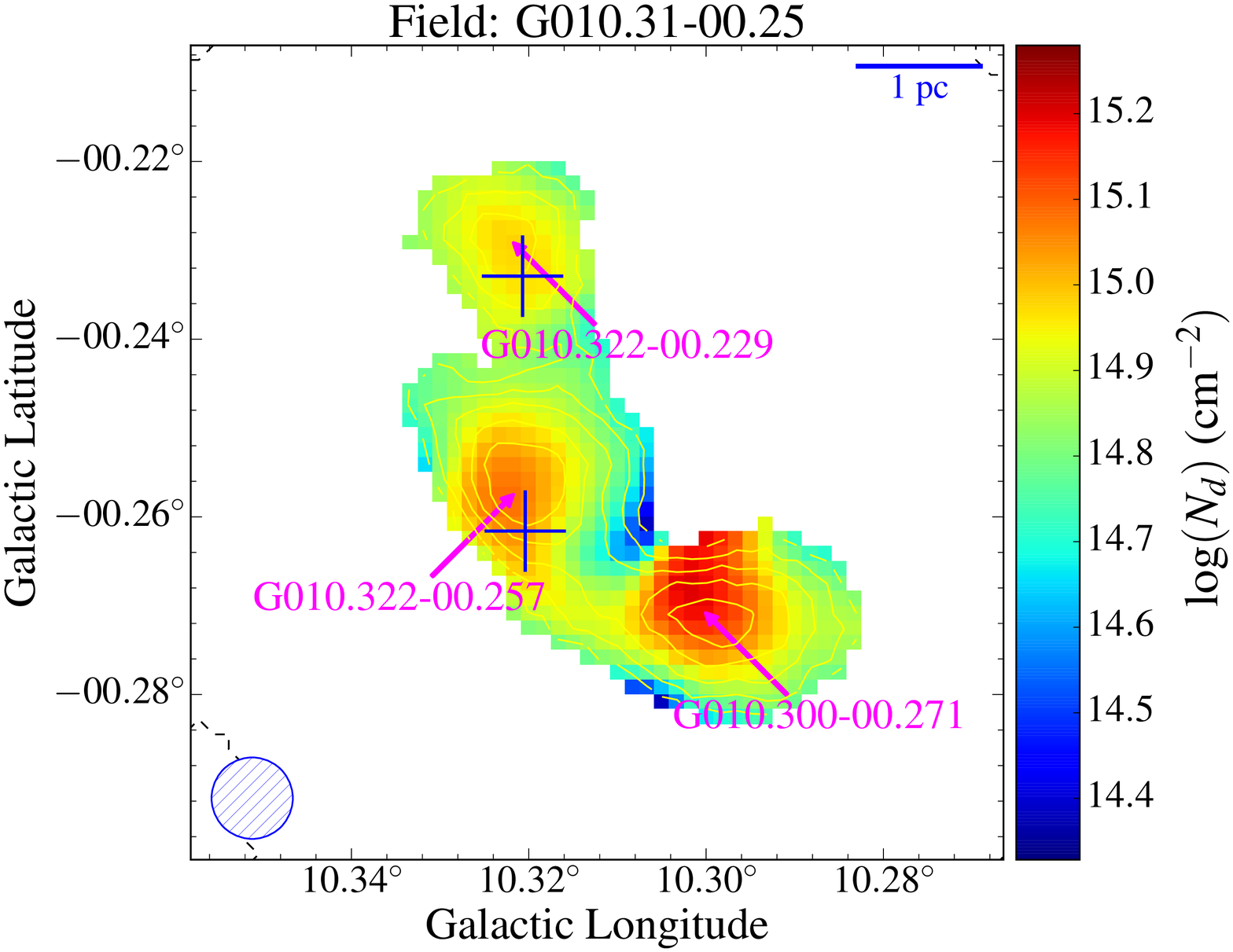}
\includegraphics[width=0.33\textwidth, trim= 0 0 0 0]{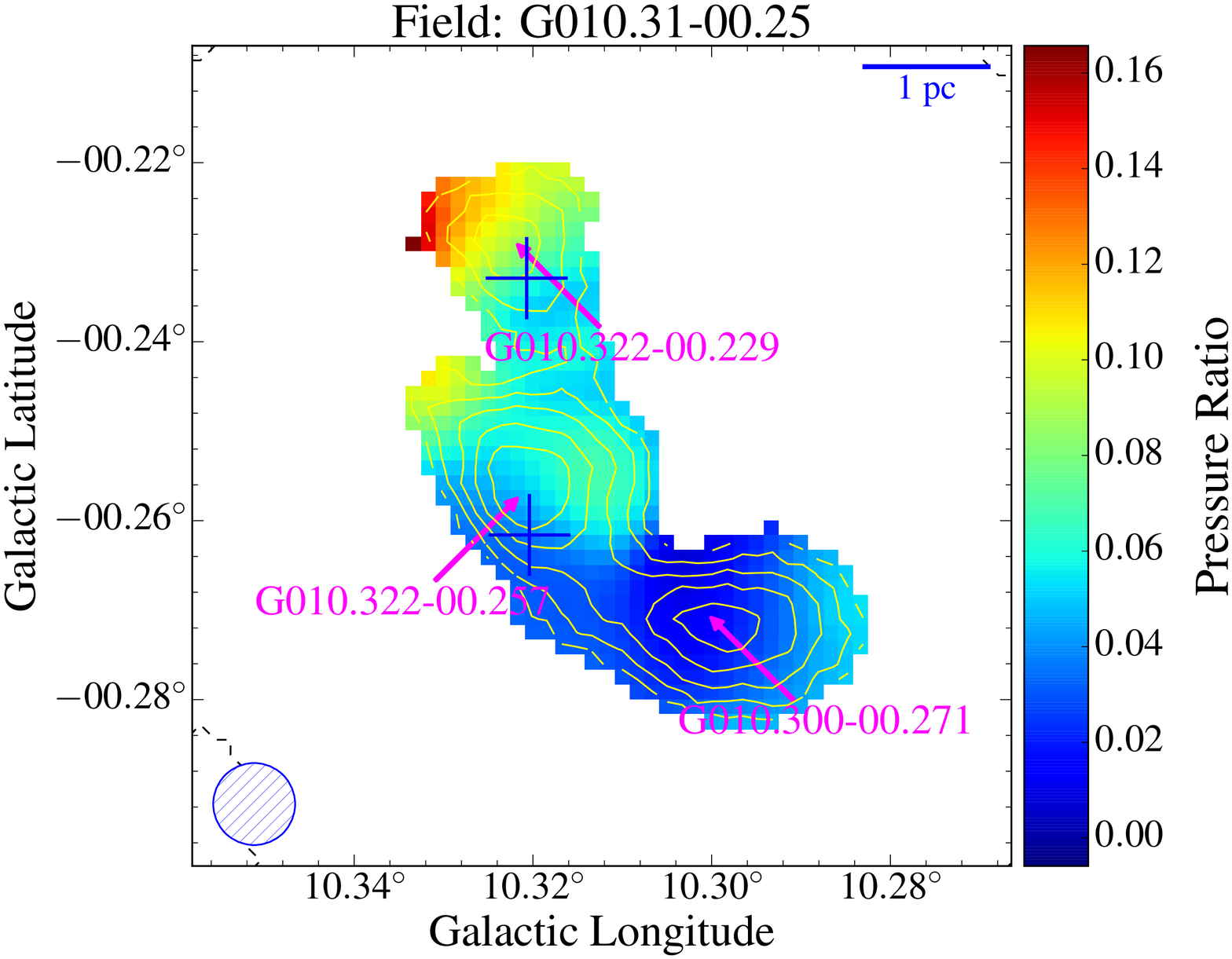}\\
\includegraphics[width=0.33\textwidth, trim= 0 0 0 0]{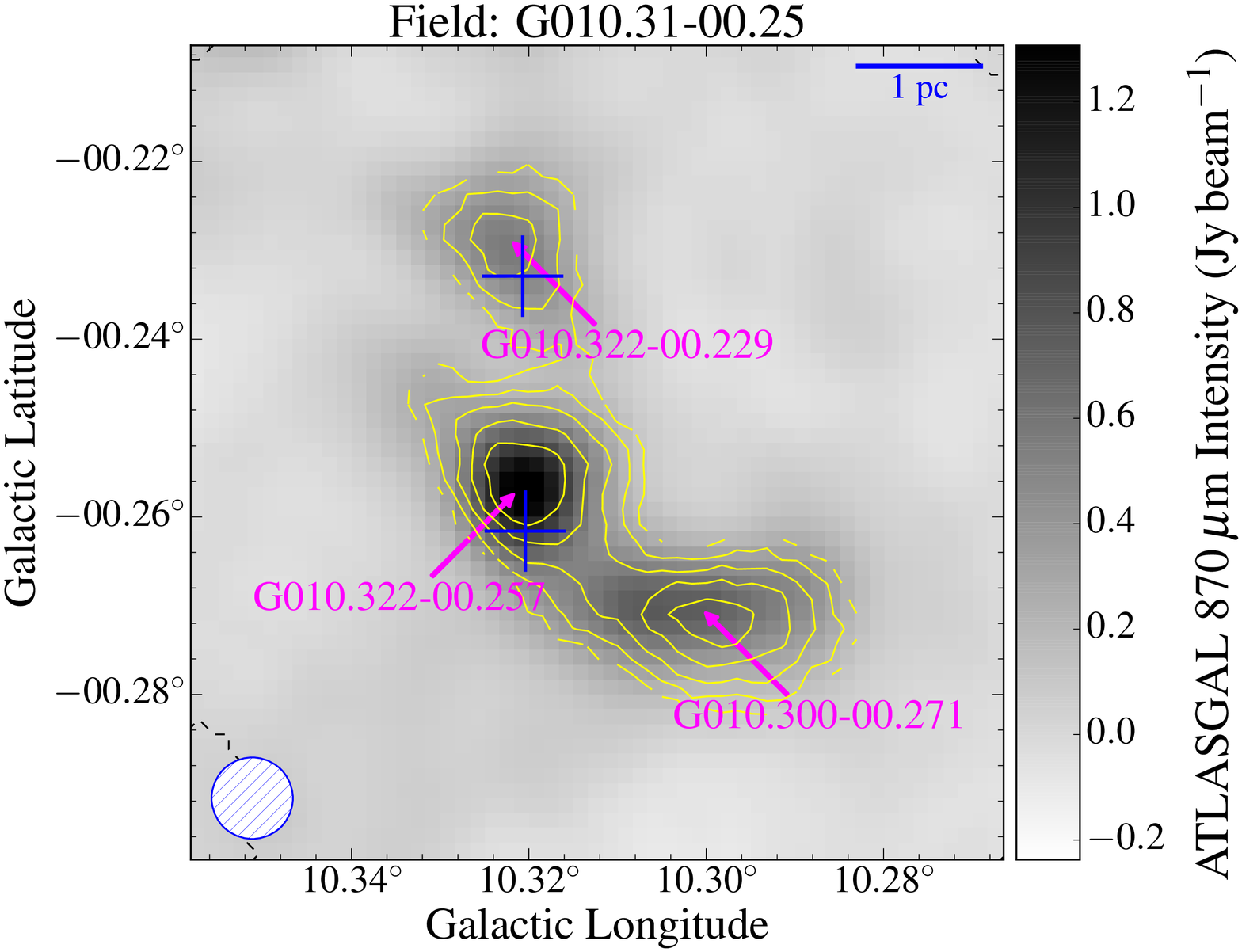}
\includegraphics[width=0.33\textwidth, trim= 0 0 0 0]{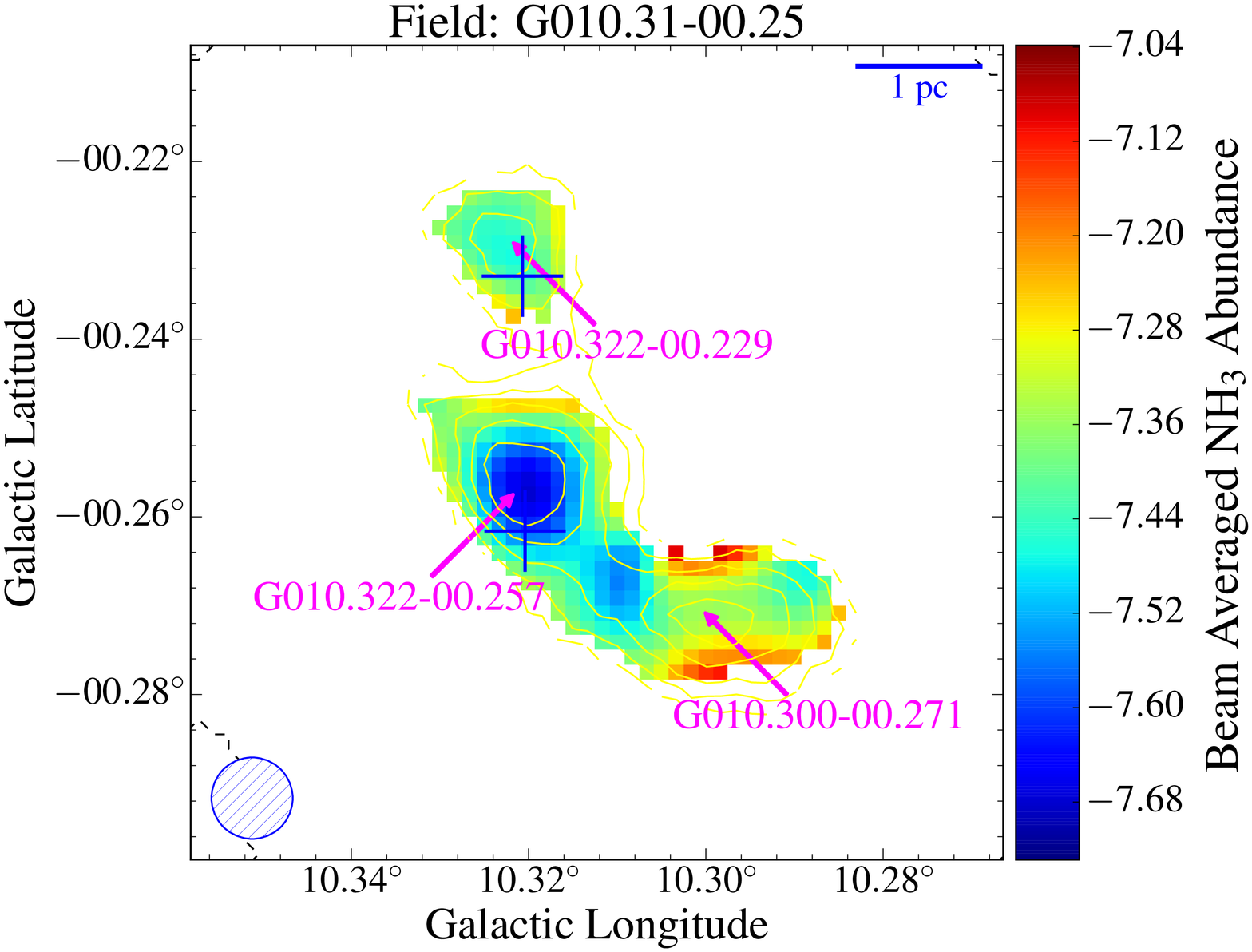}
\includegraphics[width=0.33\textwidth, trim= 0 0 0 0]{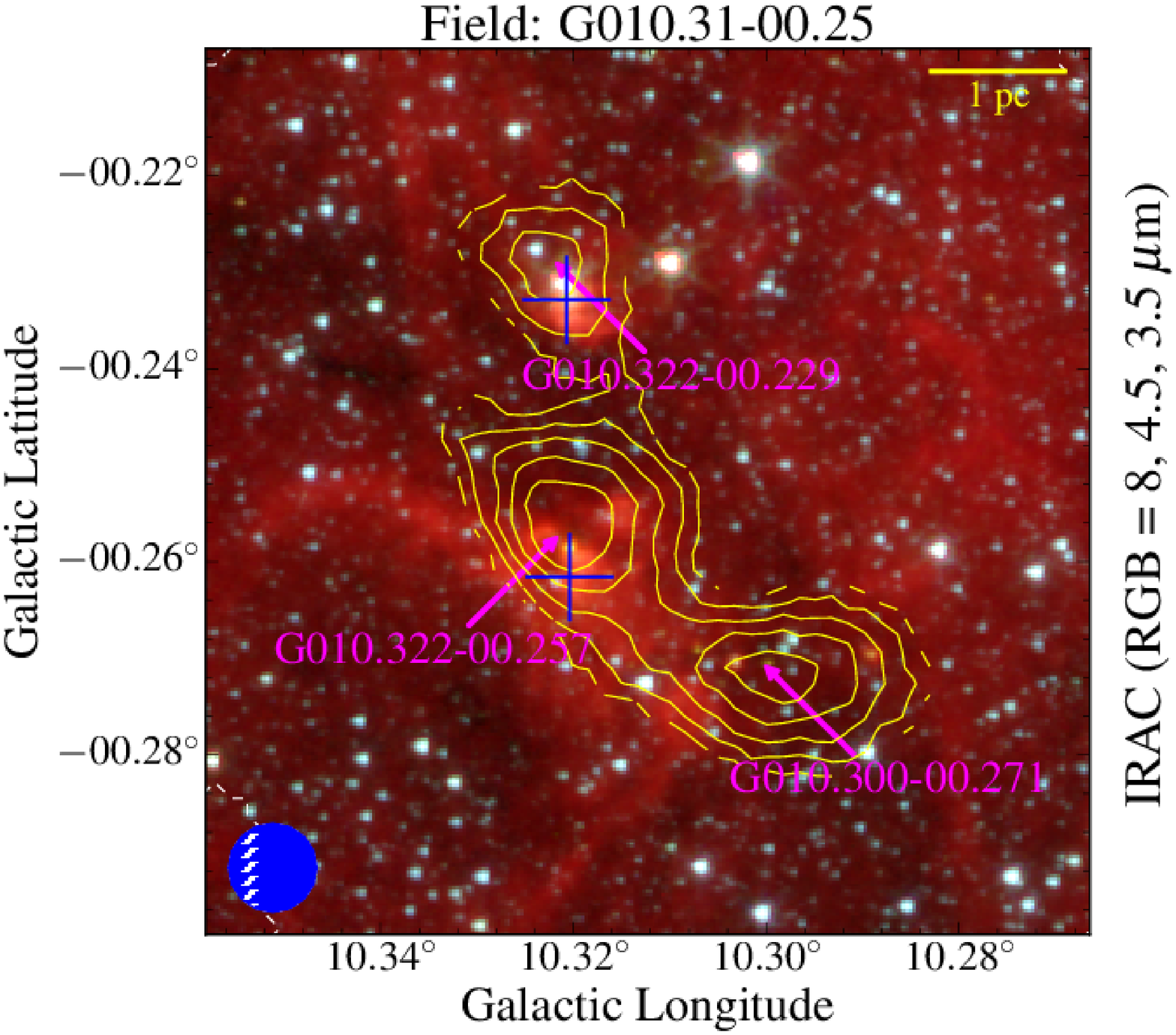}

\caption{\label{fig:distribution_maps} Distribution maps of the various parameters discussed in Section\,\ref{sect:obs}. In the upper left panel we present the integrated NH$_3$ (1,1) emission map showing the area outlined in red in Fig.\,\ref{fig:example_emission_maps}. In the upper middle and right panels we show the distribution of peak velocity and FWHM line width. The middle panels (left to right) show the kinetic temperature, ammonia column density, and gas pressure ratio ($R_{\rm{p}}$; ratio of the squares of thermal and non-thermal line-widths), while in the lower panels we present the dust emission, ammonia abundance and the three-colour IRAC image. The contours are the same in every map, and trace the integrated NH$_3$ (1,1) emission as described in Fig.\,\ref{fig:example_emission_maps}. The magenta labels identify ammonia clumps identified by \FW, while the red triangles and blue crosses identify the positions of MYSOs and compact \hii\ regions identified by the RMS survey. The angular resolution of the GBT beam at this frequency is indicated by the blue hatched circle shown in the lower left corner of each map.} 

\end{center}
\end{figure*}

The measured $FWHM$ line width is a convolution of the intrinsic line width of the source ($\Delta v_{\rm{int}}$) and the velocity resolution of the observations (i.e., spectrometer channel width). We remove the contribution of the spectrometer by subtracting the channel width (0.4\,\kms) from the measured line width in quadrature, i.e., :

\begin{equation}
\Delta v_{\rm{int}}= \sqrt{\left(\Delta v_{\rm{obs}}^2-0.16\right)}. 
\end{equation}

\noindent where $\Delta v_{\rm{int}}$ and $\Delta v_{\rm{obs}}$ are the intrinsic and observed $FWHM$ line widths in \kms. The peak $\Delta v_{\rm{int}}$ have mean and median values of $2.2\pm0.1$ and 2.1\,\kms, respectively. These values are similar to those generally found towards massive star forming regions (e.g., \citealt{sridharan2002,urquhart2011,wienen2012}) but broader than found for infrared dark clouds ($\sim$1.7\,\kms; \citealt{chira2013}). 

The line width consists of a thermal and non-thermal component. The thermal contribution to the line width can be estimated by:

\begin{equation}
 \Delta v_{\rm{th}} =  \sqrt{\left(\frac{8\ln{2}\,k_{\rm{B}}T_{\rm{kin}}}{m_{\rm{NH_3}}}\right)}
\end{equation}

\noindent where $8\ln2$ is the conversion between velocity dispersion and FWHM (i.e., 2.355), $k_{\rm{B}}$ is the Boltzmann constant, and $m_{\rm{NH_3}}$ is the mass of an ammonia molecule.  The measured line widths are significantly broader than would be expected from purely thermally driven motion ($\Delta v_{\rm{th}}\sim$0.22\,\kms\ for gas temperatures of 20\,K), and indicate that there is a significant contribution from supersonic non-thermal components such as turbulent motions, infall, outflows, rotation, shocks and/or magnetic fields (\citealt{arons1975,mouschovias1976}). Unfortunately, these observations do not have the resolution to explore these mechanisms in detail, but we are able to look at the global distribution of the thermal and non-thermal motions and infer what impact these may have on the other derived properties. We estimate the non-thermal velocity using:

\begin{equation}
\Delta v_{\rm{nt}} =  \sqrt{\left(\Delta v_{\rm{int}}^2 - \frac{8{\rm{ln}}2k_{\rm{B}}T_{\rm{kin}}}{m_{\rm{NH_3}}}\right)}
\end{equation}

We have calculated values for the thermal and non-thermal linewidths for all pixels above the 3$\sigma$ detection level in the integrated NH$_3$ (1,1) emission maps and used these to create ratio maps showing the contribution of these components to the gas pressure ratio (\citealt{lada2003b}):

\begin{equation}
R_{\rm{p}} =  \left(\frac{\Delta v_{\rm{th}}^2}{\Delta v_{\rm{nt}}^2}\right)
\end{equation}

\noindent where $\Delta v_{\rm{th}}$ and $\Delta v_{\rm{nt}}$ are as previously defined. The pressure ratio map is presented in the middle-right panel of Fig.\,\ref{fig:distribution_maps}.
 
\subsection{Beam filling factor}
\label{sect:beam_ff}

Ammonia emission is often found to be extended with respect to the beam; however, the excitation temperature of the inversion transition is commonly found to be significantly lower than the estimated  kinetic temperature of the gas, which suggests that the actual beam filling factor is less than unity ($B_{\rm{ff}}=T_{\rm{ex}}/T_{\rm{rot}}\sim0.1$-0.3; \citealt{urquhart2011,pillai2006}). There are two possible explanations for this: 1) the observed emission is the superposition of a large number of compact dense cores convolved with the telescope beam (i.e., there is structure on scales smaller than the beam); and 2) the gas is sub-thermally excited (i.e., non-LTE conditions). The latter is considered less likely, however, as the densities are sufficiently high ($>10^5$\,cm$^{-3}$; \citealt{keto2015}) for LTE between the dust and gas. 

The fitting algorithm calculates the kinetic and excitation temperatures. We calculate the rotation temperature from the derived kinetic temperature using the relationship (\citealt{walmsley1983})
\begin{equation}
  T_\text{rot} = T_\text{kin} / \left\{ {1+\frac{T_\text{kin}}{T_0}\ln \left[ {1+0.6\exp{\left (-15.7/T_\text{kin} \right)} } \right] }\right\}
\end{equation}

\noindent We have used these values to estimate the median and peak beam filling factors for the whole sample.\footnote{All peak measurements are taken toward the brightest NH$_3$  emission seen in the integrated maps presented in Fig\,\ref{fig:example_emission_maps}, which is nearly always found towards the centre of the clump.} The peak filling factors are similar to those derived in Paper\,I, although this study finds the median values are significantly lower ($\sim$0.1), which would indicate that the more of the underlying substructure is concentrated towards the centre of the clumps. At the median distance of the sample, the beam size of $\sim$30\arcsec\ corresponds to a physical area of $\sim$0.4\,pc$^2$: this is several times larger than expected for a typical core ($r\sim 0.1$\,pc (e.g., \citealt{motte2007}) and it is therefore likely that the observed clump structures consist of multiple dense cores. This is consistent with recent higher-resolution studies made with the VLA (e.g.,  \citealt{ragan2011,lu2014,battersby2014a}).

\subsection{Ammonia column density and abundance}
\label{sect:abundance}

The ammonia column density is determined as a free parameter by the fitting algorithm and uses the derived rotation temperature and so has therefore already taken the beam filling factor into account. Determination of the column density assumes that excitation conditions are homogeneous and that all hyperfine lines have the same excitation temperature. The peak NH$_3$ column densities range between 2.2-72.4$\times 10^{14}$\,cm$^{-2}$ with a median value of 11.2$\times 10^{14}$\,cm$^{-2}$. The column density in the outer envelope is significantly lower, ranging between 2.2-19.1$\times 10^{14}$\,cm$^{-2}$  with a median value of 7.2 $\times 10^{14}$\,cm$^{-2}$. The peak values for the NH$_3$ column density are consistent with studies reported towards other massive star forming regions and IRDCs that have been made at a similar resolution (e.g., \citealt{tafalla2004, pillai2006,dunham2011,wienen2012} and \citealt{morgan2014}). Although taking the beam filling factor into account produces more reliable peak NH$_3$ column densities for the unresolved substructure, we note that when estimating the abundances using the dust emission it is in fact the beam-averaged column density that is of interest; this is because no correction is made for the substructure. The beam-averaged NH$_3$ column density is therefore likely to be a factor of a few lower than that determined by the algorithm. To compensate for this, we estimate the beam-averaged NH$_3$ column density by multiplying the maps by the corresponding pixel beam filling factors.

To estimate the NH$_3$ abundance we compare the $N$(NH$_3$) to the total $N$(H$_2$) obtained from maps of the submm dust emission extracted from the APEX Telescope Large Area Survey of the Galaxy (ATLASGAL; \citealt{schuller2009}) assuming a constant gas to dust ratio. ATLASGAL has surveyed the inner parts of the Galactic plane ($300\degr < \ell < 60\degr$ and $|b| < 1.5\degr$) at 870\,\mum\ (345\,GHz) where the dust emission is optically thin and is therefore an excellent probe of column density and the total mass of the clumps. Dust maps were extracted for all but two of the fields (G075.766+00.358 and G078.977+00.363 are located outside the ATLASGAL region). A Gaussian kernel with a  FWHM of 25.6\arcsec\ was used to smooth the ATLASGAL maps to the same resolution as the GBT maps (i.e., $\sqrt{\left(32^2-19.2^2\right)}$ ). 

We use the kinetic gas temperature ($T_{\rm{kin}}$) derived from the ammonia emission for each pixel and the corresponding pixel flux from the ATLASGAL emission maps to create maps of the H$_2$ column density via:

\begin{equation}
N({\rm{H_2}}) \, = \, \frac{S_\nu \, R}{B_\nu(T_\mathrm{kin}) \, \Omega \, \kappa_\nu \, \mu
\, m_{\rm{H}}},
\end{equation}

\noindent where $\Omega$ is the beam solid angle, $\mu$ is the mean molecular weight of the interstellar medium (we take $\mu = 2.8$ assuming a 10\% contribution from helium; \citealt{kauffmann2008}), $m_{\rm{H}}$ is the mass of the hydrogen atom, $R$ is the gas-to-dust mass ratio (assumed to be 100) and $\kappa_\nu$ is the dust absorption coefficient (taken as 1.85\,cm$^2$\,g$^{-1}$  derived by \citealt{schuller2009} by interpolating to 870\,\mum\ from Table\,1, Col.\,5 of \citealt{ossenkopf1994}). We are also assuming that the kinetic temperature is roughly equivalent to the dust temperature (i.e., $T_{\rm{kin}}=T_{\rm{dust}}$; e.g., \citealt{morgan2010} and {\color{red}Ko\"nig} et al. 2015). 

\begin{figure}
\begin{center}

\includegraphics[width=0.45\textwidth, trim= 0 0 0 0]{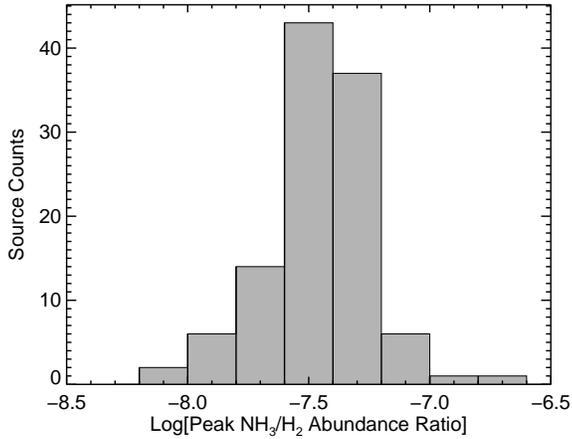}
\caption{Frequency distribution of the peak NH$_3$ abundance relative to the H$_2$ within each clump obtained as in Sect.\,\ref{sect:abundance}, for all clumps. The bin size is 0.2\,dex. } 

\label{fig:abundance_distribution}
\end{center}
\end{figure}

The H$_2$ and beam-averaged NH$_3$ column density maps have then been combined to create maps showing variation in the ammonia abundance across the clumps (i.e., $N({\rm{NH_3}})/N({\rm{H_2}})$). For the abundance analysis we only include pixels where both the ammonia and dust emission is over 5$\sigma$ in order to minimise the uncertainty in the maps (lower middle panel of Fig.\,\ref{fig:distribution_maps}). From these maps we estimate the fractional abundance in of the inner and outer envelopes and find these to be similar. The peak abundances range between 0.5-10.3$\times 10^{-8}$ with a median value of 2.5$\times 10^{-8}$ (see Fig.\,\ref{fig:abundance_distribution} for the distribution). These values are consistent with many of the studies previously mentioned.

\setlength{\tabcolsep}{4pt}

\begin{table*}


\begin{center}\caption{Detected NH$_3$ clump parameters.  The columns are as follows: (1) field ID given in Table\,1; (2) name derived from Galactic coordinates of the peak emission of each clump; (3)-(4) radial velocity and the intrinsic $FWHM$ line width; (5) optical depth of the transition; (6-8) excitation, rotation and kinetic  temperatures; (9) beam filling factor ($T_{\rm{ex}}/T_{\rm{rot}}$); (10-11) NH$_3$ column density and abundance. For columns 4-8 and 10-11 the first value is measured towards the emission peak while the value given in parentheses is the median value determined over the clump. In the final column (12) we include a flag to identify sources where broad line-widths are seen towards the centre of clumps; a value of 1 or 2 indicate whether the emission profile appears to arise from a single clump or multiple distinct clumps along the line of sight, respectively, and a value of 3 identifies sources that are associated with broad emission wings, which are themselves indicative of outflow motions. Source names that are appended by a $\dagger$ identifies clumps with warmer surface temperatures and colder centres (see Sect.\,\ref{sect:general_properties} for details).}
\label{tbl:cattable}
\begin{minipage}{\linewidth}
\begin{tabular}{cc.ccccccccc}
  \hline \hline
  \multicolumn{1}{c}{Field} &\multicolumn{1}{c}{Clump}
  &  \multicolumn{1}{c}{\vlsr} &  \multicolumn{1}{c}{$\Delta v$} &
  \multicolumn{1}{c}{$\tau_{\rm{main}}$} & \multicolumn{1}{c}{$T_{\rm{ex}}$} & \multicolumn{1}{c}{$T_{\rm{rot}}$} &
  \multicolumn{1}{c}{$T_{\rm{kin}}$} &\multicolumn{1}{c}{$B_{\rm{ff}}$} &     \multicolumn{1}{c}{log($N$(NH$_3$))} &     \multicolumn{1}{c}{log($N$(NH$_3$)/$N$(H$_2$))} & Notes \\
  
  \multicolumn{1}{c}{id} &\multicolumn{1}{c}{name}
  &  \multicolumn{1}{c}{(\kms)} &  \multicolumn{1}{c}{(\kms)}
  &  \multicolumn{1}{c}{} &  \multicolumn{1}{c}{(K)} &
  \multicolumn{1}{c}{(K)} & \multicolumn{1}{c}{(K)} & \multicolumn{1}{c}{} & \multicolumn{1}{c}{(cm$^{-2}$)} & \multicolumn{1}{c}{(cm$^{-2}$)} & \\

  \multicolumn{1}{c}{(1)} &  \multicolumn{1}{c}{(2)} &  \multicolumn{1}{c}{(3)} &  \multicolumn{1}{c}{(4)} &
  \multicolumn{1}{c}{(5)} &  \multicolumn{1}{c}{(6)} &  \multicolumn{1}{c}{(7)} &  \multicolumn{1}{c}{(8)} &
  \multicolumn{1}{c}{(9)} &\multicolumn{1}{c}{(10)}&\multicolumn{1}{c}{(11)} &\multicolumn{1}{c}{(12)} \\
  \hline
1	&	G010.283$-$00.118$\dagger$	&	14.2	&	2.4 (2.5)	&	4.0  (3.4)	&	7.0  (4.4)	&	17.0  (17.1)	&	18.8  (18.9)	&	0.26	&	15.35 (15.10)	&	$-$7.50 ($-$7.52)	&		\\
1	&	G010.288$-$00.166$\dagger$	&	12.2	&	1.8 (1.9)	&	3.9  (2.9)	&	4.8  (4.1)	&	15.7  (16.3)	&	17.1  (17.9)	&	0.25	&	15.05 (14.96)	&	$-$7.45 ($-$7.35)	&		\\
1	&	G010.296$-$00.148	&	13.6	&	4.2 (3.8)	&	1.3  (1.6)	&	7.6  (4.8)	&	21.2  (19.9)	&	24.6  (22.8)	&	0.24	&	15.22 (15.09)	&	$-$7.71 ($-$7.51)	&		\\
2	&	G010.300$-$00.271	&	29.1	&	4.8 (3.5)	&	2.3  (1.8)	&	3.9  (4.0)	&	15.7  (15.5)	&	17.1  (16.8)	&	0.26	&	15.18 (14.90)	&	$-$7.35 ($-$7.32)	&	2	\\
2	&	G010.322$-$00.229	&	32.8	&	1.9 (2.0)	&	3.8  (3.3)	&	3.9  (3.6)	&	14.9  (14.8)	&	16.1  (15.9)	&	0.24	&	14.98 (14.90)	&	$-$7.46 ($-$7.33)	&		\\
2	&	G010.322$-$00.257	&	32.8	&	2.8 (2.5)	&	2.8  (2.3)	&	4.3  (3.9)	&	16.7  (16.1)	&	18.4  (17.7)	&	0.24	&	15.08 (14.88)	&	$-$7.66 ($-$7.40)	&		\\
1	&	G010.323$-$00.165	&	12.6	&	1.7 (1.8)	&	3.2  (2.8)	&	6.2  (4.8)	&	18.3  (18.0)	&	20.6  (20.1)	&	0.27	&	15.08 (14.95)	&	$-$7.73 ($-$7.67)	&		\\
1	&	G010.346$-$00.148$\dagger$	&	12.2	&	1.8 (2.0)	&	3.2  (2.7)	&	6.8  (5.3)	&	18.1  (18.9)	&	20.2  (21.3)	&	0.28	&	15.13 (15.01)	&	$-$7.50 ($-$7.64)	&		\\
3	&	G010.440+00.003$\dagger$	&	67.3	&	3.7 (3.7)	&	1.9  (1.9)	&	3.6  (3.6)	&	14.9  (15.0)	&	16.1  (16.2)	&	0.24	&	14.94 (14.90)	&	$-$7.53 ($-$7.54)	&		\\
3	&	G010.474+00.028	&	67.2	&	5.8 (4.1)	&	5.6  (3.6)	&	5.5  (4.0)	&	21.5  (18.2)	&	25.1  (20.4)	&	0.22	&	15.84 (15.32)	&	$-$7.75 ($-$7.25)	&	2	\\
  \hline
\end{tabular}\\
\end{minipage}
Notes: Only a small portion of the data is provided here, the full table is only  available in electronic form at the CDS via anonymous ftp to cdsarc.u-strasbg.fr (130.79.125.5) or via http://cdsweb.u-strasbg.fr/cgi-bin/qcat?J/MNRAS/.
\end{center}
\end{table*}

\setlength{\tabcolsep}{6pt}

\subsection{Uncertainties on the fitted parameters}

\FW\ does not provide estimates of the uncertainties for the position or size of the extracted sources: we estimate that these values are likely to be accurate to within a few arcsec and so errors are relatively small given that most sources are well-resolved. The uncertainty in the peak flux is determined from the standard deviation of an emission-free region in each field, and is typically better than 10\,per\,cent. We assume the uncertainty is similar for the integrated flux values.

The uncertainties in the velocity, line-width, optical depth, kinetic and excitation temperatures and NH$_3$ column density are estimated by the fitting algorithm and are all relatively small. Typical uncertainties in the velocity and line-width are better than 0.1\,\kms\ and 0.3\,K for kinetic and excitation temperatures, and since the rotation temperature is derived directly from the kinetic temperature it will have a similar uncertainty. The uncertainty for the beam filling factor will be dominated by the uncertainty in the excitation temperature, which is roughly about 5\,per\,cent. Although the uncertainty in the NH$_3$ column density given by the code is small it is dominated by the uncertainty in the peak flux measurement, which as mentioned in the previous paragraph is $\sim$10\,per\,cent and we therefore adopt this value for the uncertainty for this parameter. Finally, we estimate the uncertainty in the abundance ratio to be $\sim$20\,per\,cent, which is a combination of the uncertainty in the peak fluxes of the ATLASGAL survey ($\sim$15\,per\,cent; \citealt{schuller2009}) and NH$_3$ added in quadrature; however, this is a lower limit as the dust models used to estimate the H$_2$ column densities are not well constrained and so the true uncertainty can be a factor of a few times the abundance.

\section{Results}
\label{sect:results}

\subsection{Detection statistics}

We have detected a total of \clumpnum\ clumps in the \fieldnum\ fields observed as part of this project. Integrated NH$_3$ (1,1) maps of all of the observed fields are presented in Fig\,{\color{red}A1} and all of the clump parameters determined by \FW\ are given in Table\,\ref{tbl:fw_parameters}. Example spectral profiles for the two ammonia transitions are presented in Fig.\,\ref{fig:example_spectra} and the fitted and derived parameters are given in Table\,\ref{tbl:cattable}.\footnote{A complete set of spectra are provide as online material, Fig\,{\color{red}A5}.}

In Fig.\,\ref{fig:distribution_maps} we present distribution maps of field G010.31$-$00.25 produced from the analysis described in the previous section. 31 of the 115 clumps identified have peak integrated NH$_3$ (1,1) emission below a SNR value of 10, which is too low to allow detailed analysis of the spatial distribution of the various parameters derived, particularly those that also rely on the detection of the (2,2) transition. The peak and median values for the derived properties for these lower-SNR clumps are given in the results tables, although we do not explicitly present the spatial distribution maps of these low-SNR clumps or discuss them in detail. Distribution maps are provided in  Fig\,{\color{red}A6} of all 84 clumps with a SNR of 10 of more.

\begin{table*}

\begin{center}\caption{Statistical properties for the whole sample.}
\label{tbl:derived_para}
\begin{minipage}{\linewidth}
\small
\begin{tabular}{lc......}
\hline \hline
  \multicolumn{1}{l}{Parameter}&  \multicolumn{1}{c}{Number}&	\multicolumn{1}{c}{Mean}  &	\multicolumn{1}{c}{Standard Error} &\multicolumn{1}{c}{Standard Deviation} &	\multicolumn{1}{c}{Median} & \multicolumn{1}{c}{Min}& \multicolumn{1}{c}{Max}\\
\hline
Aspect ratio &          	115&1.52&	0.04 & 0.39 & 1.44 & 1.02 & 3.56\\
$Y$-factor &          115&       4.04&      0.19 & 
       2.01 &       3.56 &       1.25 &       12.08\\
Angular offset (\arcsec) &           88&21.9&1.6 & 14.6 & 18.5 & 0.4 & 71.7\\
Distance (kpc) &           65&5.00&0.36 & 2.91 & 4.50 & 1.12 & 12.81\\
Radius (pc) &          104&0.53&0.03 & 0.33 & 0.43 & 0.09 & 1.63\\

\hline
$T_{\rm{kin}}$ (Mean MSF) (K) &           71&17.53&0.27 & 2.27 & 17.13 & 12.62 & 
24.02\\
$T_{\rm{kin}}$ (Mean Quiescent) (K) &           44&15.64&0.35 & 2.31 & 15.31 & 
12.03 & 21.18\\
$T_{\rm{kin}}$ (Peak MSF) (K) &           71&18.81&0.30 & 2.57 & 18.51 & 12.93 & 
25.82\\
$T_{\rm{kin}}$ (Median MSF) (K) &           71&17.56&0.26 & 2.21 & 17.23 & 12.82
 & 23.92\\
\hline
$FWHM$ line width (Mean) (\kms) &          115&2.08&0.07 & 0.77 & 2.02 & 0.36 & 4.58
\\
$FWHM$ line width (Mean MSF) (\kms) &           71&2.19&0.08 & 0.72 & 2.14 & 0.59
 & 4.58\\
 $FWHM$ line width (Mean Quiescent) (\kms) &           44&1.90&0.13 & 0.84 & 1.85
 & 0.36 & 4.18\\
\hline
Pressure ratio (Mean MSF) &           71&0.013&0.001 & 0.012 & 0.010 & 0.002 & 
0.092\\
Pressure ratio (Mean Quiescent) &           44&0.019&0.003 & 0.019 & 0.014 & 0.002
 & 0.113\\
\hline
Beam filling factor (Median) &          115&0.28&0.01 & 0.07 & 0.27 & 0.21 & 
0.78\\
Beam filling factor (Mean) &          115&0.30&0.01 & 0.08 & 0.28 & 0.21 & 0.85\\
Beam filling factor (Peak) &          115&0.34&0.01 & 0.11 & 0.31 & 0.22 & 1.00\\
\hline
$N$(NH$_3$) (Mean RMS) (10$^{14}$\,cm$^{-2}$) &           71&15.35&1.21 & 10.19 & 13.66 & 2.90
 & 69.49\\
 $N$(NH$_3$) (Mean Quiescent) (10$^{14}$\,cm$^{-2}$) &           44&10.85&0.95 & 6.32 & 9.63 & 
2.29 & 31.70\\
$N$(NH$_3$) (Median) (10$^{14}$\,cm$^{-2}$) &          115&14.88&0.02 & 0.18 & 14.90 & 14.43 & 
15.32\\
$N$(NH$_3$) (Peak) (10$^{14}$\,cm$^{-2}$) &          115&15.06&0.02 & 0.26 & 15.06 & 14.36 & 
15.84\\
\hline
Abundance ratio (Mean) &          110&-7.42&0.02 & 0.22 & -7.41 & -8.06 & -6.77\\
Abundance ratio (Median) &           110&-7.46&0.02 & 0.22 & -7.45 & -8.15 & -6.73\\
Abundance ratio (Peak) &          110&-7.58&0.02 & 0.23 & -7.55 & -8.32 & -7.00
\\
\hline
Log[Clump mass] (\msun) &          106&2.86&0.05 & 0.56 & 2.89 & 1.49 & 4.06\\
Log[Clump mass] MSF (\msun) &           64&3.04&0.07 & 0.55 & 3.01 & 1.49 & 4.06\\
Log[Clump mass] Quiescent (\msun) &     42&2.59&0.07 & 0.46 & 2.68 & 1.58 & 3.47\\
\hline
$N$(H$_2$) (Peak) (cm$^-2$) &          110&22.14&0.03 & 0.31 & 22.10 & 21.51 & 
23.01\\
$N$(H$_2$) (Median) (cm$^-2$) &          110&21.75&0.02 & 0.19 & 21.75 & 21.09
 & 22.16\\
\hline
Bolometric luminosity (\lsun) &           85&4.67&3.92 & 4.89 & 4.22 & 2.52 & 5.60\\
\hline\\
\end{tabular}\\

\end{minipage}

\end{center}
\end{table*}

\subsection{RMS associations}

We have matched these clumps with the RMS catalogue to distinguish those associated with massive star formation from those likely to be less active (quiescent). A match was made between a clump and an RMS source if the RMS source was located within clump boundaries as defined by the lowest contour shown in Fig.\,\ref{fig:example_emission_maps} (i.e., 3$\sigma$) and possessed a similar \vlsr\ as the clump.  In 21 cases, multiple RMS sources have been associated with the same clump. We find \rmsclumpnum\ clumps are associated with RMS sources while the star formation is less evolved towards the remaining \nonrmsclumpnum\ clumps (approximately 40\,per\,cent of the clumps identified); here we are making the assumption that infrared faint/dark clumps are less evolved. We therefore have a useful sample of relatively quiescent clumps (i.e., not currently associated with an MYSO or \hii\ region) with which to compare the results from our more evolved sample. We also find there is no difference in the proportions of RMS sources associated with spherical and filamentary clumps.

\setlength{\tabcolsep}{3pt}

\begin{table}


\begin{center}\caption{RMS associations.}
\label{tbl:rms_associations}

\begin{minipage}{\linewidth}
\begin{tabular}{ccl.l}
  \hline \hline
   \multicolumn{1}{c}{Field} &\multicolumn{1}{c}{Clump}
  &  \multicolumn{1}{c}{RMS} &  \multicolumn{1}{c}{Offset}&
  \multicolumn{1}{c}{Source} \\
   
  \multicolumn{1}{c}{Id.} &\multicolumn{1}{c}{name}  &  \multicolumn{1}{c}{name} &
  \multicolumn{1}{c}{(\arcsec)} &  \multicolumn{1}{c}{type} \\
 
  \hline

1	&	G010.296$-$00.148	&	G010.3040$-$00.1466	&	28.6	&	Diffuse HII region	\\
1	&	G010.323$-$00.165	&	G010.3208$-$00.1570A	&	32.6	&	HII region	\\
1	&	G010.323$-$00.165	&	G010.3208$-$00.1570B	&	17.7	&	YSO	\\
2	&	G010.322$-$00.229	&	G010.3207$-$00.2329	&	8.6	&	HII region	\\
2	&	G010.322$-$00.257	&	G010.3204$-$00.2616	&	6.2	&	HII region	\\
3	&	G010.440+00.003	&	G010.4413+00.0101	&	27.7	&	HII region	\\
3	&	G010.474+00.028	&	G010.4616+00.0327	&	47.0	&	HII region	\\
3	&	G010.474+00.028	&	G010.4718+00.0206	&	36.1	&	HII region	\\
3	&	G010.474+00.028	&	G010.4718+00.0256	&	6.2	&	HII region	\\
4	&	G010.625$-$00.339	&	G010.6291$-$00.3385	&	16.5	&	HII region	\\
  \hline
\end{tabular}\\
\end{minipage}
Notes: Only a small portion of the data is provided here, the full table is only  available in electronic form at the CDS via anonymous ftp to cdsarc.u-strasbg.fr (130.79.125.5) or via http://cdsweb.u-strasbg.fr/cgi-bin/qcat?J/A+A/.
\end{center}
\end{table}

\setlength{\tabcolsep}{6pt}

\begin{figure}\begin{center}
\includegraphics[width=0.45\textwidth, trim= 0 0 0 0]{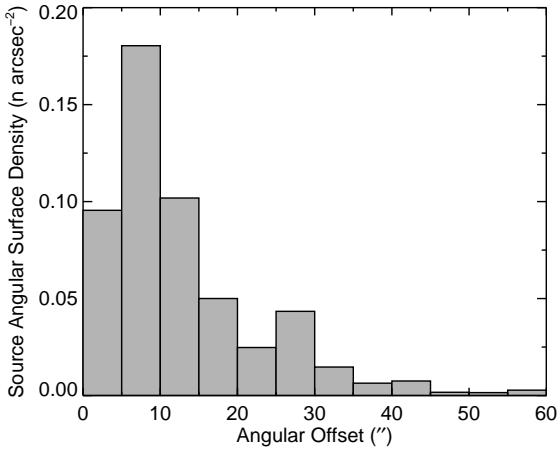}

\caption{Distribution of the surface density of sources with a given angular separation between the peak emission from the clump and the nearest RMS association. The bin size is 5\arcsec. } 
\label{fig:gbt_rms_offset}
\end{center}\end{figure}

In Table\,\ref{tbl:rms_associations} we present a list of RMS and clump associations, the RMS classification, and the angular separation between the peak emission position of the ammonia and the position of the embedded \hii\ region or MYSO. In Fig.\,\ref{fig:gbt_rms_offset} we show the distribution of angular separations between the embedded RMS sources and the peak of the integrated NH$_3$ (1,1) emission. This plot reveals a strong positional correlation between the embedded massive stars and the peak ammonia emission seen towards the centre of the clumps,  with the vast majority of embedded objects having offsets of less than 10\arcsec. We find no significant difference in the separations between the MYSOs and the \hii\ regions. The star formation  is therefore primarily taking place towards the centres of centrally condensed clumps where the column densities are highest. 

We note that the peak of the offset distribution is shifted from zero, with the embedded sources typically found between 5-10\arcsec\ from the ammonia emission peak. This is larger than the nominal pointing error ($\sim$4\arcsec) and is a little surprising given that a tighter correlation between the embedded sources and the peak of the dust emission has been previously observed (i.e., \citealt{urquhart2014a, urquhart2014c}). A similar angular offset was recently reported by \citet{morgan2014} from a comparison of the dust and ammonia emission peaks from clumps associated with the Perseus molecular cloud. These authors suggested that these two tracers may be sensitive to different conditions or structure in the clumps. In many cases the ammonia emission is clearly much more extended than the dust emission (e.g., G011.918$-$00.618 and G014.328$-$00.646), which would support this hypothesis. Another possibility is that feedback from the embedded massive stars is starting to alter the structure and composition of their local environments, resulting in a shift in position of the peak column density away from the massive stars (\citealt{thompson2006}). 

Although the offset between the peak of the ammonia emission and the embedded source is significantly larger than that found between the embedded source and the dust emission, the difference is only a fraction of the GBT beam and so is relatively small. Furthermore, although the angular offset is noticeable we will see in Sect.\,3.5 that the actual physical offset is less obvious once the distance to the source has been taken into account. 

\subsection{Mid-infrared imaging}

To investigate the embedded protostellar population and evaluate the influence of the local environment we have extracted mid-infrared images from the GLIMPSE legacy survey (\citealt{benjamin2003_ori,churchwell2009}). We combined the 3.5, 4.5 and 8\,\mum\ band images obtained with the IRAC instrument (\citealt{fazio2004}) to produce three-colour images of the mid-infrared environment. 

These images are very sensitive to embedded objects such as MYSOs and compact \hii\ regions, and provide a census of the embedded stellar content of these clumps (see lower right panel of Fig.\,\ref{fig:distribution_maps}). Extinction  results in a greater reddening for the more deeply-embedded objects in these three-colour images. As a result, these images not only reveal the stellar content and their luminosity but also provide hints to their evolutionary stage. 

The 8\,\mum\ band is sensitive to emission from polycyclic aromatic hydrocarbons (PAHs) that are excited by UV-radiation from embedded compact \hii\ regions as well as large nearby \hii\ regions, and thus is an excellent tracer of the interaction regions between ionized and molecular gas. The 4.5\,\mum\ band includes excited H$_2$ and CO transitions which are thought to trace shocked gas driven by powerful molecular outflows. Excess emission seen in this band is therefore considered a good tracer of massive star formation (e.g., \citealt{cyganowski2008,cyganowski2009}). These images can therefore be useful to understand some of the temperature and velocity structure observed in the clumps with respect to the position of the embedded MYSOs and \hii\ regions.

\subsection{General properties of the sample}
\label{sect:general_properties}

The statistics for the parameters derived in the previous section for the whole sample are given in Table\,\ref{tbl:derived_para} along with bolometric luminosities taken from the RMS survey; where the median and peak values are significantly different we provide both values.

\begin{figure*}\begin{center}

\includegraphics[width=0.33\textwidth, trim= 0 0 0 0]{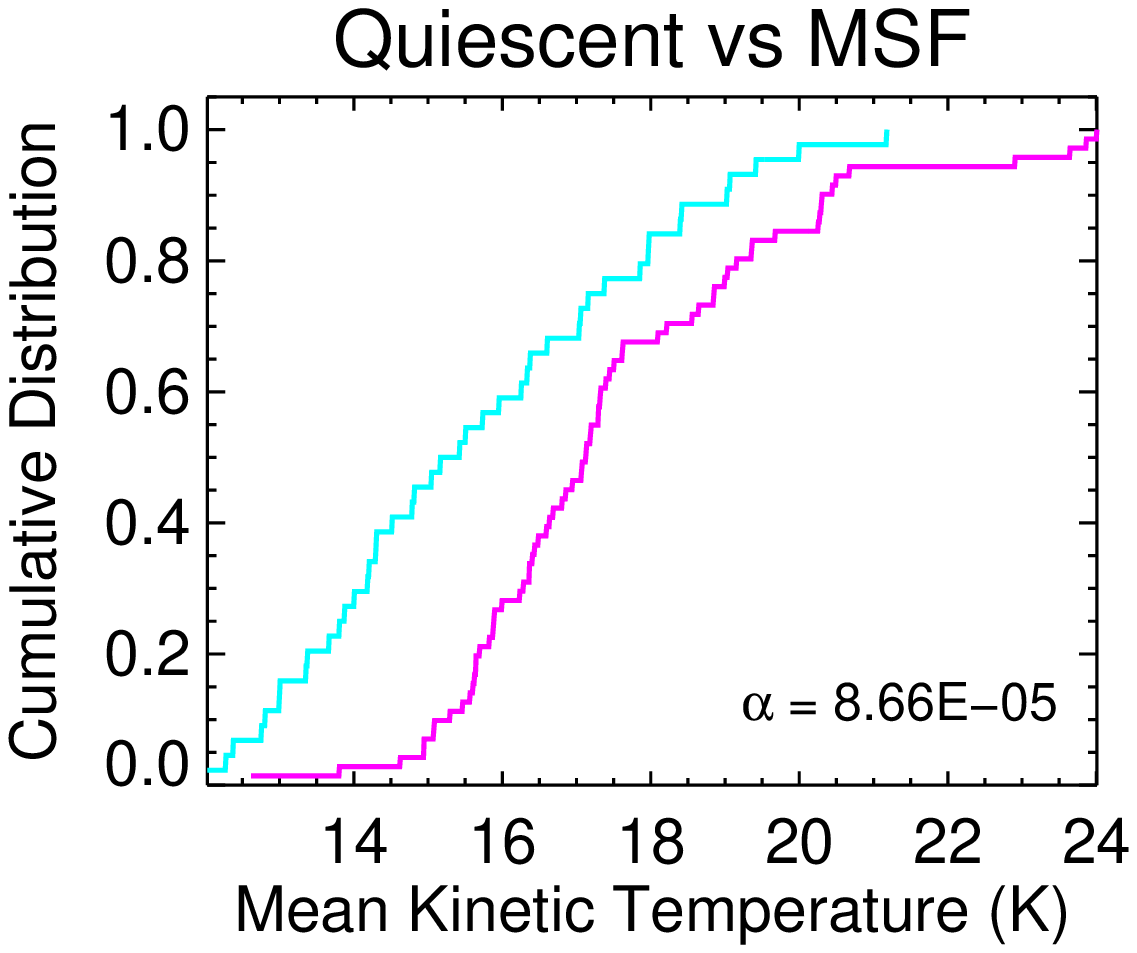}
\includegraphics[width=0.33\textwidth, trim= 0 0 0 0]{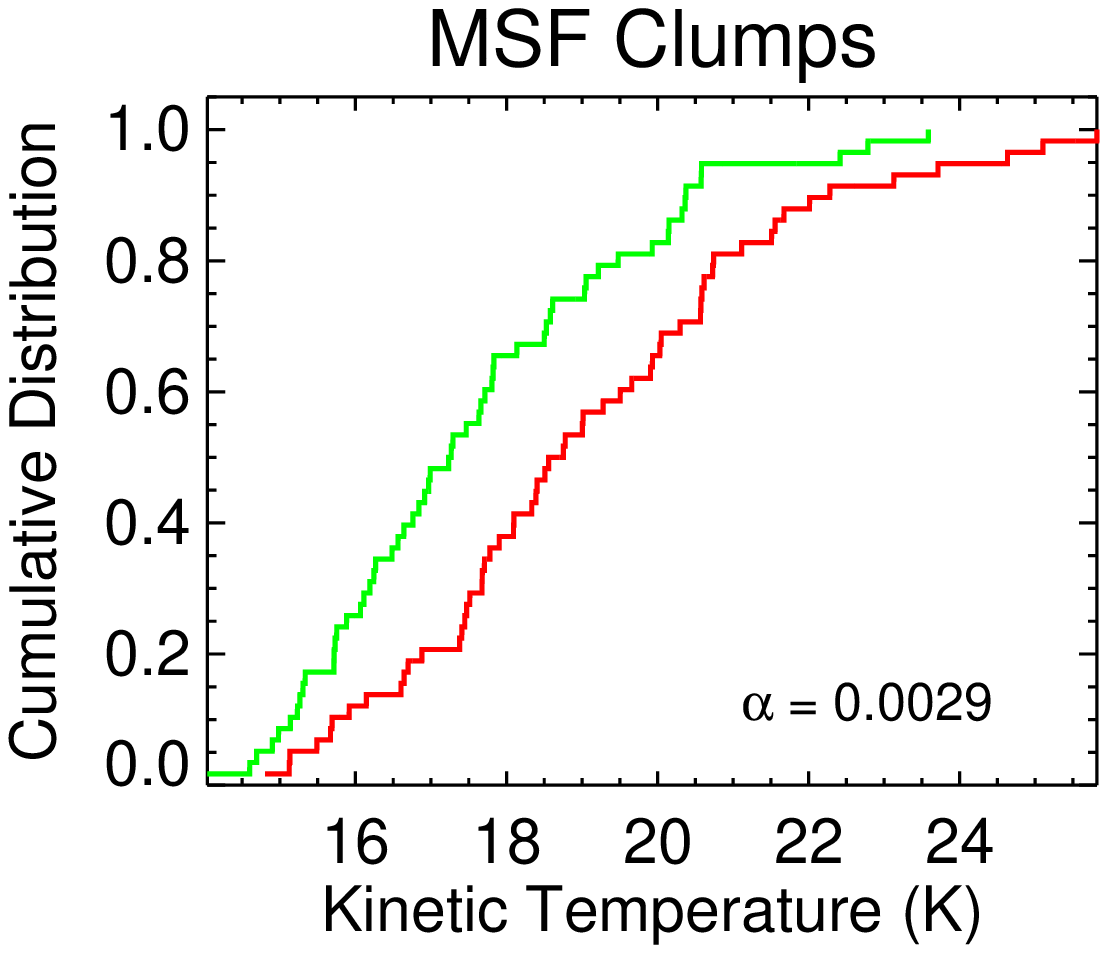}
\includegraphics[width=0.33\textwidth, trim= 0 0 0 0]{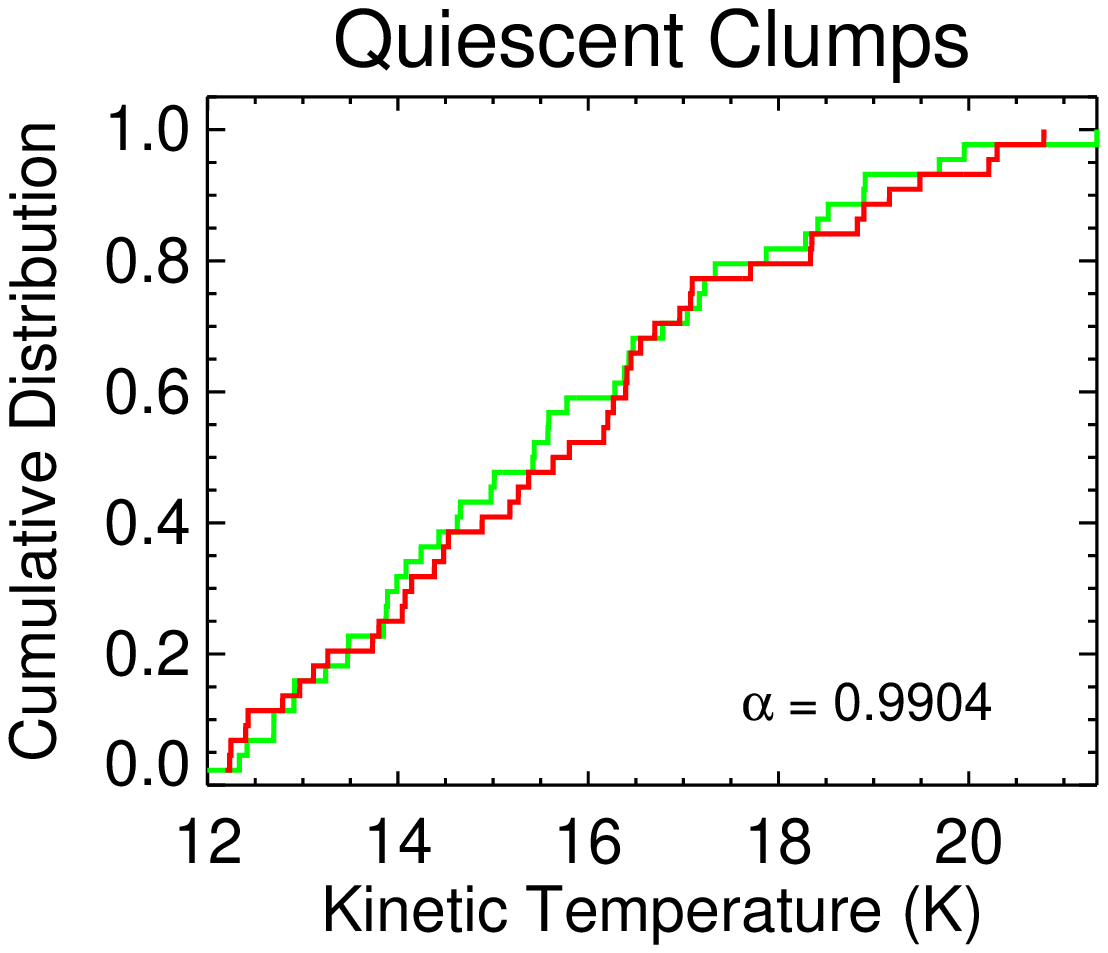}\\
\includegraphics[width=0.33\textwidth, trim= 0 0 0 0]{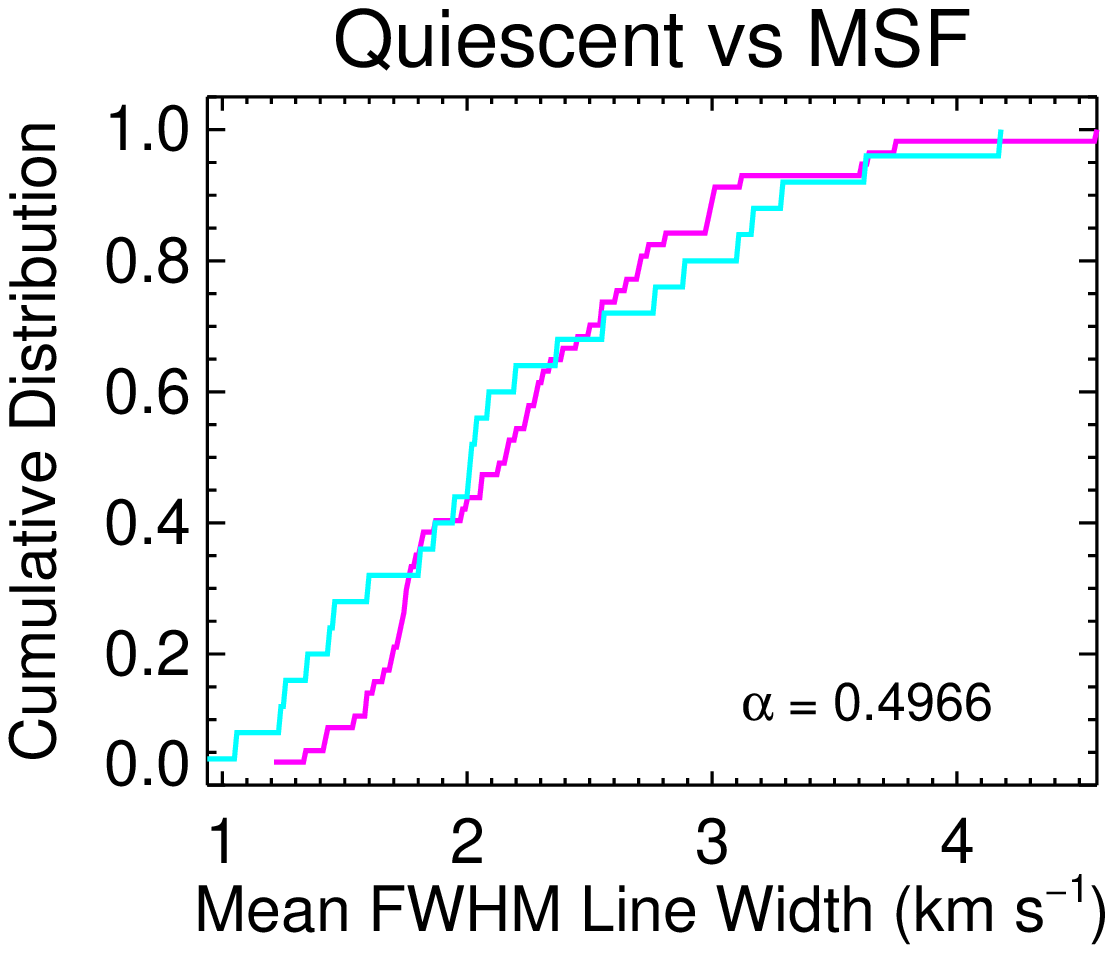}
\includegraphics[width=0.33\textwidth, trim= 0 0 0 0]{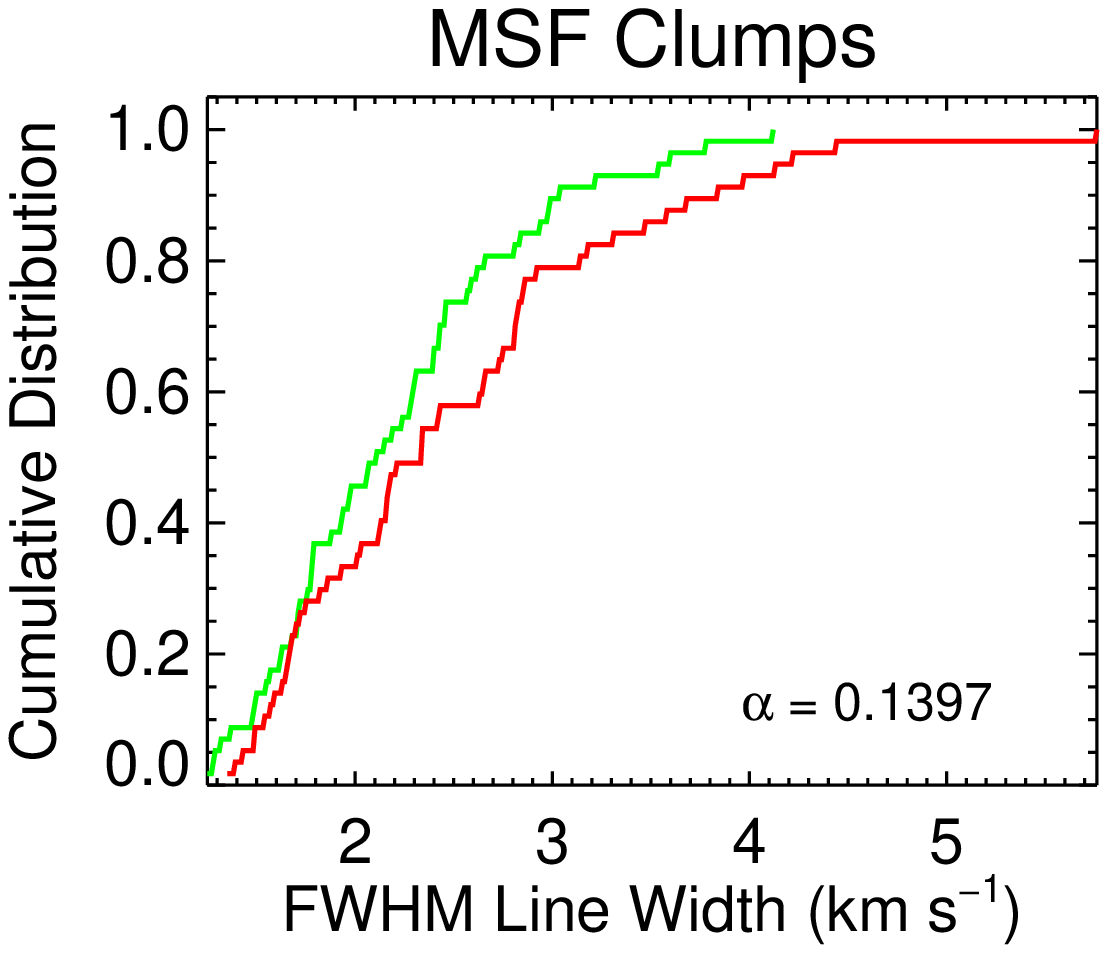}
\includegraphics[width=0.33\textwidth, trim= 0 0 0 0]{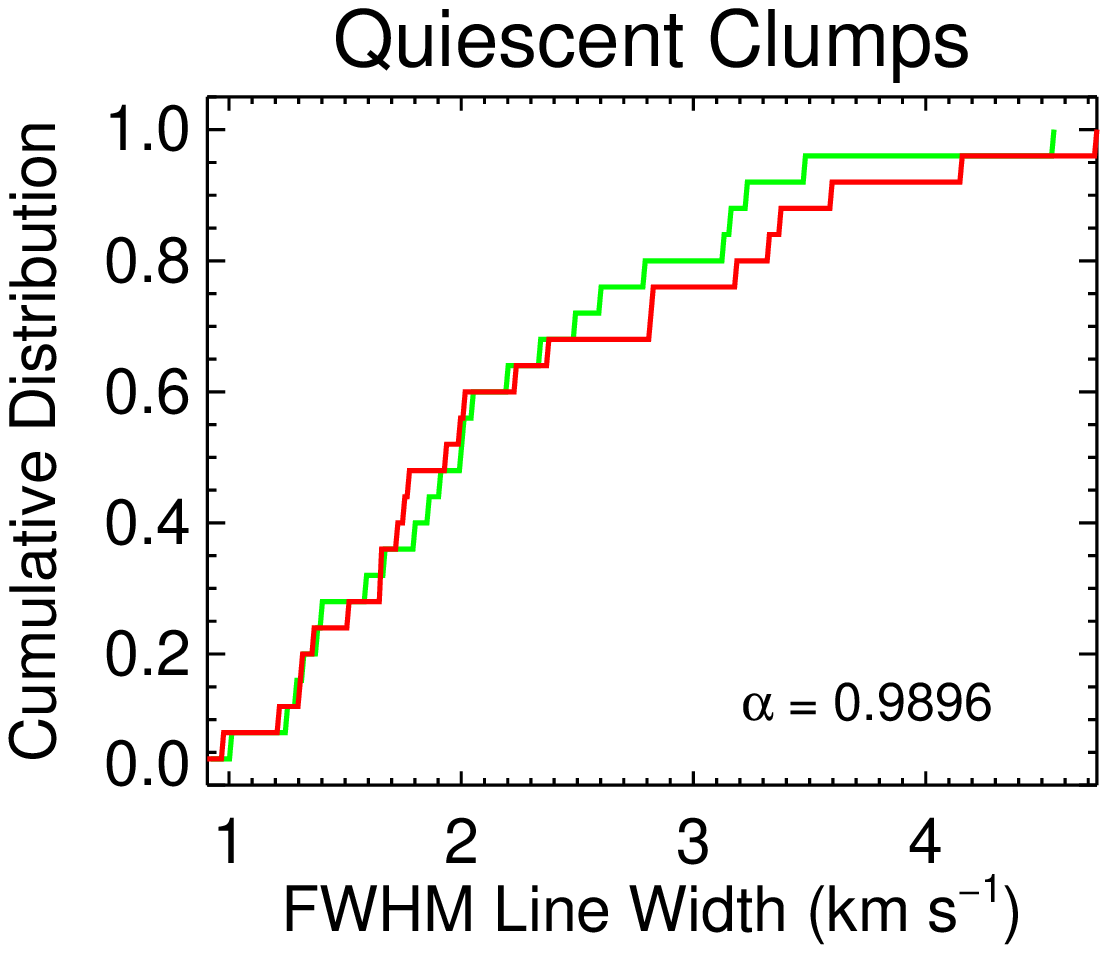}\\
\includegraphics[width=0.33\textwidth, trim= 0 0 0 0]{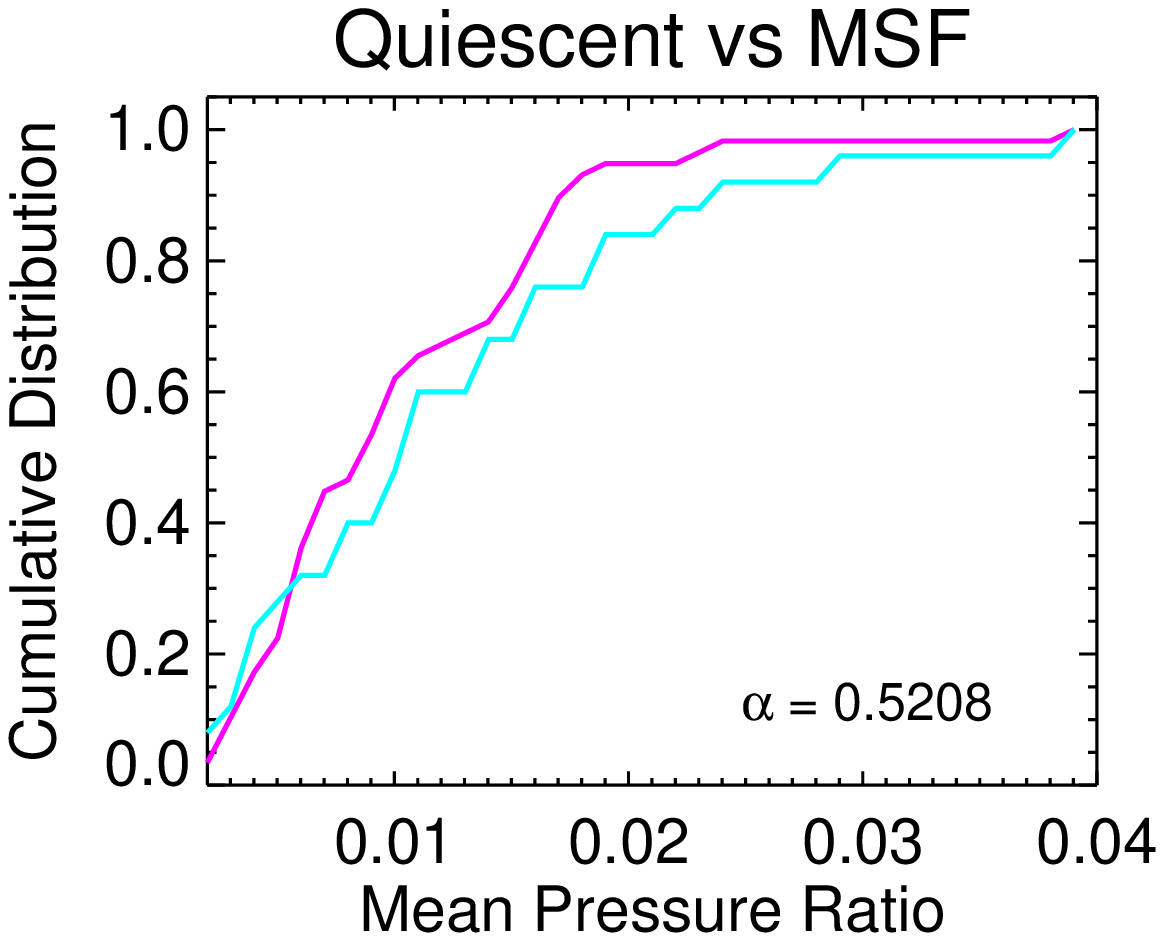}
\includegraphics[width=0.33\textwidth, trim= 0 0 0 0]{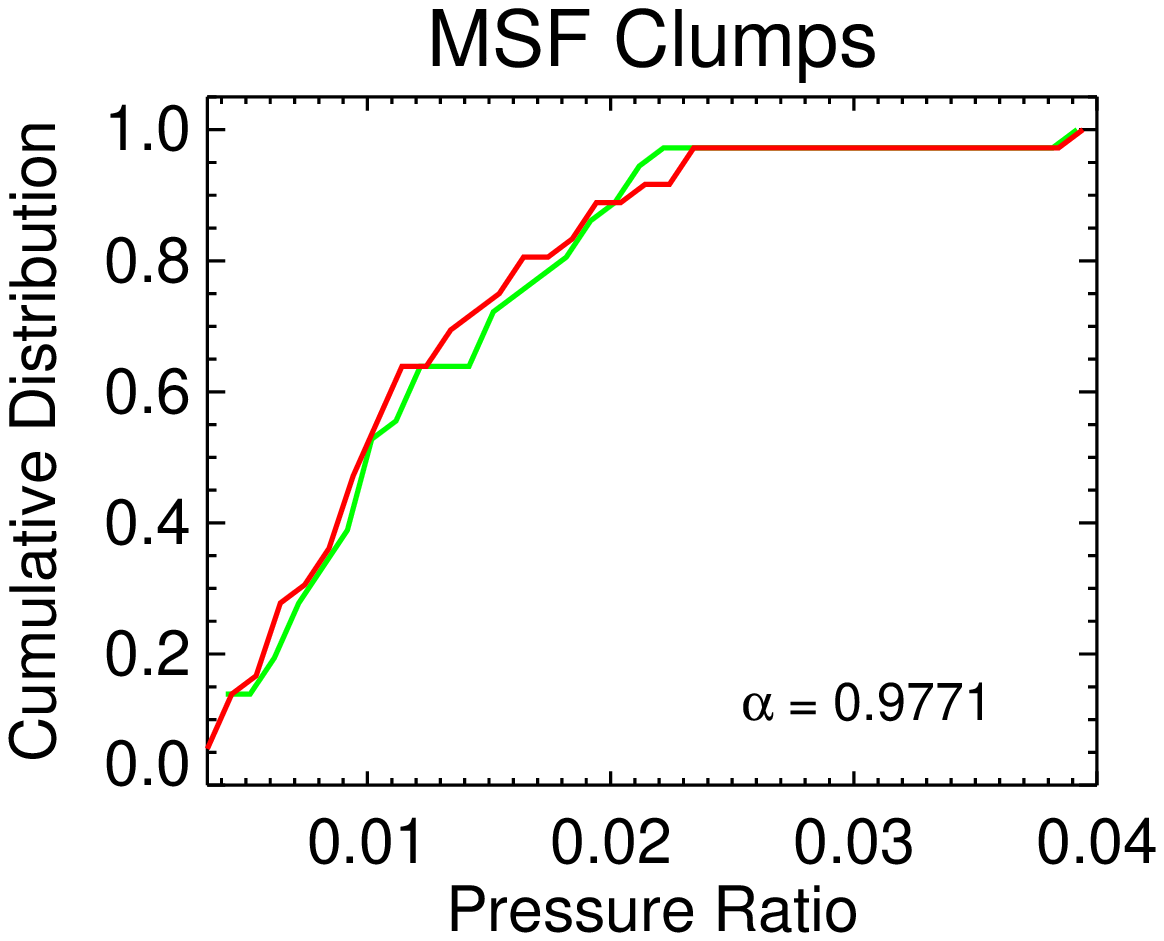}
\includegraphics[width=0.33\textwidth, trim= 0 0 0 0]{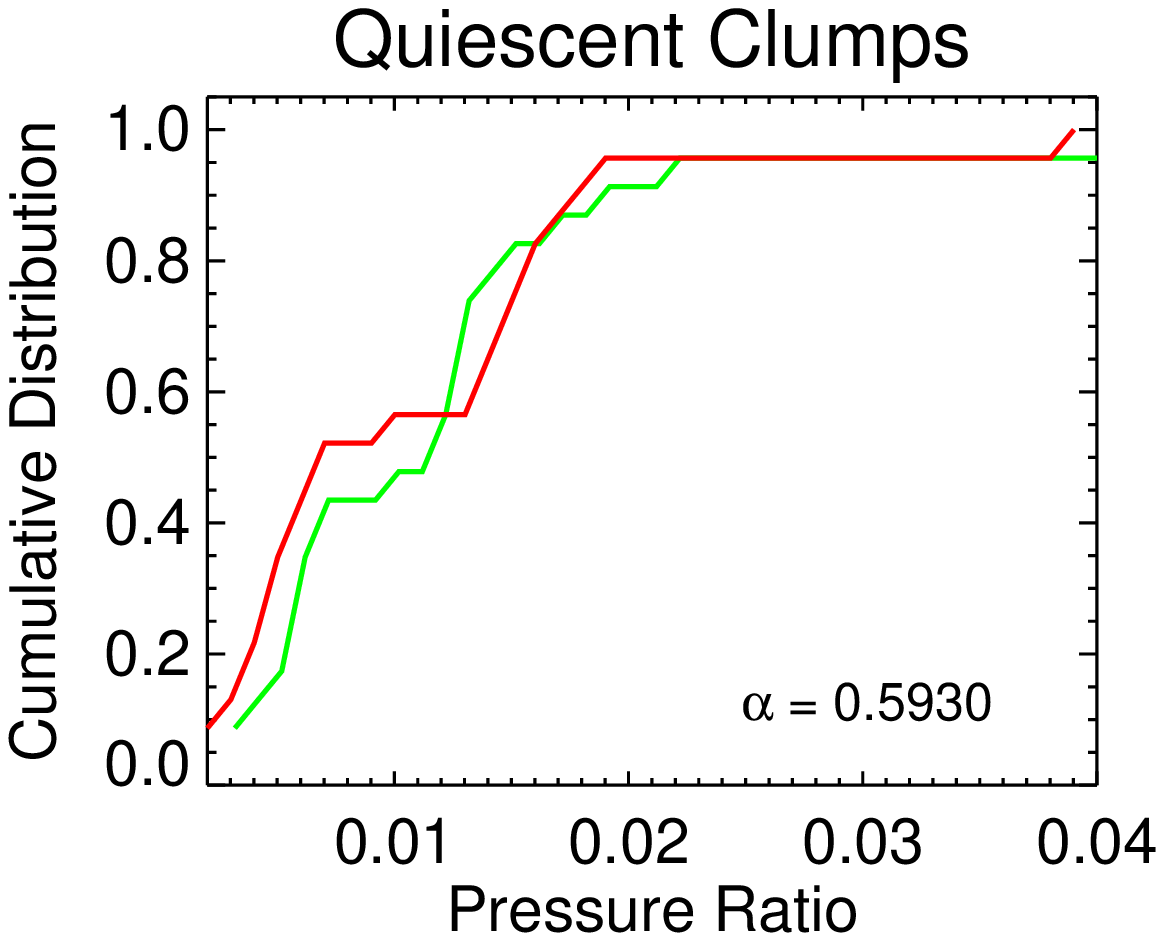}\\

\caption{\label{fig:full_sample_CDF} Left panels: Cumulative distribution plots comparing the mean values for the kinetic temperature, $FWHM$ line width and pressure ratio for the MSF and quiescent clumps (cyan and magenta curves, respectively). Middle and right panels: Cumulative distribution plots comparing the peak and median values (shown in red and green, respectively) for the MSF and quiescent clumps  for the same parameters. These allow us to identify any significant differences between the internal and external structure of the clumps. The results of KS tests comparing the peak and median values is given in the lower right corner of each plot. For these plots and the discussion in the text we only include sources with a SNR $>$10.} 

\end{center}\end{figure*}

\begin{figure*}\begin{center}

\includegraphics[width=0.32\textwidth, trim= 0 0 0 0]{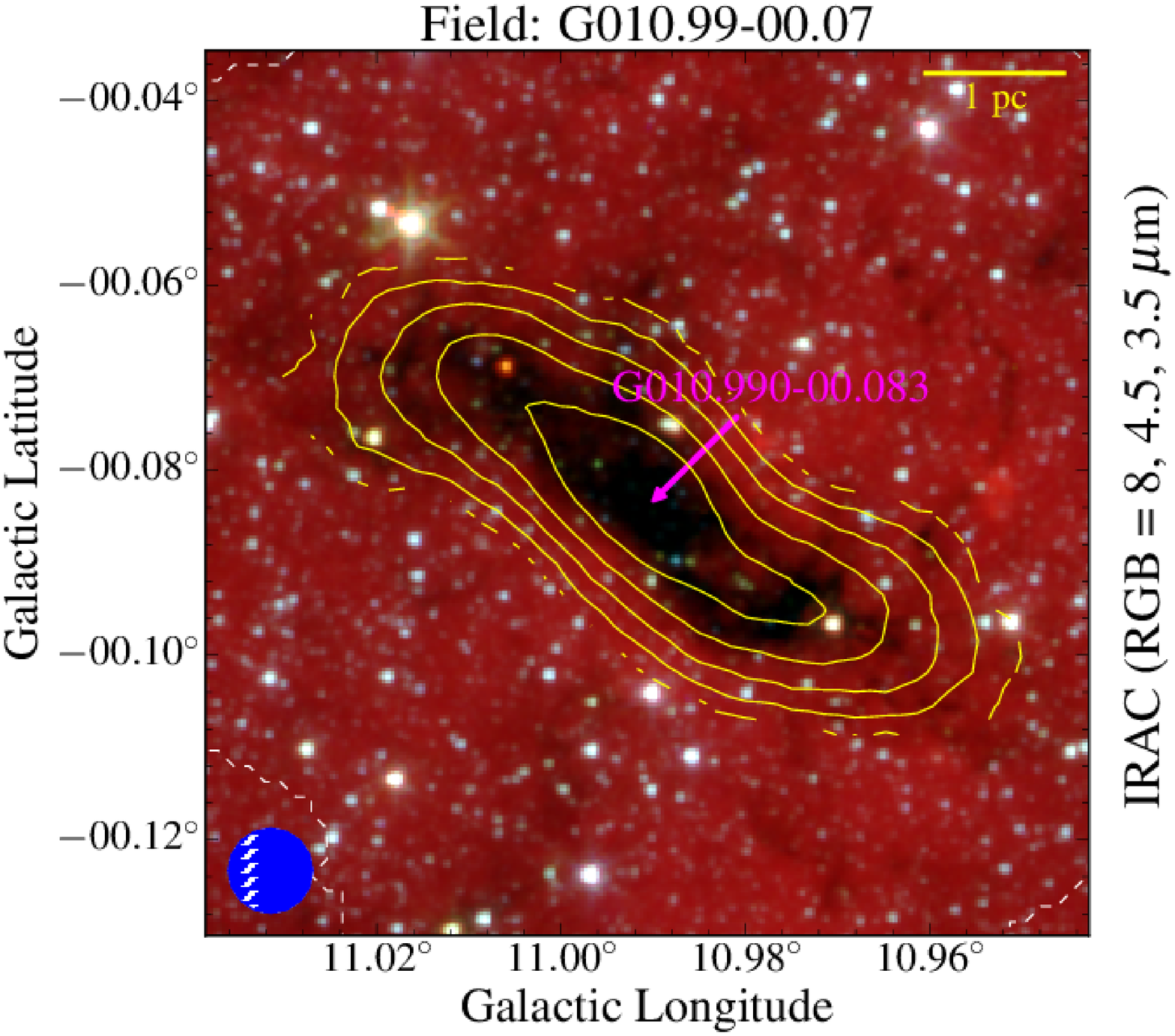}
\includegraphics[width=0.32\textwidth, trim= 0 0 0 0]{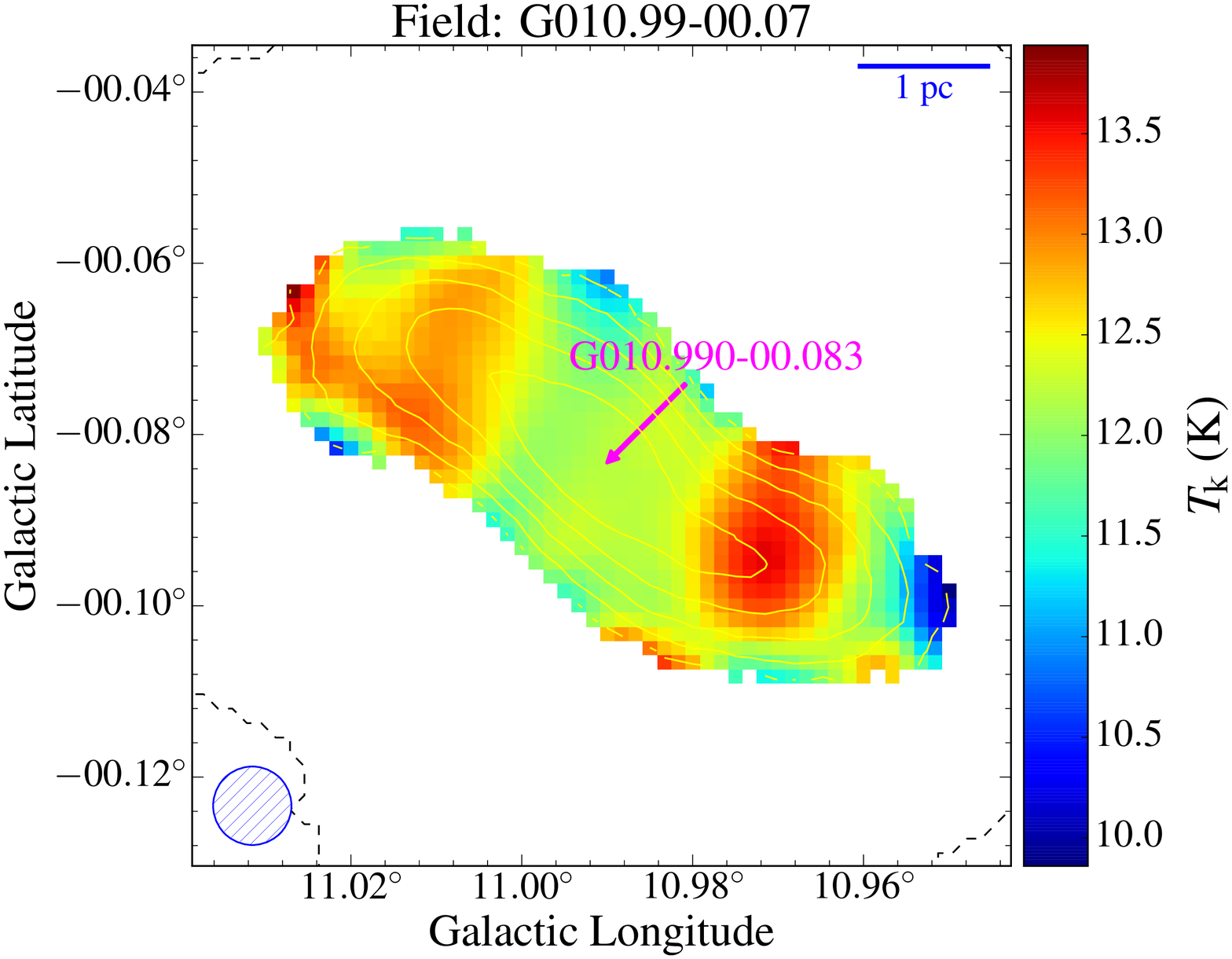}
\includegraphics[width=0.32\textwidth, trim= 0 0 0 0]{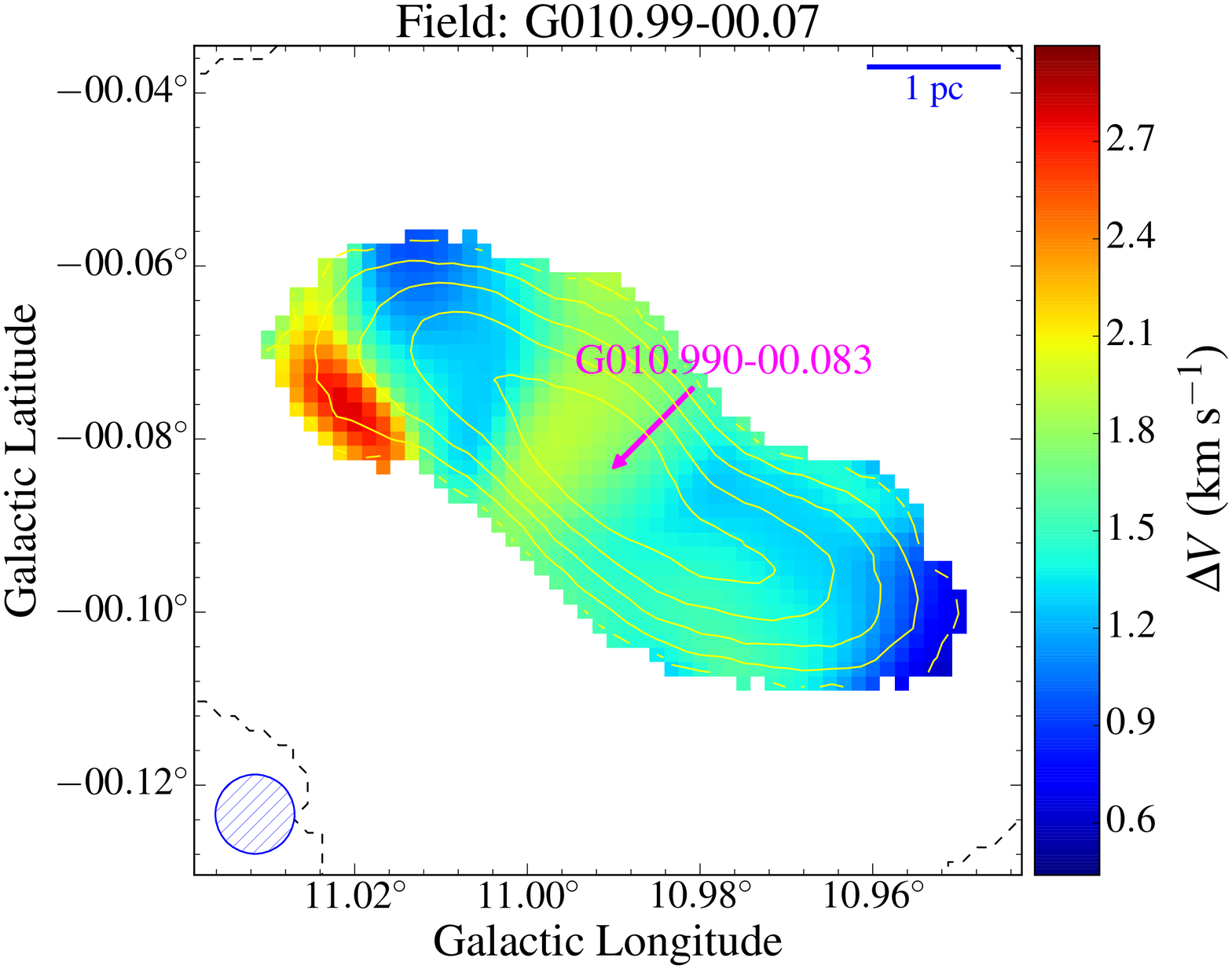}\\
\includegraphics[width=0.32\textwidth, trim= 0 0 0 0]{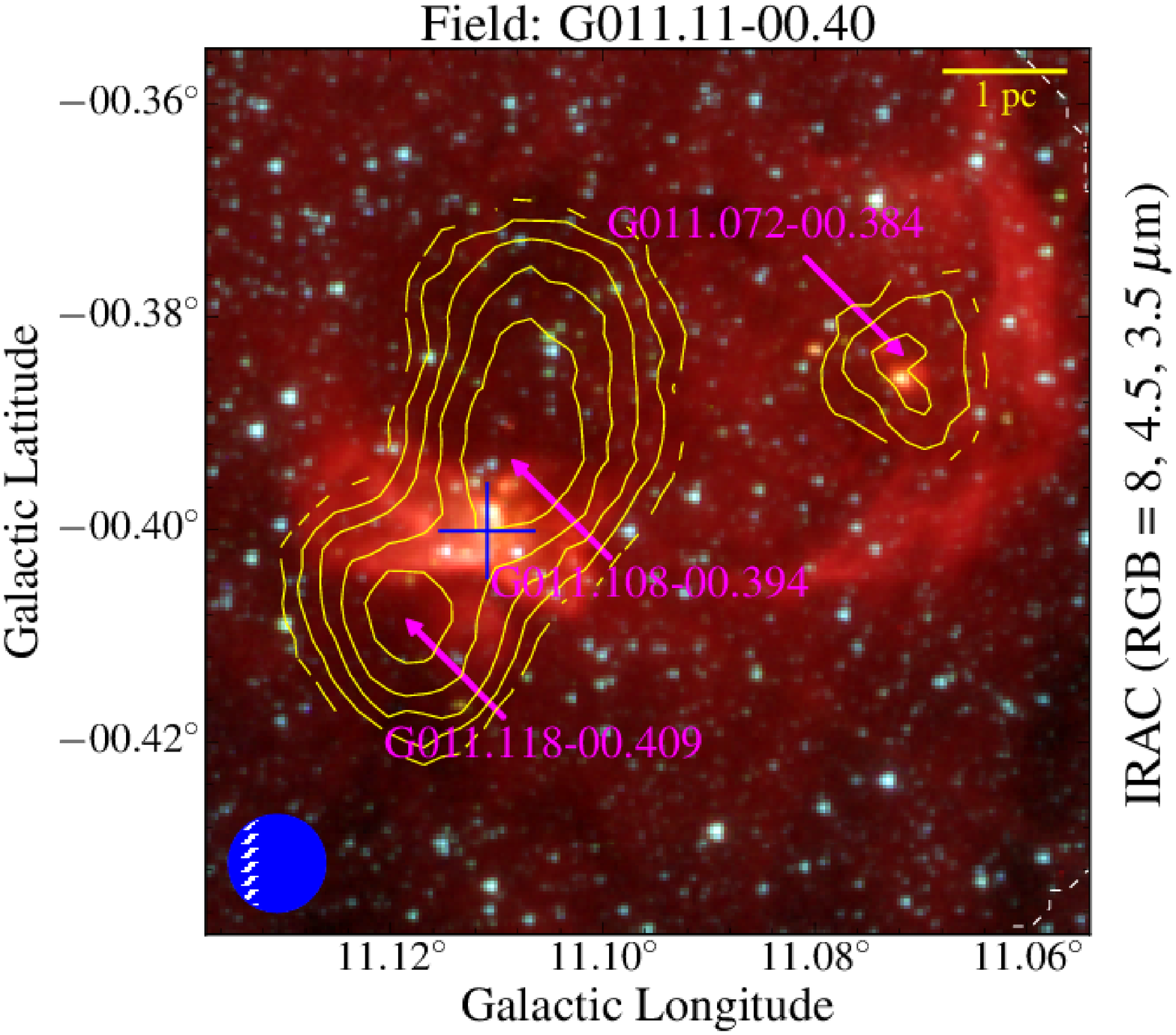}
\includegraphics[width=0.32\textwidth, trim= 0 0 0 0]{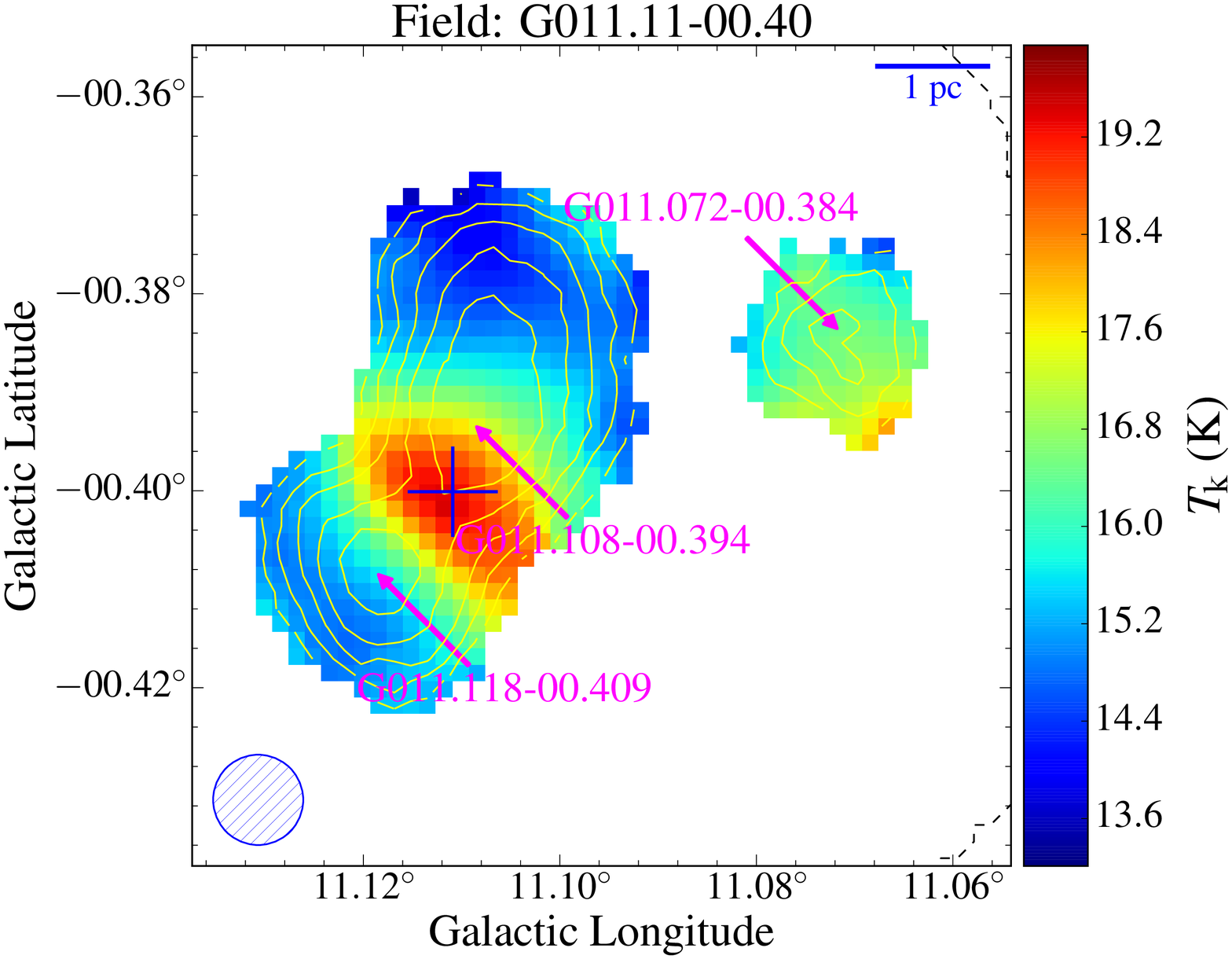}
\includegraphics[width=0.32\textwidth, trim= 0 0 0 0]{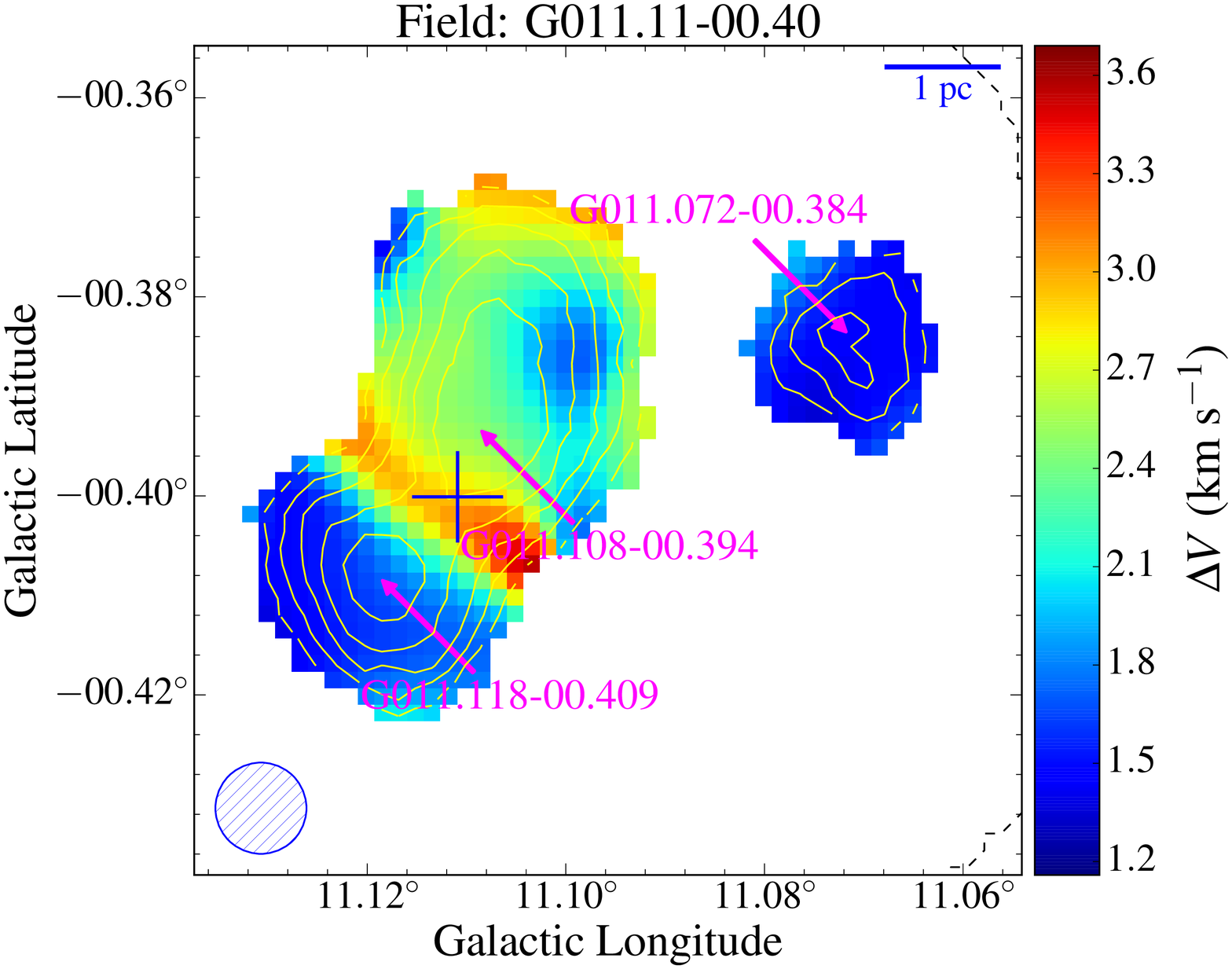}\\
\includegraphics[width=0.32\textwidth, trim= 0 0 0 0]{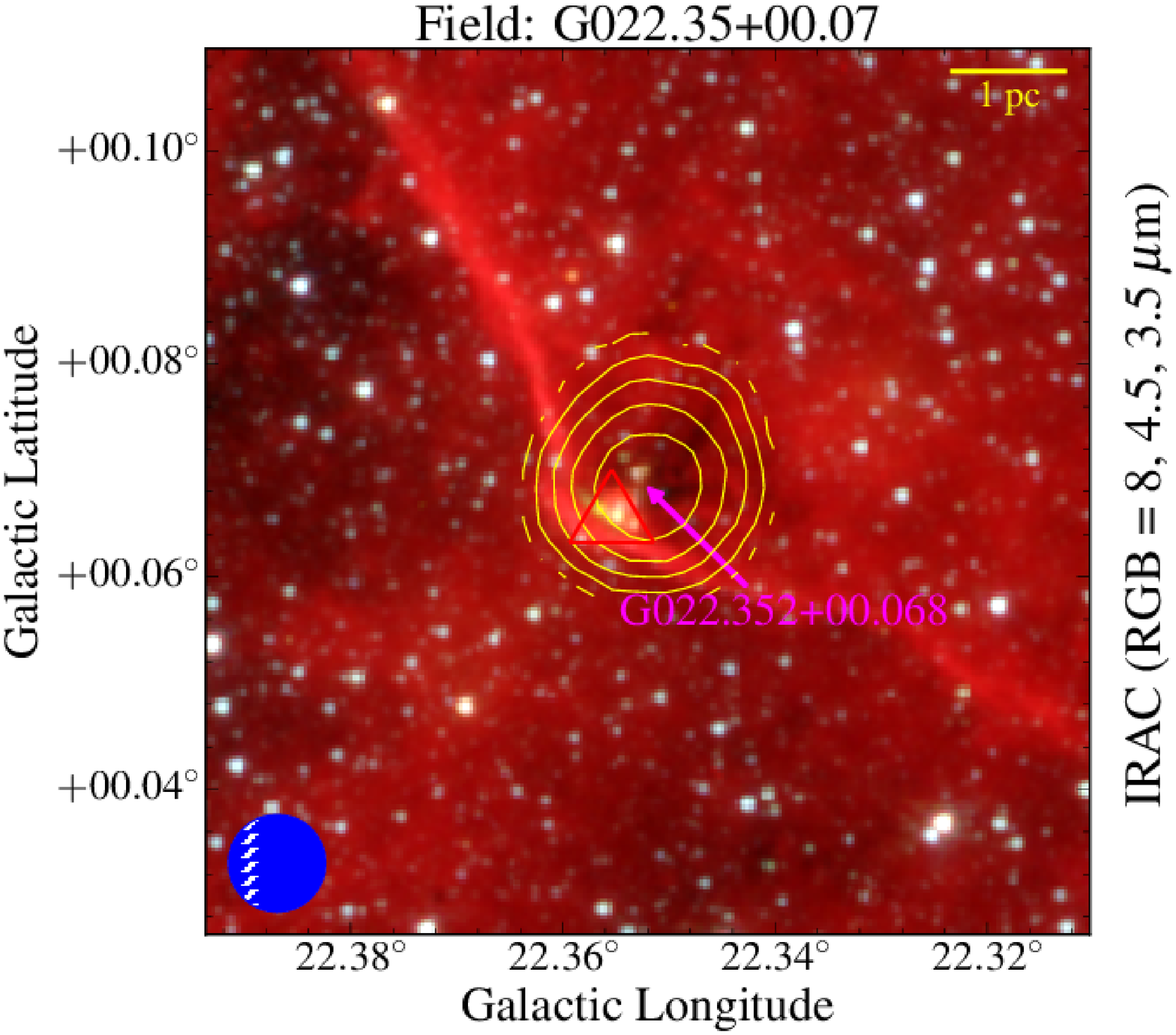}
\includegraphics[width=0.32\textwidth, trim= 0 0 0 0]{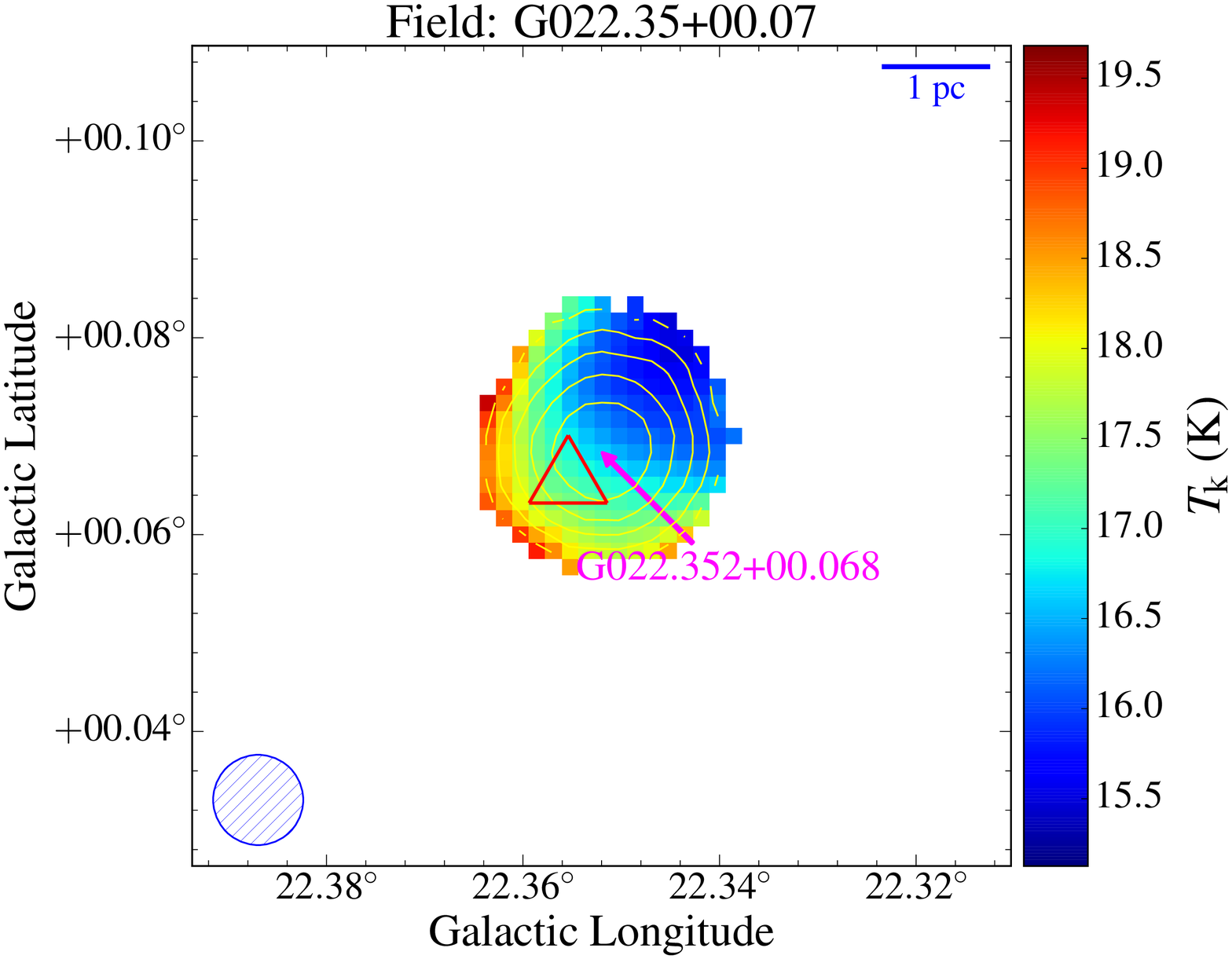}
\includegraphics[width=0.32\textwidth, trim= 0 0 0 0]{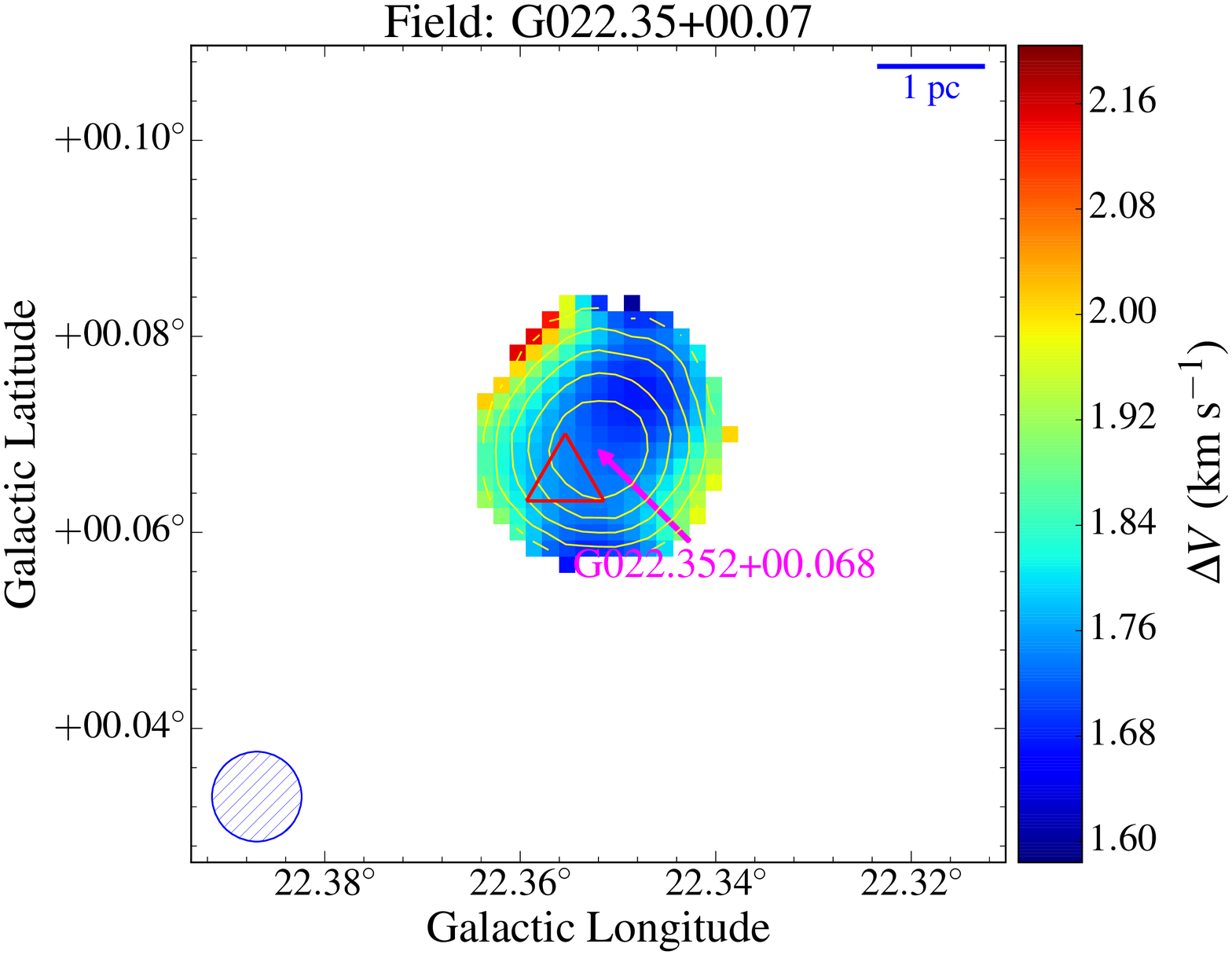}\\
\includegraphics[width=0.32\textwidth, trim= 0 0 0 0]{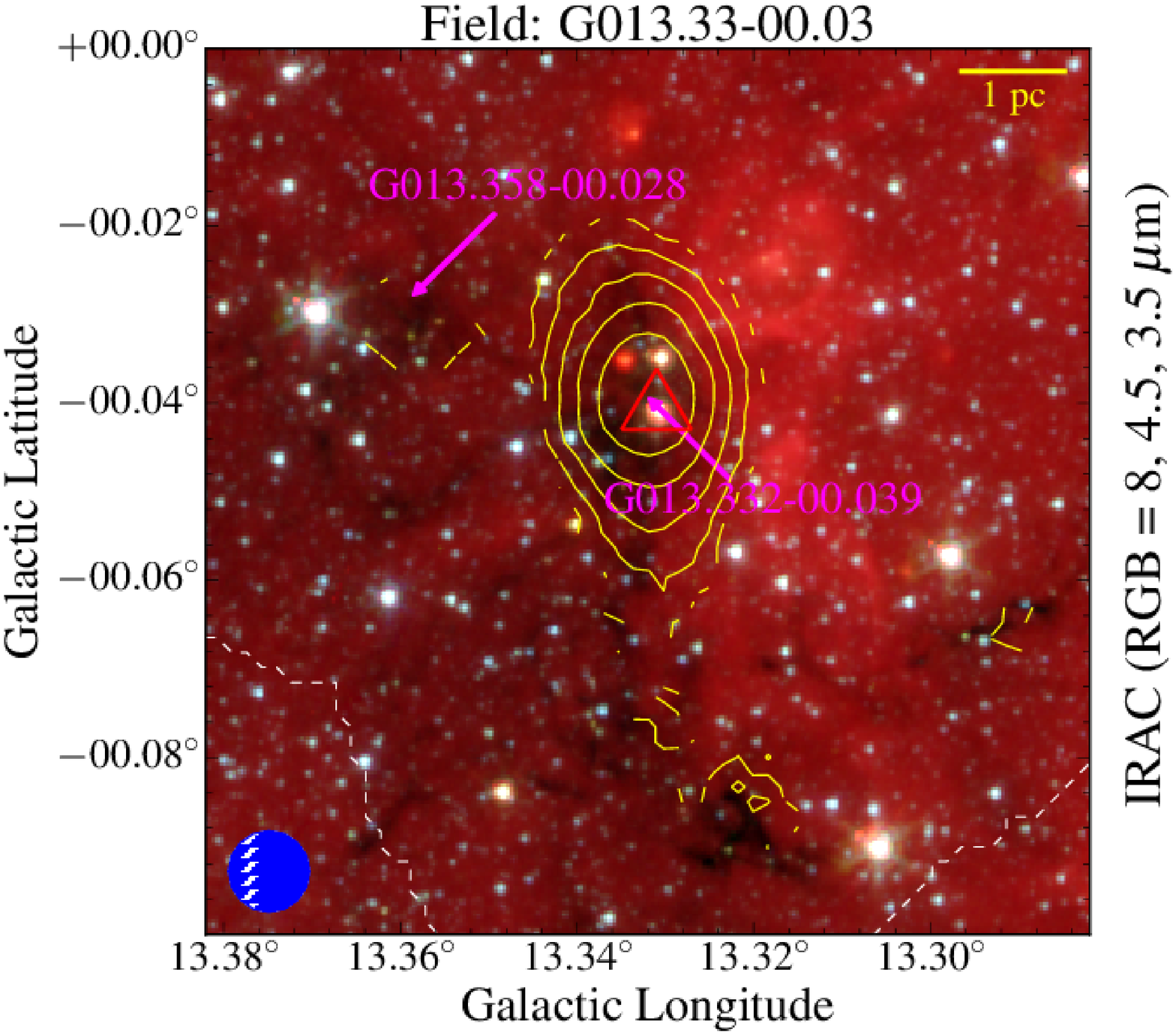}
\includegraphics[width=0.32\textwidth, trim= 0 0 0 0]{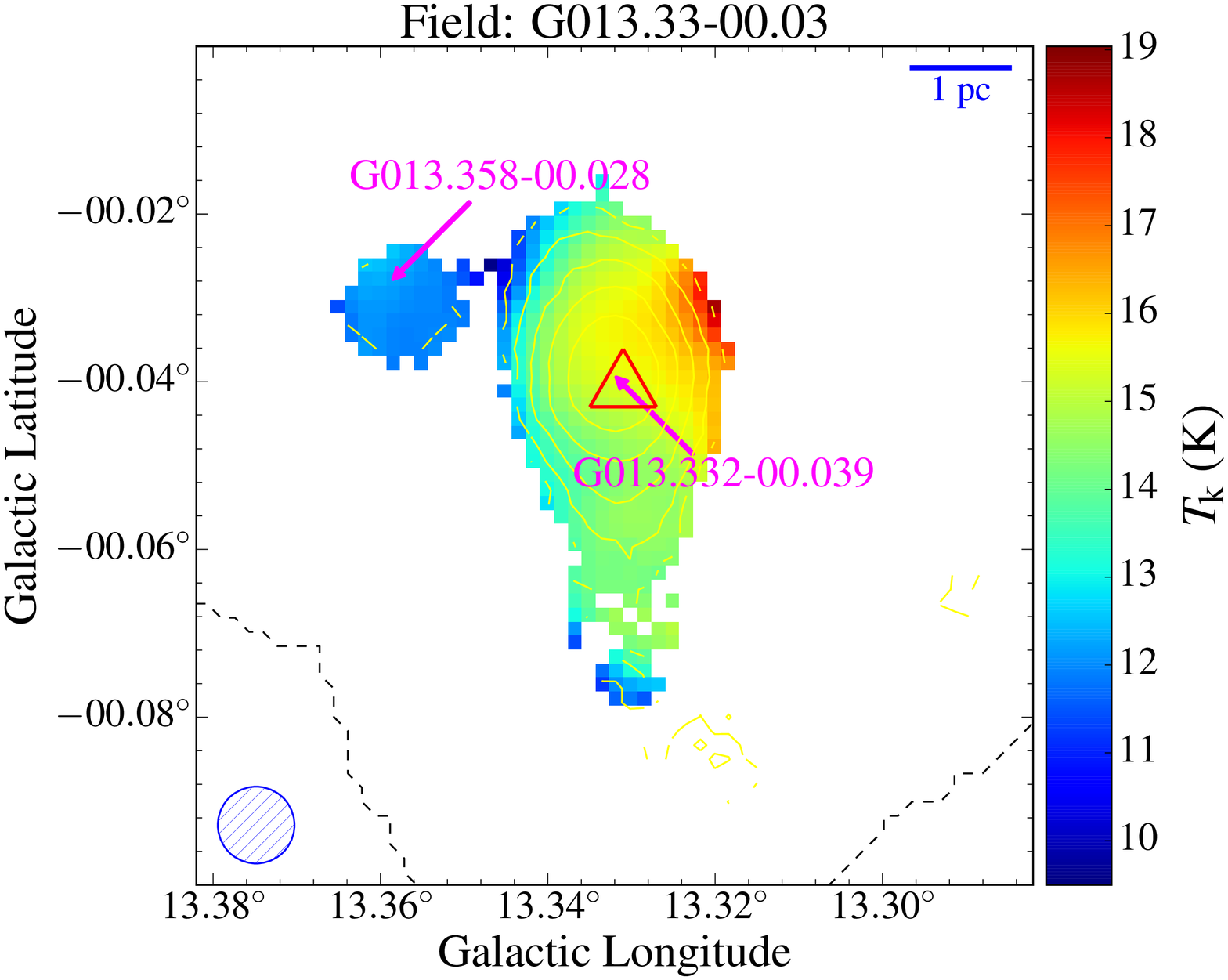}
\includegraphics[width=0.32\textwidth, trim= 0 0 0 0]{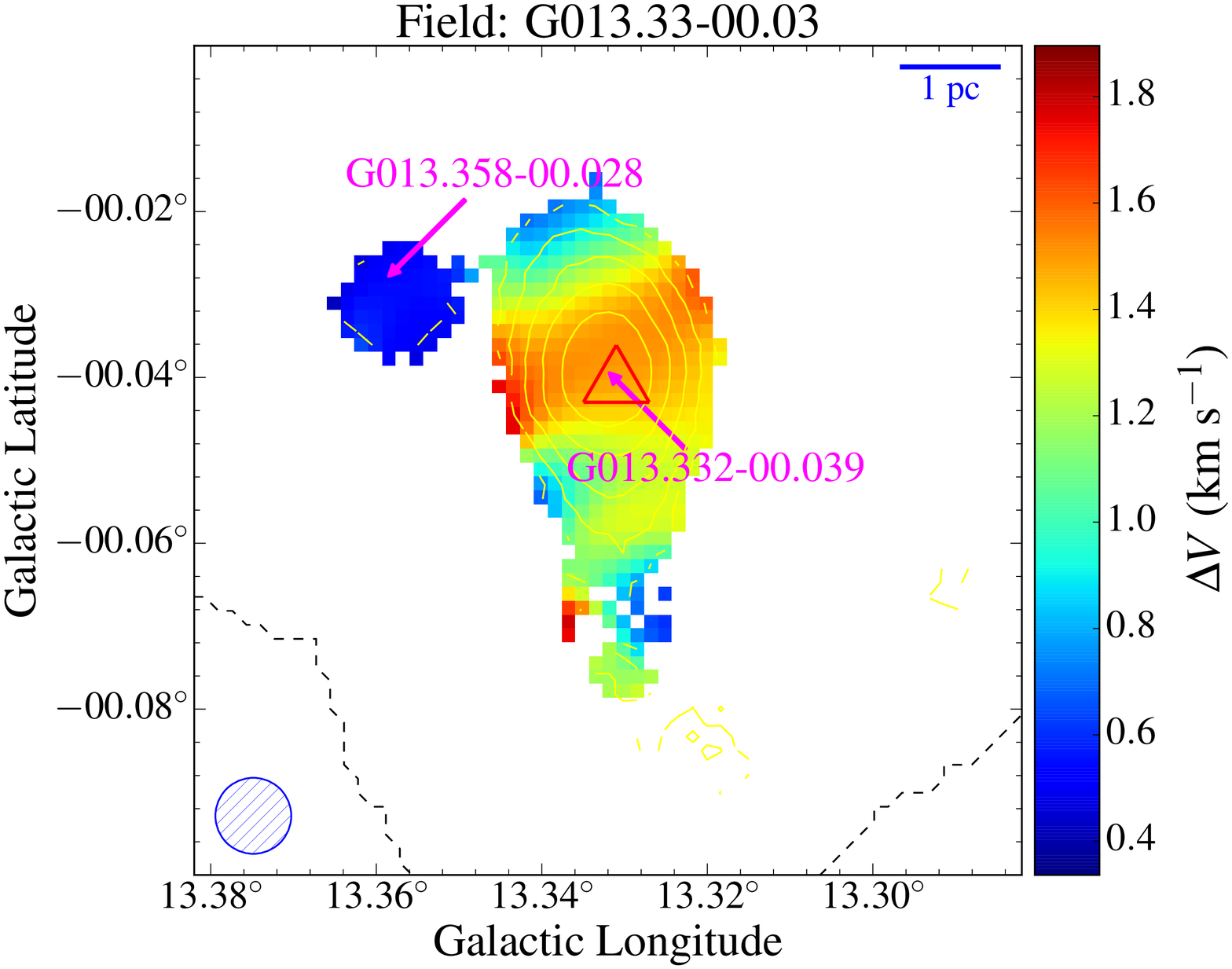}\\

\caption{\label{fig:temp_linewidth_maps} Three-colour mid-infrared images and distribution maps of the kinetic temperature and line width for a range of clump types. The contours are the same in every map and trace the integrated NH$_3$ (1,1) emission as described in Fig.\,\ref{fig:example_emission_maps} while the symbols used and labels are as described in Fig.\,\ref{fig:distribution_maps}. In the upper, upper-middle, lower-middle and lower panels, respectively, we present examples of: a quiescent clump showing a relatively smooth temperature and velocity distribution; an example of a clump associated with an RMS source that is coincident with a localised increase in both the line-width and temperature, which is indicative of feedback from the embedded \hii\ region; a clump located in the edge of a \hii\ region, which is having an impact on the exposed side of the clump; and a clump associated with MYSOs that is coincident with a localised enhancement of the line-widths, which may be linked to the molecular outflow.} 

\end{center}\end{figure*}

Nearly all of the clumps are extended with respect to the beam and have a relatively simple morphology with their emission distribution being reasonably well described by an ellipse with aspect ratio $\sim$1.5. The internal structures and dynamics of the gas of many of the clumps, however, are complicated and hard to interpret. This is made more difficult by the external environment that is heating the surface layers of the clumps and affecting the kinematics. As noted by \citet{ragan2011}, a single model is unlikely to account for the large range of properties observed; however, in this section we attempt to give an overview of the general properties of the clumps and will endeavour to provide a more detailed discussion of some of more interesting differences in later sections.

We find a correlation between the position angles of the clumps and the Galactic plane, such that the semi-major axis of the clumps are preferentially aligned parallel to the Galactic mid-plane (see Fig.\,\ref{fig:pa_hist}). Furthermore, examination of the velocity information reveals typical velocity gradients of a couple of \kms\ and these tend to also be aligned along the semi-major axis. There are relatively few examples of clumps that show no sign of a velocity gradient (e.g., G022.412+00.316).

We present in Fig.\,\ref{fig:full_sample_CDF} the cumulative distribution plots comparing the median and peak values for the kinetic temperature, FWHM line width, and the gas pressure ratio ($R_{\rm{p}}$) for the whole sample and the MSF and quiescent subsamples. We use the peak and median values of these parameters to compare the conditions towards the centres of the clumps with their outer envelopes, and refer to these as `inner' and `outer envelope' values, respectively.  In the lower right corner of these plots we give the results of \KS\ (KS) tests, $\alpha$, which is the probability of the two samples being drawn from the same population. In order to reject the null hypothesis that any two distributions are drawn from the same parent distribution with greater than 3$\sigma$ confidence, the value of $\alpha$ must be lower than 0.0013. In Fig.\,\ref{fig:temp_linewidth_maps} we present the mid-infrared image and temperature and line width distribution maps for a selection of clumps to help illustrate some of the features discussed in the following paragraphs.

Comparison of the MSF and quiescent distributions for these three parameters reveals that the only significant difference between them is their kinetic temperatures, which are clearly higher for the MSF clumps. 

Inspection of the kinetic temperature and line width cumulative distribution plots (Fig.\,\ref{fig:full_sample_CDF}) for the quiescent clumps indicates that there is no significant difference between the inner and outer envelopes of these sources. This is also seen in the example distribution map presented in the upper panel of Fig.\,\ref{fig:temp_linewidth_maps}, where the quiescent clump displays a relatively flat distribution for these parameters. This is consistent with the interpretation that these clumps are in a more homogeneous (`pristine') state prior to the onset of star formation. The distributions of these two parameters are very different for the MSF clumps where the internal temperatures and line widths are clearly seen to be higher than in the outer envelope. 

A temperature gradient is clearly seen in many of the distribution maps presented in Fig.\,\ref{fig:distribution_maps}, with the peaks often coincident with the positions of the embedded infrared sources seen in the IRAC images (see also Fig.\,\ref{fig:temp_linewidth_maps}). The peak temperatures are $\sim$20\,K and decrease to $\sim$12\,K towards the edges of the clumps. This suggests that feedback from the central object is having a significant impact on the temperature and dynamics of the surrounding gas. The KS test confirms that the temperature difference between the inner and outer envelope is statistically significant. Furthermore, the temperature distributions within the MSF clumps display sharper spatial variation and more complex distribution patterns, which is possibly evidence of feedback from the embedded massive stars.

Comparison of the temperature distribution maps reveals a subsample of 34 clumps where the outer envelope is higher than found towards the central region (e.g., G012.887+00.492, G013.332$-$00.039 and G019.080$-$00.290; all of these are identified in Table\,\ref{tbl:cattable}). The majority of these (21) are members of the quiescent sample, but approximately one-third (13) are members of the MSF sample. A negative temperature gradient is not unexpected for the quiescent clumps because their exteriors are exposed to the interstellar radiation field while their inner regions are shielded (e.g., \citealt{wang2008,peretto2010}), but is perhaps a little surprising for the MSF clumps. Examination of the mid-infrared images reveals that many are located near the periphery of large \hii\ regions where the strong radiation fields and expanding ionisation fronts are clearly having an impact on their temperature and velocity structure (see lower middle panels of Fig.\,\ref{fig:temp_linewidth_maps} for an example).

\begin{figure}\begin{center}

\includegraphics[width=0.45\textwidth, trim= 0 0 0 0]{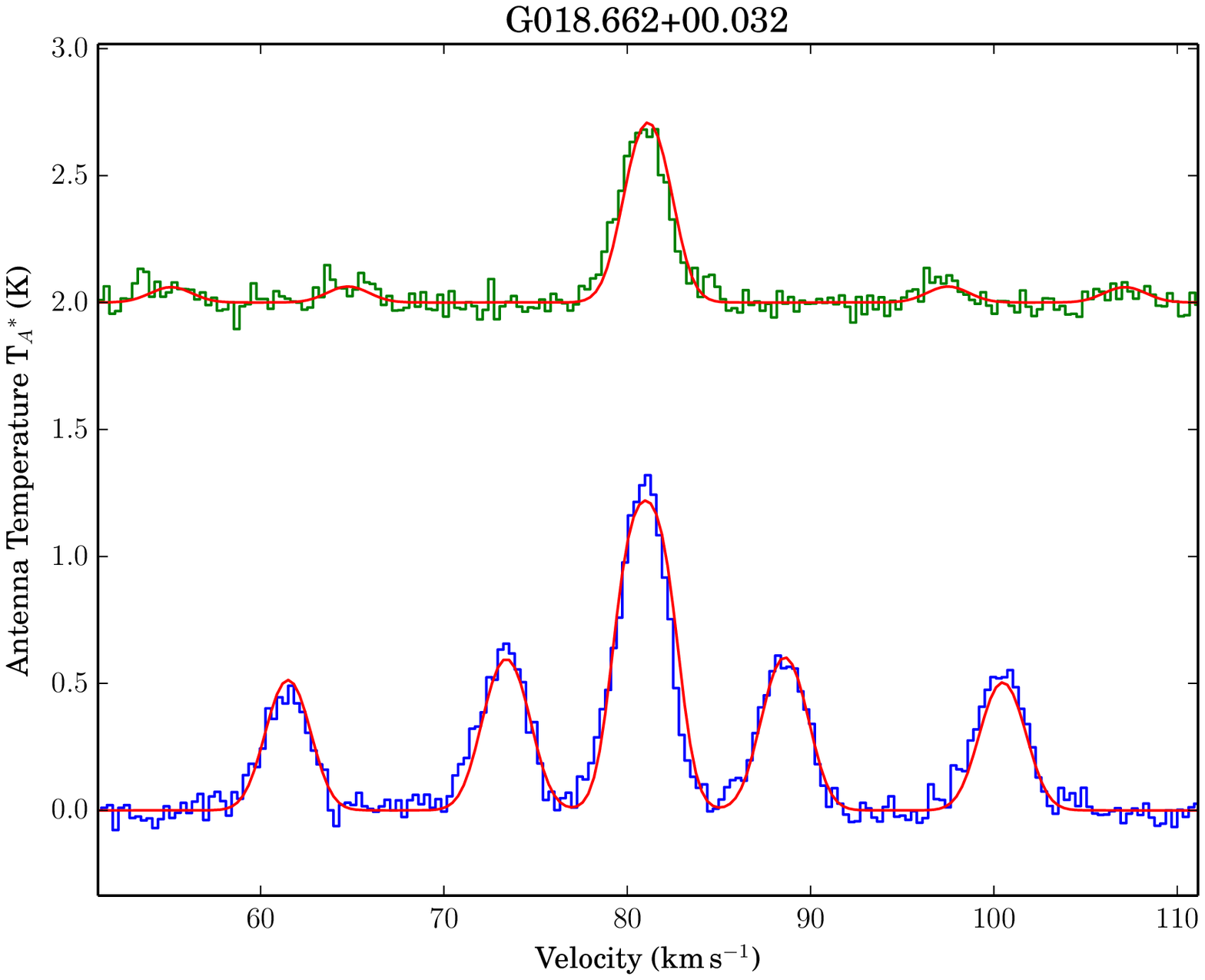}
\includegraphics[width=0.45\textwidth, trim= 0 0 0 0]{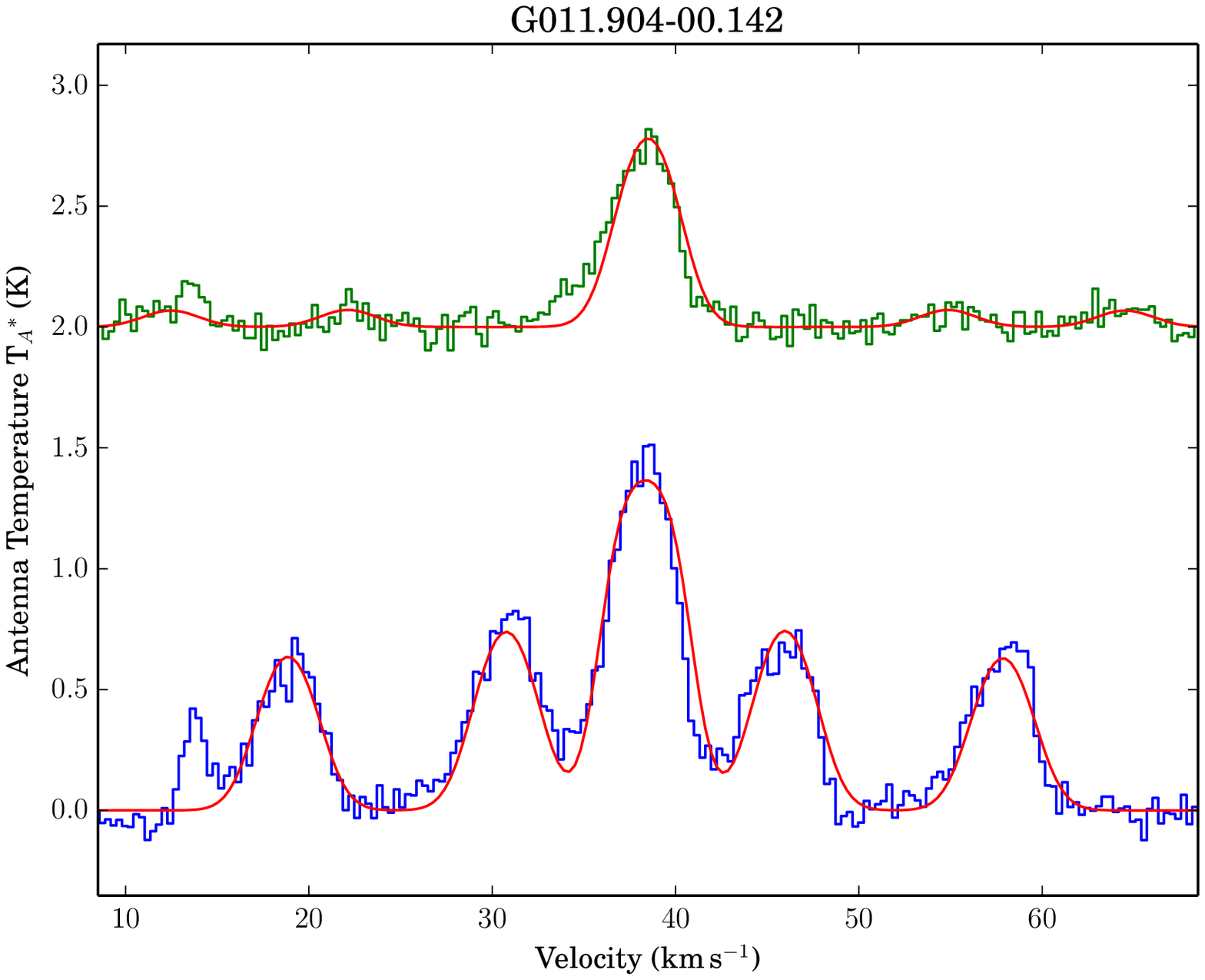}
\includegraphics[width=0.45\textwidth, trim= 0 0 0 0]{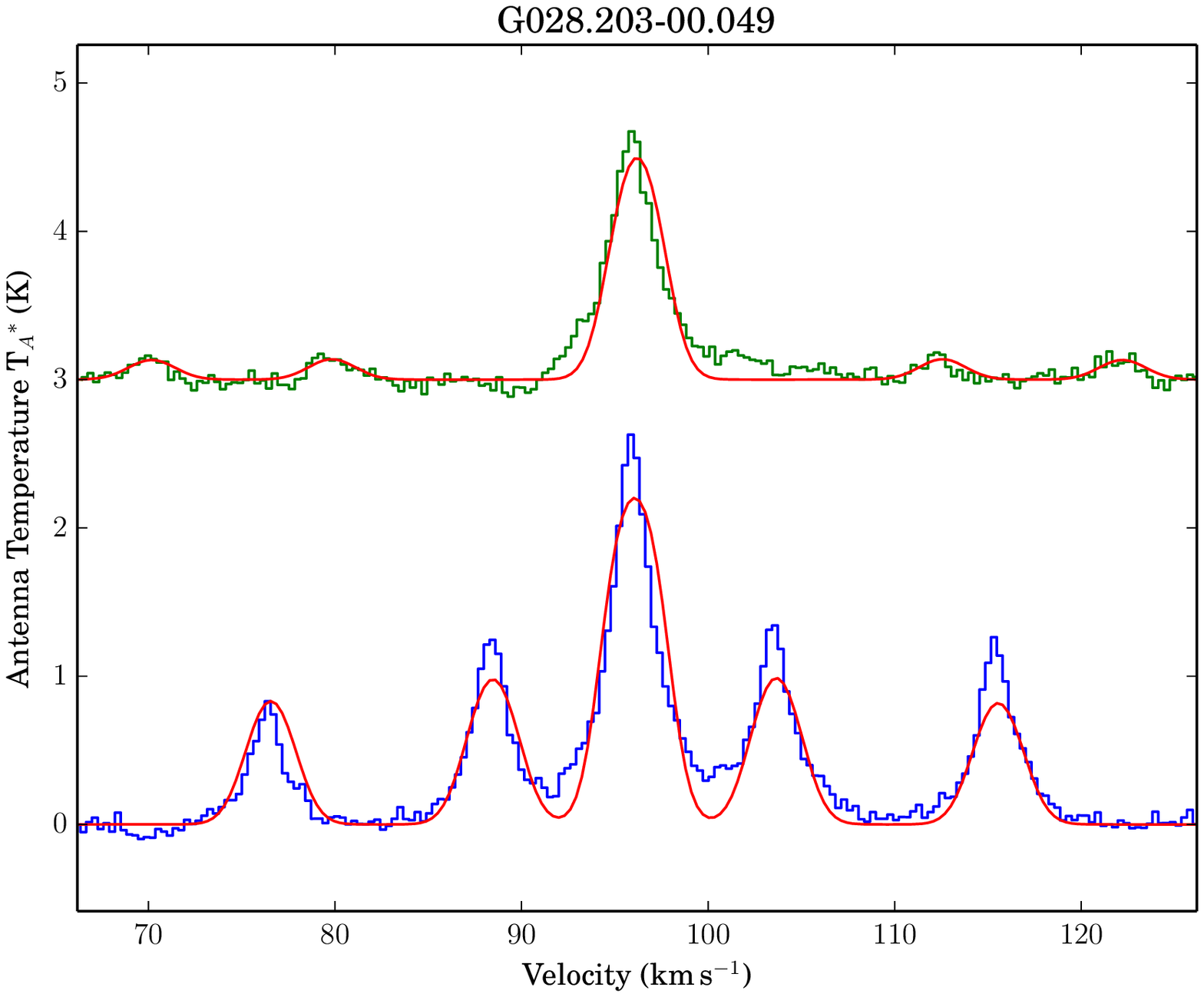}

\caption{\label{fig:spectra_types} Examples of NH$_3$ spectra towards regions of high velocity dispersion. The upper, middle and lower panels present an example of single component (40\,per\,cent), multiple component (23\,per\,cent) and wing component (13\,per\,cent) profiles, respectively. The NH$_3$ (1,1) and (2,2) lines are shown as blue and green histograms, respectively, and the fits to these transitions are shown in red.} 

\end{center}\end{figure}

The observed line widths are more complex than a simple radial profile expected from a single velocity component. There is evidence of an increase in the line width towards the centres of approximately two-thirds of the clumps (a note in Table\,\ref{tbl:cattable} identifies these sources and provided an indication of their nature). In total, 54 clumps are associated with significantly higher velocity dispersions towards their centres with the majority being associated with massive star formation (40). For the massive star forming clumps the broadest line widths are not only coincident with the peak of the integrated NH$_3$ emission but also closely correlated with the position of the embedded source. A KS test does not find a significant difference betweeen the line widths of the inner and outer envelopes of the MSF clumps. This may be due to the small sample size and because the line profiles can result from the blended emission from multiple clumps and can include infall and outflow motions, making the interpretation of the distribution maps difficult in some cases. 

Inspection of the line profiles reveals evidence for the presence of multiple components towards $\sim$23\,per\,cent of the clumps (19/82), with the emission profile seen toward $\sim$40\,per\,cent appearing to be consistent with a single source (34/82), while one source is classified as ambiguous (G045.467+00.046). Eleven clumps show evidence of line wings, which are usually indicative of the presence of molecular outflows. The remaining clumps (24\,per\,cent of the sample) show no significant increase in their line widths towards the centre of the clumps. In Fig.\,\ref{fig:spectra_types} we show an example of the three types of profiles discussed. 

The problem of multiple components is a difficult one to deal with as often the velocity components are blended and cannot be separated in a reliable way. Caution should be exercised when interpreting the line width, temperature and column density distributions in these cases, as these quantities have been derived from a single fit to blended profile and therefore the peak values may be a little less reliable, however, the mean values should not be significantly affected. This is only likely to affect $\sim$20\,per\,cent of the sample and so is unlikely to impact the statistical results; however, it is also possible that the emission seen towards other sources is also the result of the blending of multiple components that cannot be easily identified by eye. One interesting aspect of this is that the star formation is often associated with the overlapping clump regions, which brings up the intriguing idea that the two are somehow related.

We find upon examination of the single component sources that the broad line widths are often found to be at the apex of a cone shape that extends to the edges of the clump, and in a smaller number of cases this is mirrored in the opposite direction to produce an hourglass-shaped region of warmer and/or more turbulent gas (e.g., G013.332$-$00.028 and G028.203$-$00.049). The line profiles of these sources show no evidence of a deviation from a smooth Gaussian at the central velocity and so there is no reason to suspect that the spatial distribution could be due to overlapping clumps. The observed morphology is therefore suggestive of the presence of a bipolar molecular outflow that is clearing a cavity and injecting kinetic energy into the gas along its path. In some cases this interpretation is supported by the presence of emission wings seen in the NH$_3$ emission. Six of the clumps with detected emission wings have been recently been associated with outflows identified by a JCMT CO (3-2) survey {\color{red}Maud et al. 2015}; G013.661$-$00.595, G017.632+00.157, G028.203$-$00.049,  G028.285$-$00.355, G037.555+00.201 and G048.989$-$00.301).

\begin{figure}
\begin{center}

\includegraphics[width=0.45\textwidth, trim= 0 0 0 0]{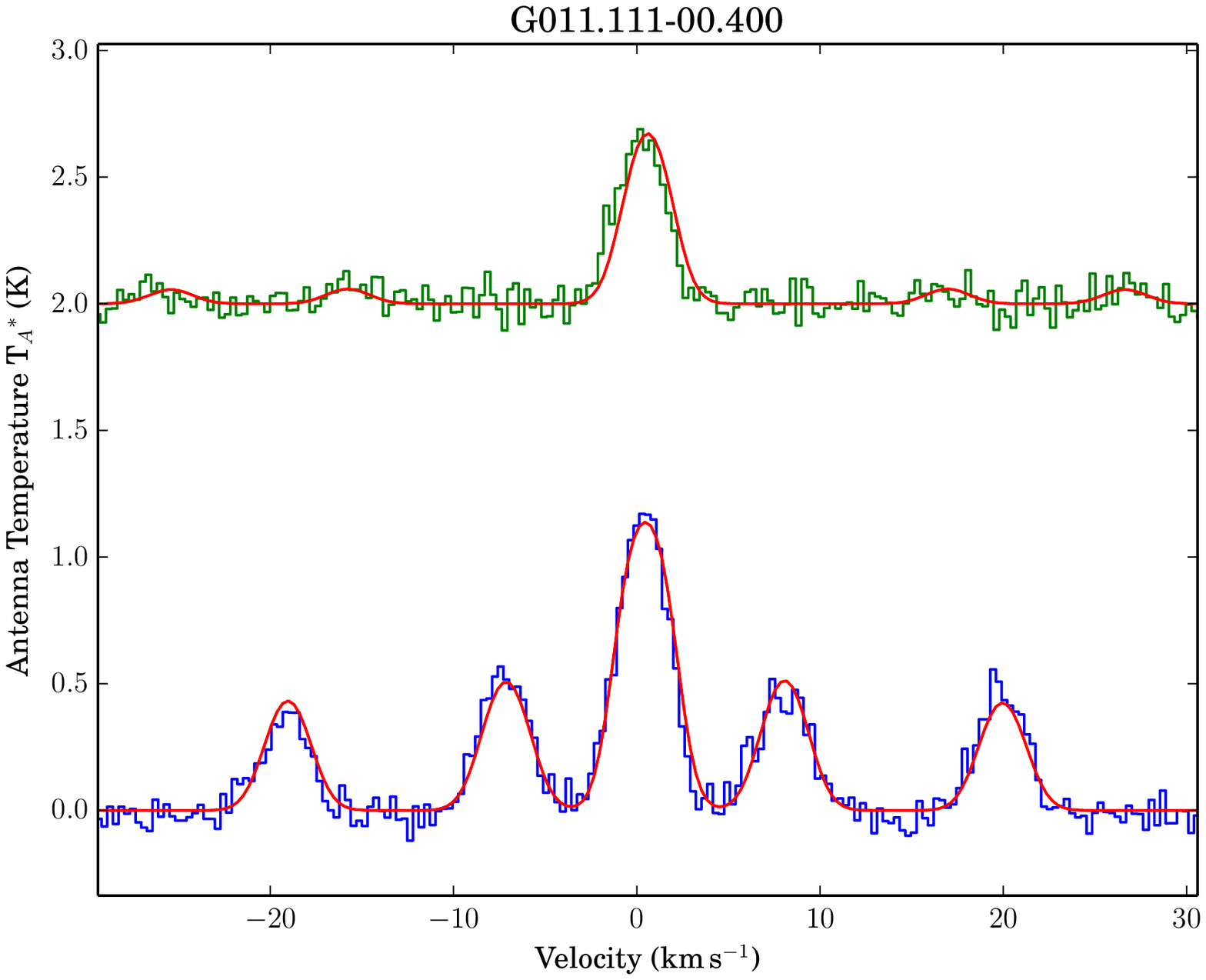}
\includegraphics[width=0.45\textwidth, trim= 0 0 0 0]{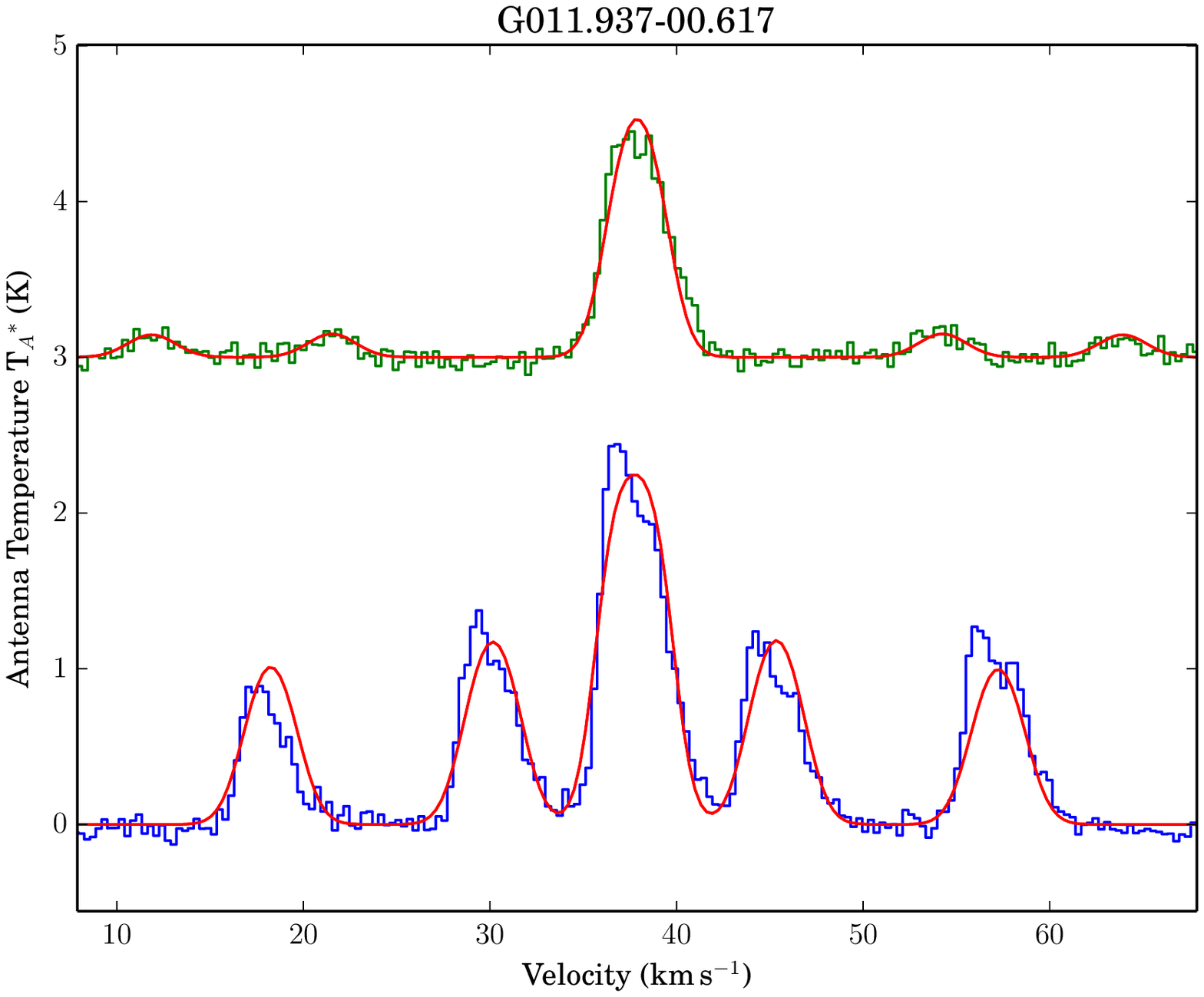}

\caption{\label{fig:offset_spectra} Examples of NH$_3$ spectra towards regions of high velocity dispersion that are offset from the centre of their clumps but coincident with the position of the embedded source. The NH$_3$ (1,1) and (2,2) lines are shown as blue and green histograms, respectively, and the fits to these transitions are shown in red. The spectra presented in the lower panel reveals that the larger line-width seen towards this source is likely due to the blending of multiple components.} 

\end{center}
\end{figure}

We also find two examples of regions of significant increase in the line width that are noticeably offset from their associated clump centres. One example is the intersection of G011.108$-$00.394 and G011.118$-$00.409, which shows a sharp rise that is coincident with the associated \hii\ region (G011.1109$-$00.4001). The velocity difference between the two clumps is $\sim$1.6\,\kms\ and so it is quite feasible that the increased velocity dispersion could be the result of the blending of these two clumps along the line of sight, however, there is no evidence of two separate components in the spectra seen towards the \hii\ region and so this explanation cannot be confirmed (see upper panel of Fig.\,\ref{fig:offset_spectra}). 

A second example is found towards G011.918$-$00.616. In this case the spectrum taken towards the centre of the clump shows a single component with evidence of associated wings, indicative of the presence of an outflow. However, the associated \hii\ region (G011.9373$-$00.6165) is offset to the east and is coincident with the peak of the velocity dispersion, and the spectrum taken at this position does show evidence for multiple components (see lower panel of Fig.\,\ref{fig:offset_spectra}). In these two cases the overlap between nearby clumps and resulting increase in the velocity dispersion is not likely to significantly affect the derived bulk properties of the clumps; however, the values calculated towards these interface regions should be considered unreliable.

We find no significant variation in the pressure ratio across the observed fields for the whole sample or either of the two subsamples, suggesting that the non-thermal motions are relatively homogeneous across the clumps. This implies that sources that have regions of higher line-widths have coincident areas of higher temperatures, which is in general what is observed in the spatial distribution maps presented in Fig.\,\ref{fig:distribution_maps}. \citet{ragan2011} studied similar cores using combined data from VLA and GBT observations, and found peak pressure ratios of $\sim$0.07.  This is a factor of a few higher than we find ($\sim$0.02), suggesting that the contribution from the non-thermal emission becomes less important on smaller core scales; however, the pressure is still clearly dominated by non-thermal motions.

\begin{figure*}\begin{center}

\includegraphics[width=0.33\textwidth, trim= 0 0 0 0]{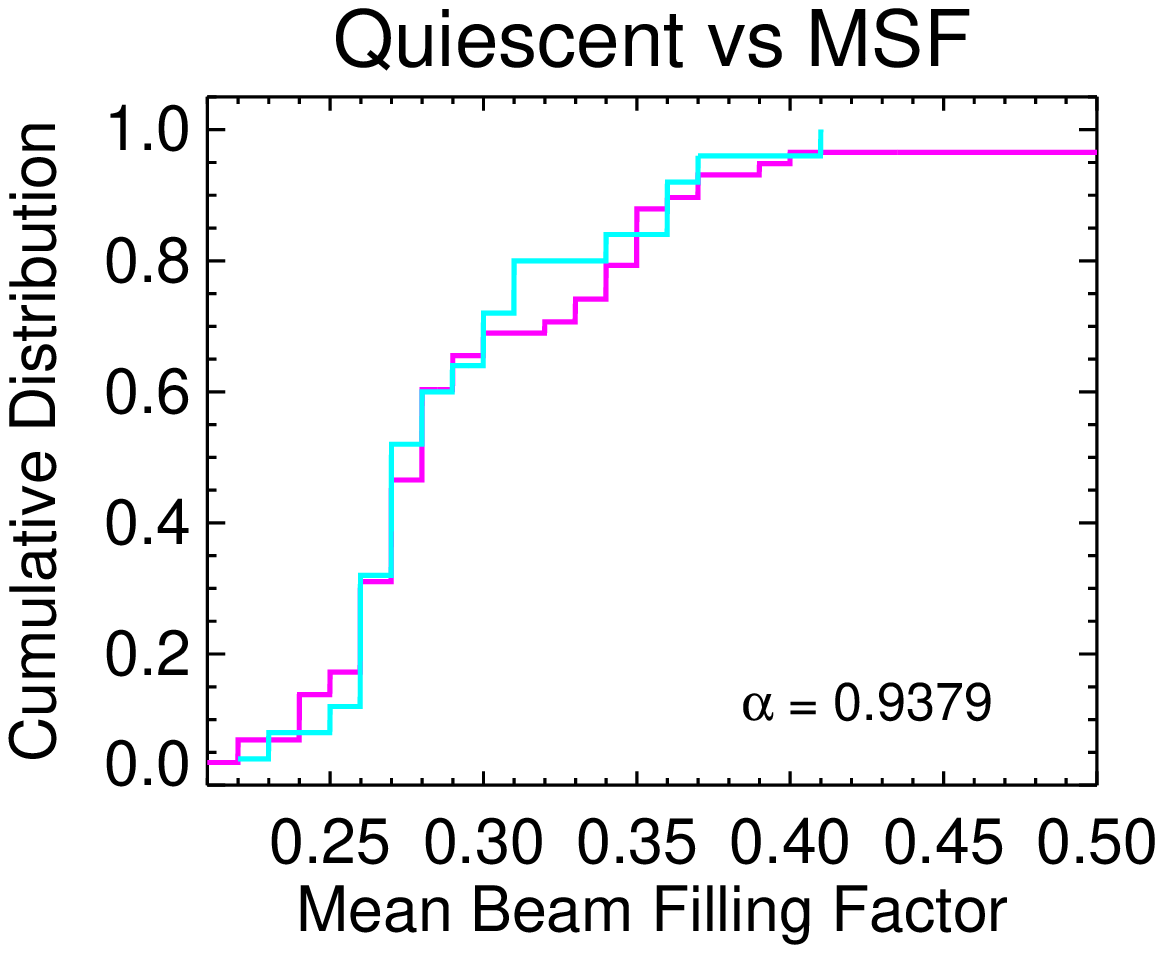}
\includegraphics[width=0.33\textwidth, trim= 0 0 0 0]{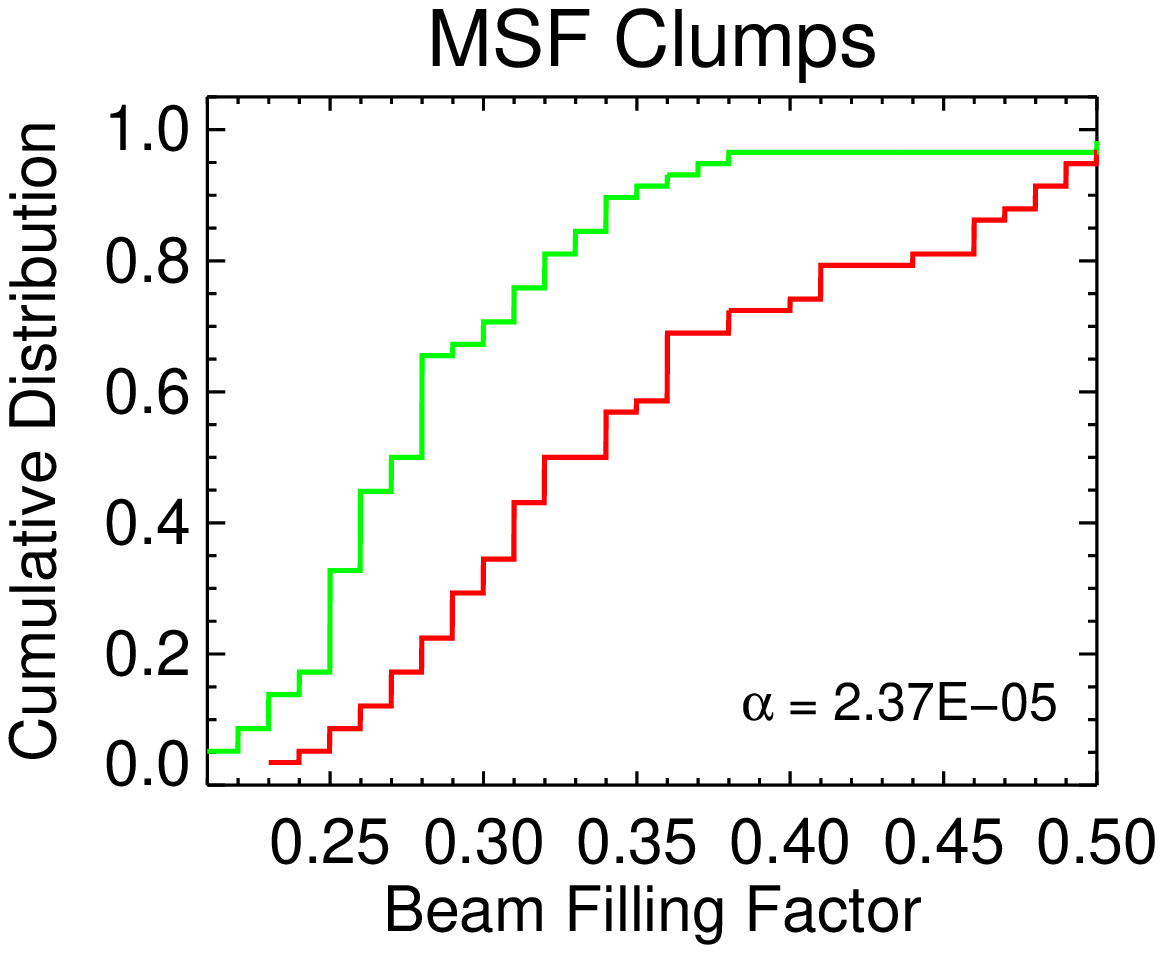}
\includegraphics[width=0.33\textwidth, trim= 0 0 0 0]{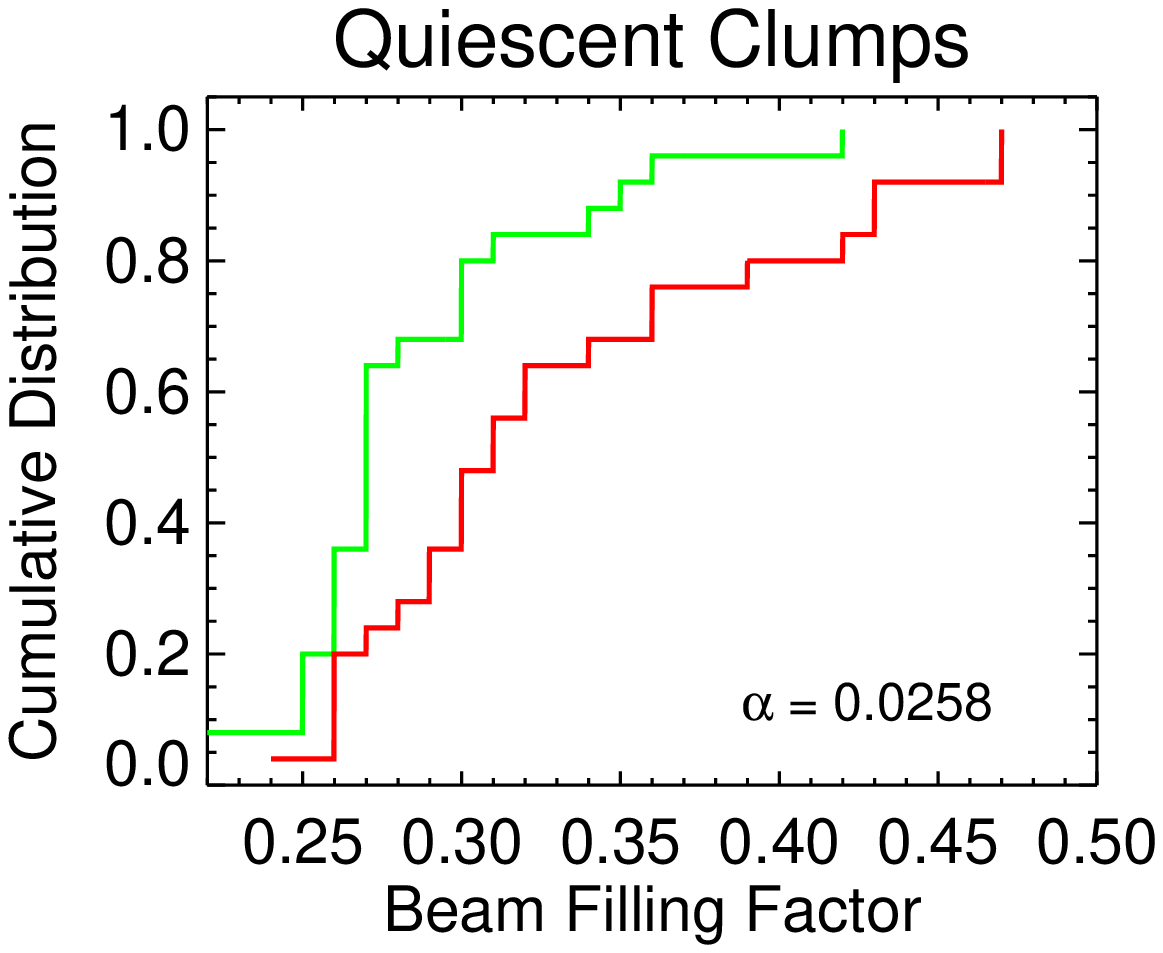}
\includegraphics[width=0.33\textwidth, trim= 0 0 0 0]{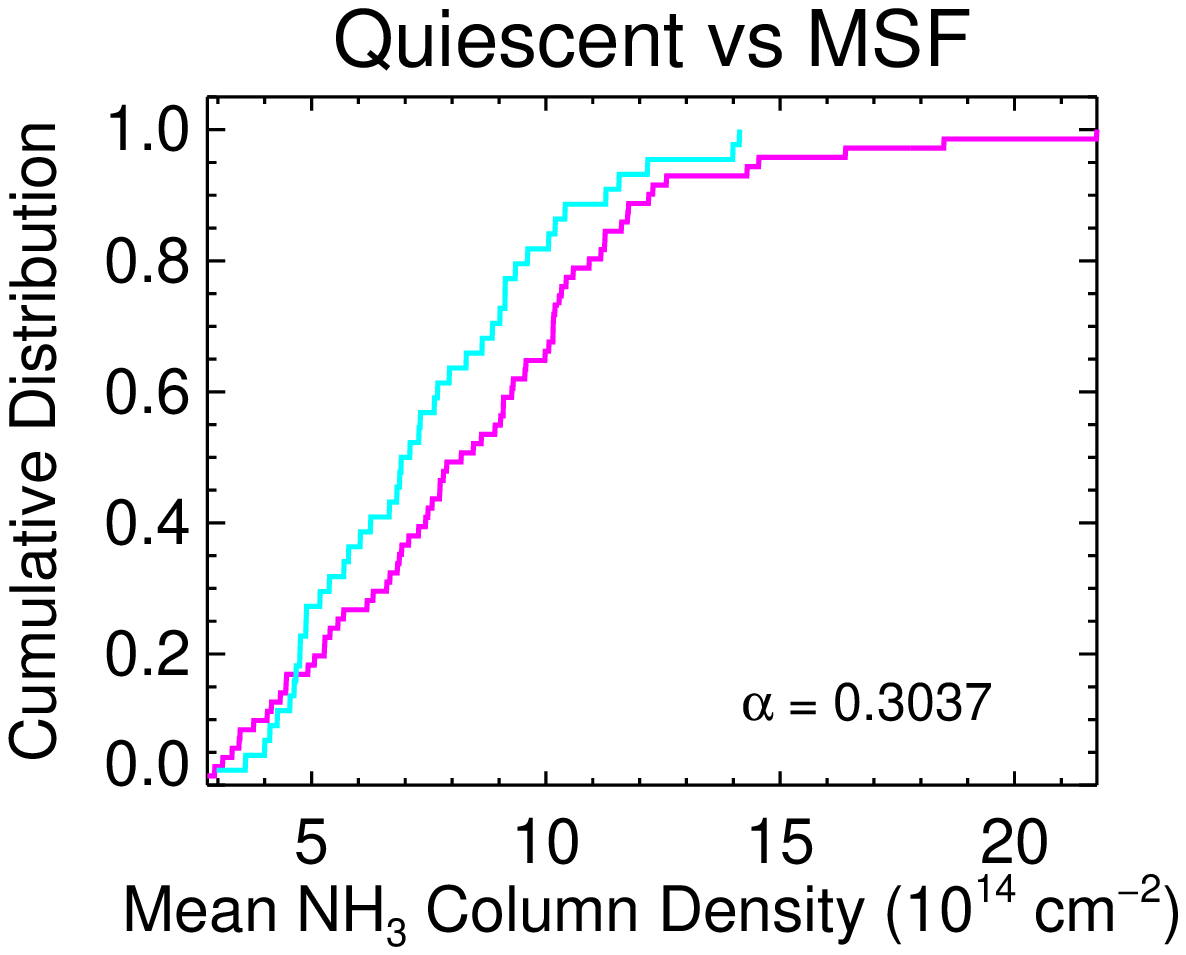}
\includegraphics[width=0.33\textwidth, trim= 0 0 0 0]{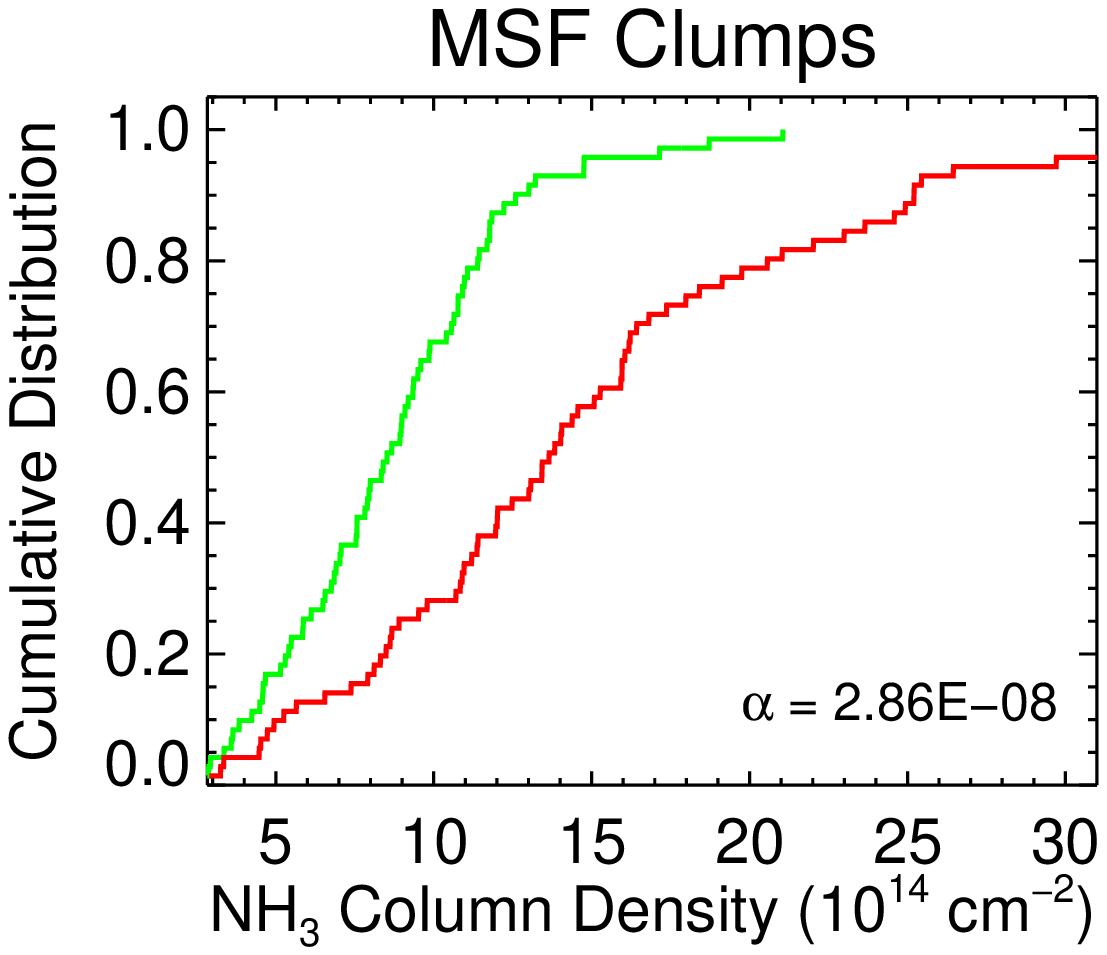}
\includegraphics[width=0.33\textwidth, trim= 0 0 0 0]{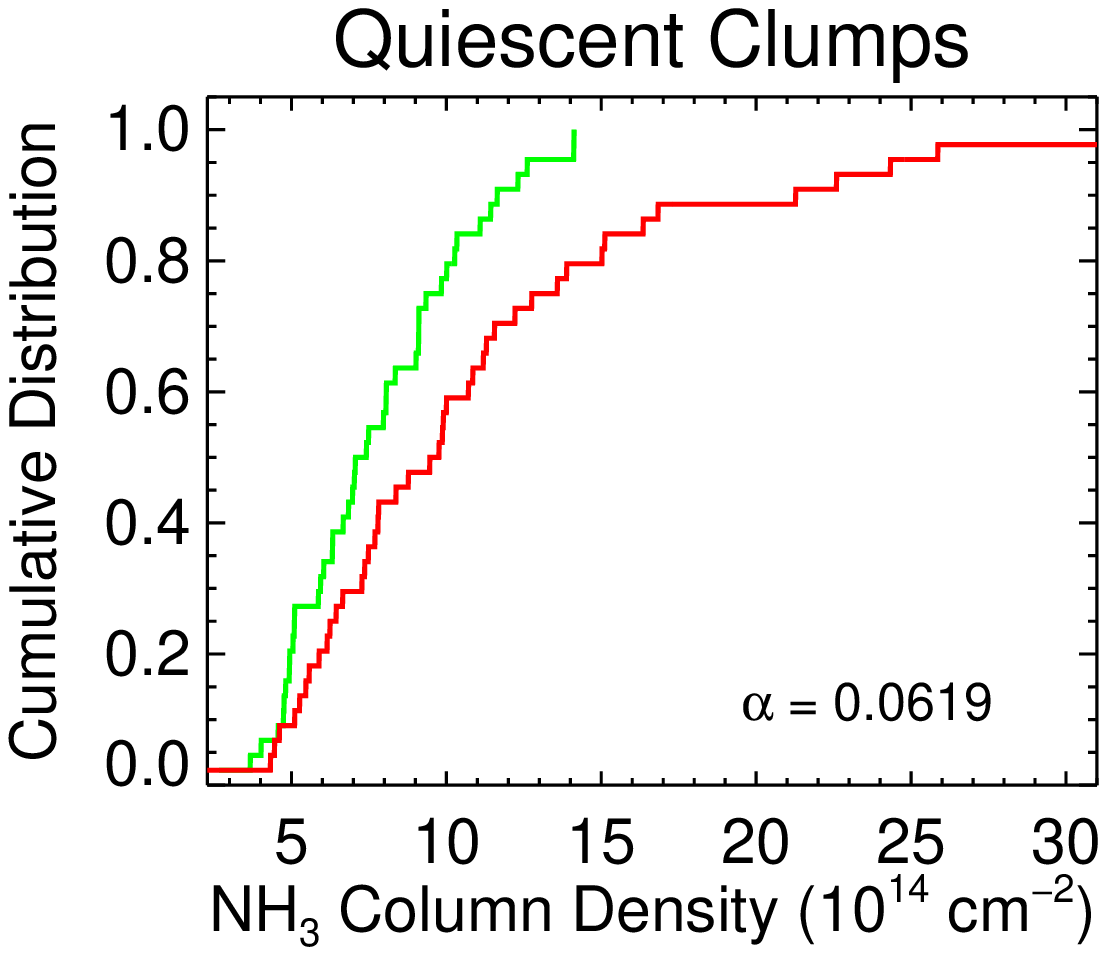}
\includegraphics[width=0.33\textwidth, trim= 0 0 0 0]{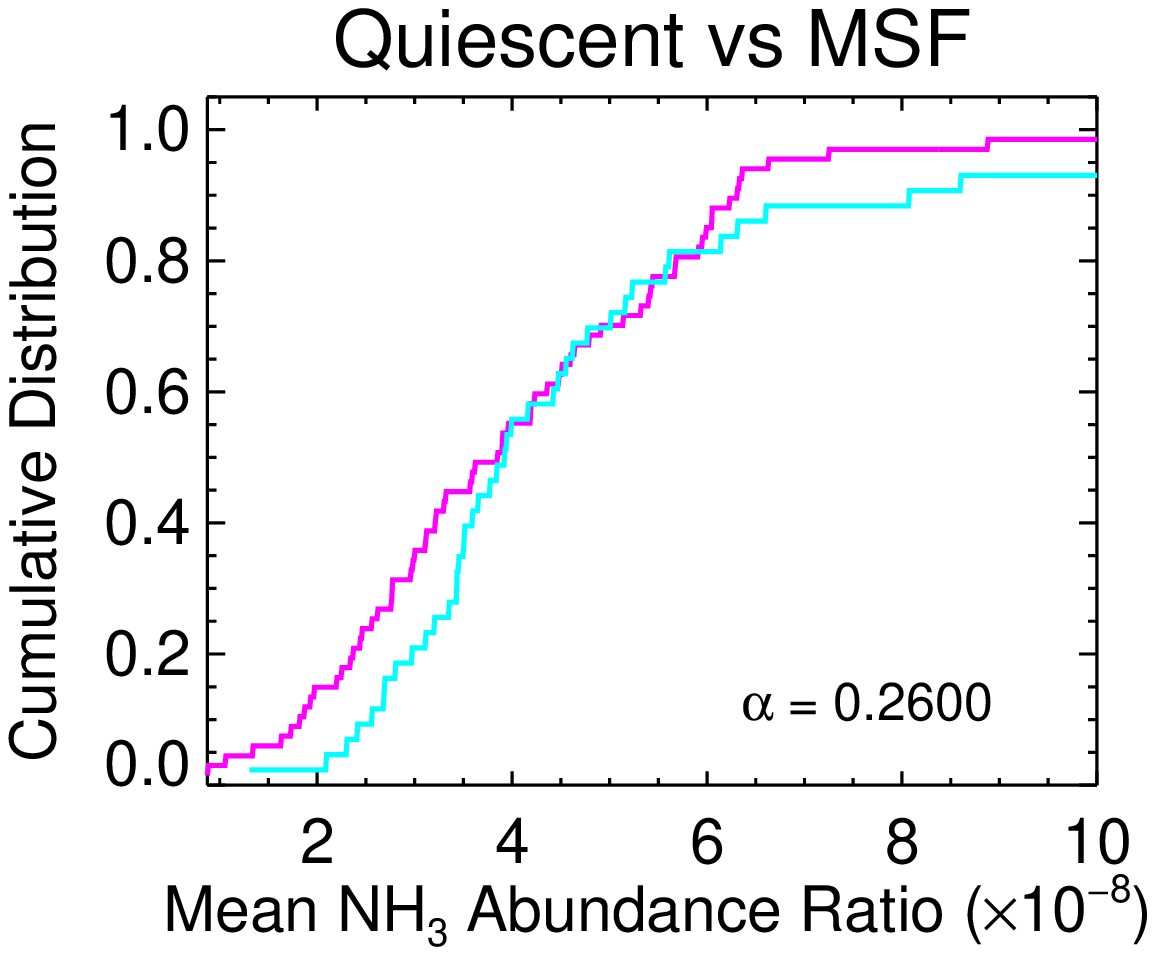}
\includegraphics[width=0.33\textwidth, trim= 0 0 0 0]{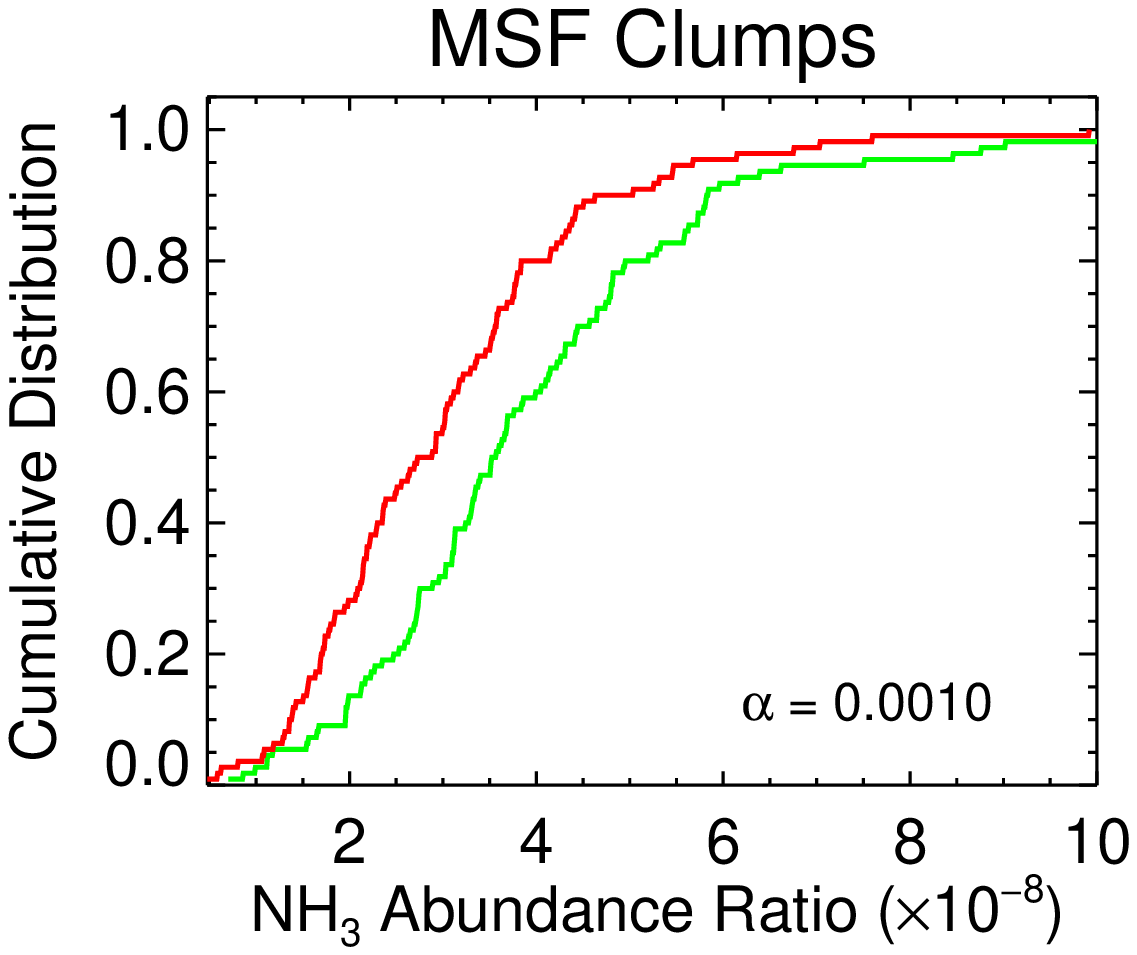}
\includegraphics[width=0.33\textwidth, trim= 0 0 0 0]{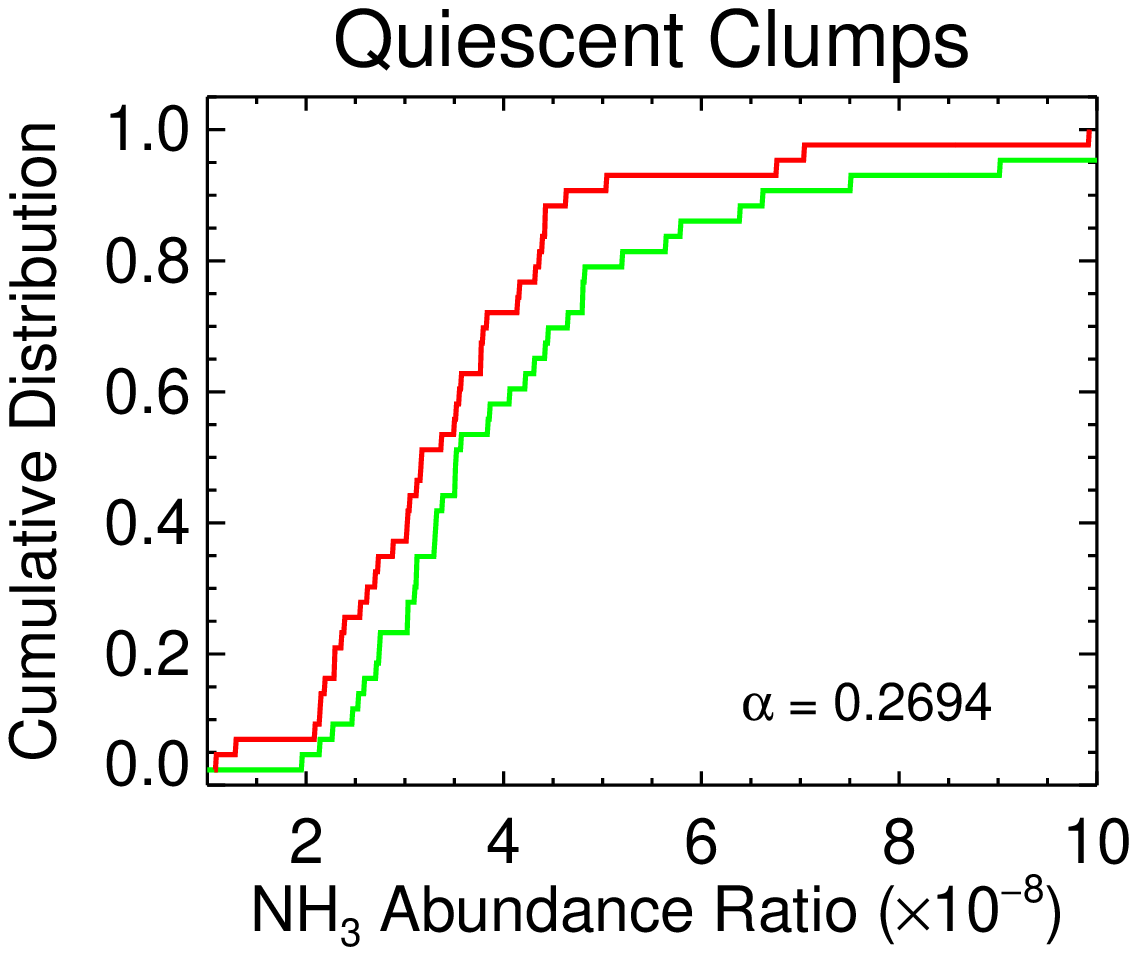}

\caption{\label{fig:full_sample_CDF2} Left panels: Cumulative distribution plots comparing the mean values for the  beam filling factor, NH$_3$ column density and NH$_3$/H$_2$ abundance ratio for the MSF and quiescent clumps (cyan and magenta curves, respectively). Middle and right panels: Cumulative distribution plots comparing the peak and median values (shown in red and green, respectively) for the MSF and quiescent clumps  for the same parameters. These allow us to identify any significant differences between the internal and external structure of the clumps. The results of KS tests comparing the peak and median values is given in the lower right corner of each plot. For these plots and the discussion in the text we only include sources with a SNR $>$10.} 

\end{center}\end{figure*}

In Fig.\,\ref{fig:full_sample_CDF2} we present cumulative distribution plots comparing the median and peak values for the beam filling factor, the NH$_3$ column density, and fractional abundance of ammonia. We find that there are no significant differences between the quiescent and MSF clumps. However, we do see a difference between the inner and outer envelopes of the MSF clumps for all three of these parameters, which is confirmed to be statistically significant by the KS test. Although there does appear to be a difference between the inner and outer envelopes for the beam filling factor and column densities for the quiescent clumps it falls short of 3$\sigma$ confidence required for these to be considered significant; however, the sample size is relatively small and this may contribute. 

The beam filling factor and the column density are each higher towards the centres of the MSF sample, which is consistent with the clumps having a centrally condensed structure. The positions of the peak temperatures are roughly coincident with the highest column density regions, but can often be a slightly offset (see Fig.\,\ref{fig:gbt_rms_offset}). 

The abundance distribution for the MSF clumps is quite different from that of the beam filling factor and the column density in that it reveals that the abundance is significantly lower at the centre of the clumps compared to the outer envelope. Similar lower abundances have been found towards the centres of dense star forming clumps by other studies (e.g., \citealt{morgan2014,friesen2009}). Inspection of the abundance maps reveals that the abundance can be clearly seen to rise towards the edges of the clumps; however, this may be the result of photoionisation or chemical fractionation in the outer regions of the clumps where the exposure to uv-radiation is expected to be higher. We do not see any difference in the abundance distribution for the quiescent clumps, however, which could suggest that photoionisation or fractionation are not major factors and that other processes may be responsible.  For example, depletion in the high-density central region is possible, but not likely given the temperatures measure.  Higher optical depths toward the centres, however, would lead to an underestimate of the temperature leading to an overestimate of the H$_2$ column density in those locales.

In summary, the whole sample appears to consist of roughly spherical centrally-condensed structures with line widths of $\sim$2\,\kms\ and kinetic temperatures of $\sim$20\,K. The MSF clumps tend to be slightly warmer, have marginally broader line widths, higher column densities and a steeper column density gradient when compared to the quiescent clumps. These differences are consistent with those expected from the presence of an embedded thermal source. It is therefore no surprise that the associated RMS sources are found towards the centre of their host clumps where the gravitational potential well is deepest. However, it is also noteworthy that many of the properties of the quiescent clumps are very similar to those of the MSF clumps and may be examples of massive pre-stellar clumps. A more detailed examination of the mid-infrared images reveals the presence of embedded protostellar objects that are associated with quiescent clumps. We conclude from this that although the quiescent clumps are not yet associated with a MYSO or compact \hii\ region, they may not be devoid of star formation activity altogether; but any  star formation present is either lower-mass or in a younger evolutionary phase.

\subsection{Deriving physical properties}

In this subsection we will describe the methods used to determine some key physical values such as the masses and sizes of the clumps and the bolometric luminosity of the embedded objects. The results of this analysis are given in Table\,\ref{tbl:derived_clump_para}.

\setlength{\tabcolsep}{3pt}

\begin{table*}


\begin{center}\caption{Derived clump parameters. The bolometric luminosity given is the sum of all RMS sources embedded in each clump.}
\label{tbl:derived_clump_para}
\begin{minipage}{\linewidth}
\begin{center}
\small
\begin{tabular}{lcccccccc}
\hline \hline
  \multicolumn{1}{c}{Field id.}& \multicolumn{1}{c}{Clump}&  \multicolumn{1}{c}{RMS}&\multicolumn{1}{c}{Distance}&\multicolumn{1}{c}{Radius}&\multicolumn{1}{c}{Log[Peak $N$(H$_2$)]}  &        \multicolumn{1}{c}{Log[$M_{\rm{clump}}$]} & \multicolumn{1}{c}{Log[$M_{\rm{vir}}$]} &\multicolumn{1}{c}{Log[$L_{\rm{bol}}$]}\\
  
    \multicolumn{1}{c}{id.}&\multicolumn{1}{c}{name}&  \multicolumn{1}{c}{Association} & \multicolumn{1}{c}{(kpc)}  &\multicolumn{1}{c}{(kpc)} &\multicolumn{1}{c}{(pc)}&\multicolumn{1}{c}{(cm$^{-2}$)} &      \multicolumn{1}{c}{(\msun)}&\multicolumn{1}{c}{(\lsun)}  \\
\hline

1	&	G010.283$-$00.118	&	no	&	2.2	&	0.32	&	1.35	&	2.7	&	2.62	&	\multicolumn{1}{c}{$\cdots$}	\\
1	&	G010.288$-$00.166	&	no	&	2.0	&	0.11	&	1.34	&	1.8	&	1.97	&	\multicolumn{1}{c}{$\cdots$}	\\
1	&	G010.296$-$00.148	&	yes	&	2.4	&	0.21	&	1.35	&	2.6	&	2.67	&	4.22	\\
2	&	G010.300$-$00.271	&	no	&	4.0	&	0.35	&	1.34	&	2.5	&	2.83	&	\multicolumn{1}{c}{$\cdots$}	\\
2	&	G010.322$-$00.229	&	yes	&	4.0	&	0.34	&	1.34	&	2.3	&	2.32	&	3.62	\\
2	&	G010.322$-$00.257	&	yes	&	4.0	&	0.37	&	1.35	&	2.6	&	2.60	&	3.73	\\
1	&	G010.323$-$00.165	&	yes	&	3.5	&	0.38	&	1.35	&	2.9	&	2.43	&	4.58	\\
1	&	G010.346$-$00.148	&	no	&	3.5	&	0.24	&	1.35	&	2.8	&	2.29	&	\multicolumn{1}{c}{$\cdots$}	\\
3	&	G010.440+00.003	&	yes	&	8.5	&	0.31	&	1.34	&	2.9	&	2.91	&	4.39	\\
3	&	G010.474+00.028	&	yes	&	8.6	&	0.94	&	1.36	&	4.1	&	3.49	&	5.58	\\

\hline\\
\end{tabular}\\
Notes: Only a small portion of the data is provided here, the full table is available in electronic form at the CDS via anonymous ftp to cdsarc.u-strasbg.fr (130.79.125.5) or via http://cdsweb.u-strasbg.fr/cgi-bin/qcat?J/MNRAS/.
\end{center}
\end{minipage}

\end{center}
\end{table*}
\setlength{\tabcolsep}{6pt}

\subsubsection{Distances, physical offsets and clump radii}
\label{sect:distance}

The majority of source distances have been drawn from \citet{urquhart2014a} and references therein. A few have been modified from those given in that paper as more reliable distances have recently become available. These changes only affect the following three fields.  

G052.20+00.72 and G011.11$-$00.40 have velocities close to zero and so their distances are unreliable; however, G011.11$-$00.40 is very close to the W31 complex and has a similar velocity and so we have adopted the distance to this complex for this field  (4.95\,kpc; \citealt{sanna2014}). The G018.61$-$00.07 field has been previously placed at the far distance due to a lack of any evidence of self-absorption seen in \hi\ at the same velocity as the source, which is generally expected for sources located at the near distance (\citealt{wienen2015a}). This source is, however, associated with a prominent infrared dark cloud complex and is therefore more likely to be located at its kinematic near distance of 3.6\,kpc. We do not allocate a distance for sources located in the G052.20+00.72 field as their kinematic distance is not reliable and there is no association that can be used.

We reported in Sect.\,\ref{sect:clump_structure} that two fields include multiple sources that have significantly different velocities: these are G024.18+00.12 and G028.29$-$00.38. The G024.18+00.12 field has two clumps (G024.143+00.129 and G024.161+00.085) with a \vlsr\ $\sim$ 53\,\kms\ and a third clump with a \vlsr\ of 113.5\,\kms\ (G024.183+00.122). The first of these clumps is associated with the RMS source G024.1328+00.1213 and we have therefore adopted the RMS distance of 3.8\,kpc to both clumps with velocities $\sim$ 53\,\kms. The third clump in this field is associated with the RMS source G024.1838+00.1198 and we have adopted this distance for the clump (9\,kpc). The second field contains two clumps, G028.315$-$00.397 and G028.285$-$00.355 with radial velocities of 85.4 and 48.8\,\kms, respectively. The first clump is not associated with an RMS source and so no distance is assigned, while the second is associated with the RMS source G028.2875$-$00.3639 and we have adopted the distance to that source for that clump (11.6\,kpc).

\begin{figure}
\begin{center}
\includegraphics[width=0.45\textwidth, trim= 0 0 0 0]{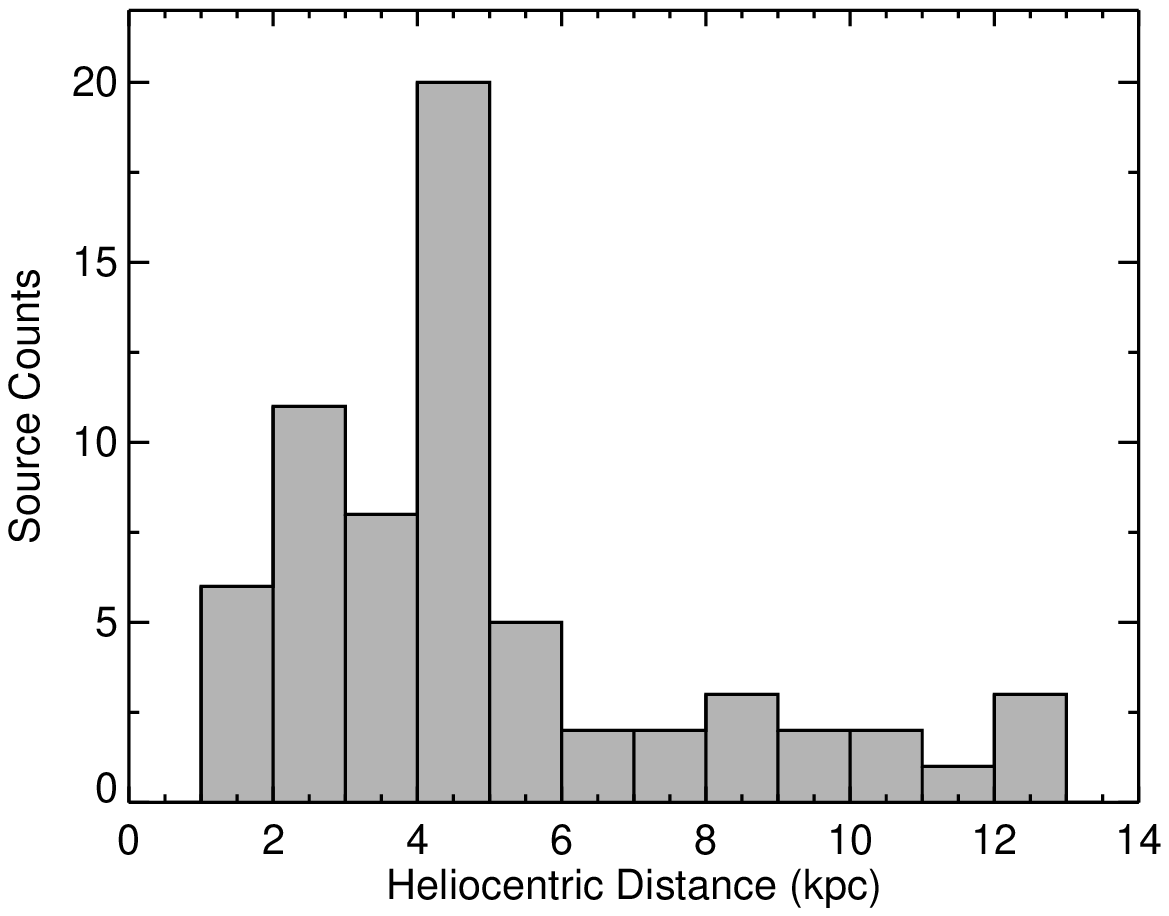}
\includegraphics[width=0.45\textwidth, trim= 0 0 0 0]{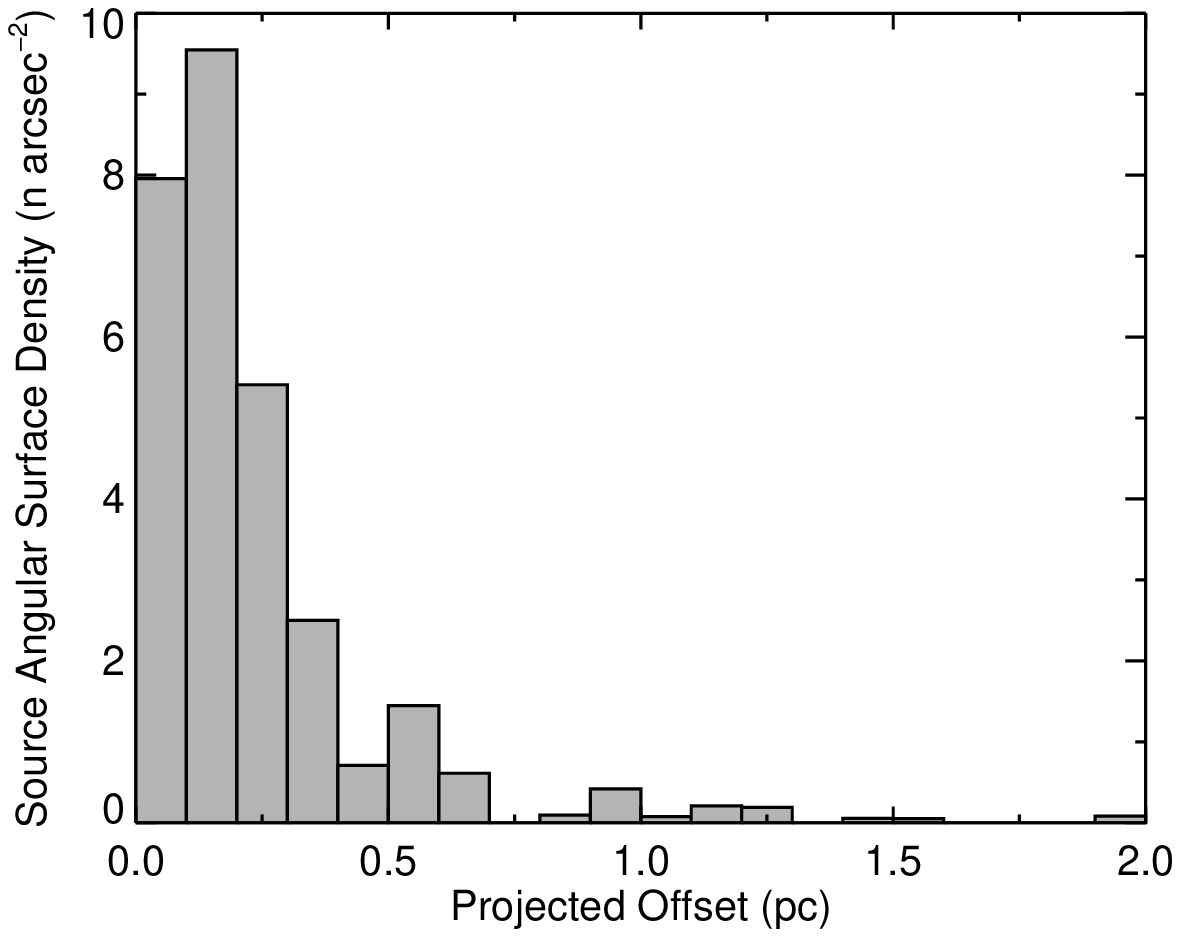}
\caption{Heliocentric distance distribution for the observed fields is shown in the upper panel and the projected physical separation between NH$_3$ peak and RMS source is shown in the lower panel. The distribution has been truncated at offset $>$2\,pc, however, only 9 matches have larger offsets. The bin sizes used for the upper and lower panels are 1\,kpc and 0.1\,pc, respectively.} 
\label{fig:distance_hist}

\end{center}
\end{figure}

\begin{figure}
\begin{center}
\includegraphics[width=0.45\textwidth, trim= 0 0 0 0]{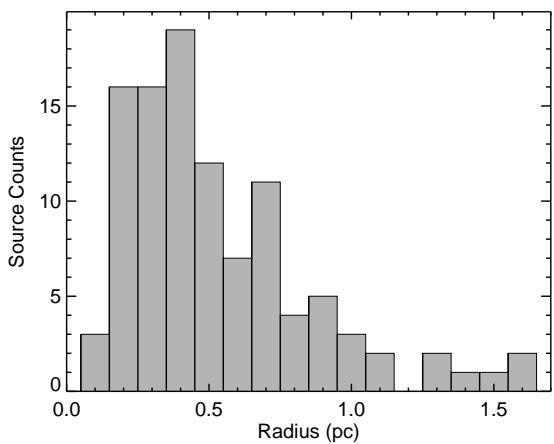}

\caption{Radius distribution for the complete set of resolved clumps for which a distance has been determined. This sample consists of 104 sources. The bin size is 0.1\,kpc.} 
\label{fig:radius_hist}

\end{center}
\end{figure}

In the upper panel of Fig.\,\ref{fig:distance_hist} we show the distance distribution of the sources observed. In the lower panel of Fig.\,\ref{fig:distance_hist} we show the projected physical offset between the peak of the integrated NH$_3$ emission and embedded objects, which clearly shows the two are tightly correlated with each other. These distances have been combined with the effective radii determined in Sect.\,\ref{sect:clump_structure} to determine the physical sizes of the clumps. Fig.\,\ref{fig:radius_hist} shows the size distribution of the sample for the resolved clumps. The angular resolution of these observations corresponds to a physical diameter of $\sim$0.7\,pc at the median distance of the sample (4.5\,kpc) and are therefore probing the more global properties of the host clumps. The median diameter is $\sim$0.8\,pc and we find no significant difference between the RMS associated clumps and the quiescent subsamples (KS test finds $\alpha=0.08$). The rather narrow distribution of distances means that these observations are probing similar spatial scales and NH$_3$ abundances, and therefore minimising the uncertainties associated with comparing large samples.

\subsubsection{Column densities and clump masses}
\label{sect:clump_mass}

The large variations in the ammonia abundance across the clumps and from source-to-source found in Sect.\,2.5 suggests that the ammonia-derived clumps masses are not reliable. We therefore use the submillimetre dust emission to estimate the total clump masses, which is regarded as one of the most reliable tracer of column density and mass  (\citealt{schuller2009}).

\begin{figure}
\begin{center}

\includegraphics[width=0.45\textwidth, trim= 0 0 0 0]{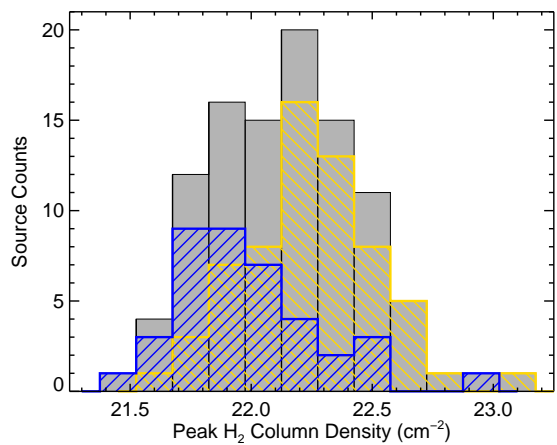}
\includegraphics[width=0.45\textwidth, trim= 0 0 0 0]{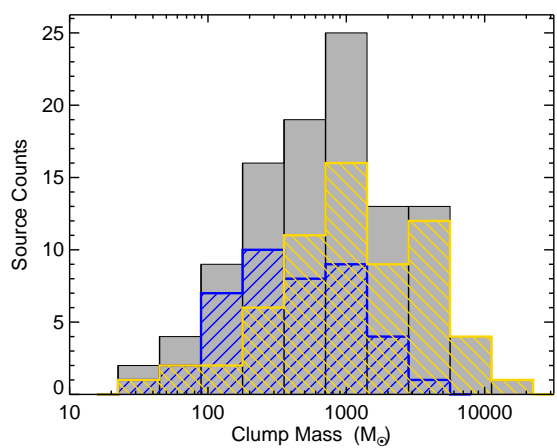}

\caption{Peak column density and clump mass distributions for the complete set of clumps for which a distance has been determined are shown in the upper and lower panels, respectively. This sample consists of the 106 clumps for which distances are available. The distributions of the MSF-associated and quiescent subsamples are shown by yellow and blue hatching and consist of 64 and 42 sources, respectively. The bin size is 0.15 and 0.3\,dex, respectively.} 
\label{fig:mass_hist}

\end{center}
\end{figure}

The H$_2$ column density maps were produced by smoothing dust emission maps and using the kinetic temperature derived from the ammonia data assuming that the gases are coupled and that $T_{\rm{kin}}=T_{\rm{dust}}$ (see Sect.\,\ref{sect:abundance} for more details). The column densities range from $\sim$3-$100\times10^{21}$\,cm$^{-2}$ and the peak of the distribution is $\sim$16$\times10^{21}$\,cm$^{-2}$. The distributions of the whole sample as well as the MSF and quiescent subsamples are shown in the upper panel of Fig.\,\ref{fig:mass_hist}. It is clear from comparison of the two subsamples that the quiescent clumps have a significantly lower column density than the MSF clumps; this is confirmed by a KS test ($\alpha = 10^{-5}$).

We estimate the total mass of the clumps by integrating the mass contained within each pixel:

\begin{equation}
M \, = \, \frac{d^2 \, S_\nu \, R}{B_\nu(T_{\rm{kin}}) \, \kappa_\nu},
\end{equation}

\noindent where $S_\nu$ is the integrated 870\,\mum\ flux per pixel, $d$ is the heliocentric distance to the source and $R$, $B_\nu$, and $\kappa_\nu$ are as previously defined. As before, the $T_{\rm{kin}}$ is the temperature of each pixel as derived from the ammonia analysis.

In these calculations we have assumed that all of the measured flux arises from thermal dust emission, that the emission is optically thin, and that molecular line emission and/or free-free emission from embedded ionised nebulae is likely to be small and in the majority of cases will be negligible (\citealt{schuller2009}). We also assume that the dust and gas are in local thermodynamical equilibrium (LTE) and that the kinetic temperature is a reliable measure of the gas temperature.

Clump mass distributions are shown in the lower panel of Fig.\,\ref{fig:mass_hist}. The clump masses of the complete sample range from a few tens of \msun\ to 10$^4$\,\msun\ with a mean and median value of approximately 1000\,\msun, which matches the peak of the distribution. The RMS-associated and quiescent clumps appear to be different, and their mean values differ by almost a factor of $\sim$3 (1100$\pm$200\,\msun\ and 400$\pm$70\,\msun, respectively). A KS test is unable to reject the null hypothesis that the two subsamples are drawn from the same population at more than the required 3$\sigma$ confidence ($\alpha = 0.0015$), however, it only just fails this criterion may be satisfied if the sample size was larger.

\subsubsection{Virial masses}
\label{sect:virial_mass}

The viral mass is an estimate of the bulk kinetic energy of the clump and can be calculated via

\begin{equation}
\left (\frac{M_{\rm{vir}}}{\rm{M}_\odot} \right) \; = \; \frac{209}{a_1 a_2} \left(\frac{R_{\rm{eff}}}{\rm{pc}}\right) \left(\frac{\Delta v_{\rm{avg}}}{\rm{km\,s}^{-1}}\right)^{2}
\end{equation}

\noindent where $R_{\rm{eff}}$ is the effective radius of the clump and $a_1$ and $a_2$ are corrections for the assumptions of uniform density and spherical geometry, respectively (\citealt{bertoldi1992}). For aspect ratios less than 2 (which is satisfied by the majority of our sample of clumps), $a_2 \sim 1$ and $a_1 \sim 1.3$. $\Delta v^{2}_{\rm{avg}}$ is the corrected value of the measured ammonia (1,1) line width so that it better reflects the average velocity dispersion of the total column of gas (i.e., \citealt{fuller1992}):

\begin{equation}
\Delta v^{2}_{\rm{avg}} \; = \; \Delta v^{2}_{\rm{int}}+8ln2\times \frac{k_{\rm{B}}T_{\rm{kin}}}{m_{\rm{H}}}\left(\frac{1}{\mu_{\rm{p}}}-\frac{1}{\mu_{\rm{NH_3}}}\right),
\end{equation}

\noindent where  $\Delta v_{\rm{int}}$ is the intrinsic line width, $k_{\rm{B}}$ is the Boltzmann constant, $T_{\rm{kin}}$ is the kinetic temperature of the gas and $\mu_{\rm{p}}$ and $\mu_{\rm{NH_3}}$  are the mean molecular masses of molecular hydrogen and ammonia, taken as 2.33 and 17, respectively.

\subsubsection{RMS luminosities}

 \begin{figure}
\begin{center}
\includegraphics[width=0.45\textwidth, trim= 0 0 0 0]{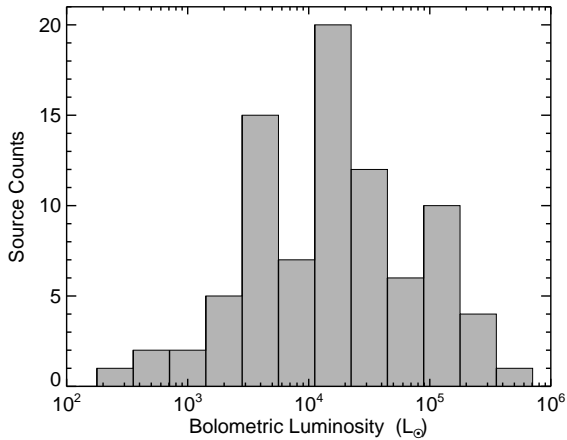}

\caption{Luminosity distribution for the complete set of the MSF-associated resolved clumps for which a distance has been determined. This sample consists of 85 sources. The bin size is 0.3\,dex.} 
\label{fig:lum_hist}

\end{center}
\end{figure}

The bolometric luminosities of all RMS sources have been determined either from model fits to their spectral energy distributions (SEDs) or by scaling their MSX 21\,\mum\ flux (\citealt{mottram2011a, urquhart2014a}). 

The source SEDs for 65 sources were fitted with the \citet{robitaille2006} code to estimate their bolometric flux.   21\,\mum\ fluxes were used to determine bolometric luminosities for an additional {\color{red}23} sources.  These were scaled by $F_{\rm{bol}}/F_{\rm{21 \umu m}} = 26.9$ (as determined by \citealt{mottram2011a}) to obtain bolometric fluxes.  All fluxes were then combined with their corresponding distances to obtain a value for their bolometric luminosities. We present the luminosity distribution for all associated MYSOs and compact \hii\ regions in Fig.\,\ref{fig:lum_hist}.

\subsubsection{Uncertainties in the derived parameters}

	Most of the distances are kinematic and as such are associated with an uncertainty of approximately $\pm$1\,kpc; this is primarily the due to affect of streaming motions of the clouds through the spiral arms causing them to deviate from the rotation models ($\pm7$\,\kms; \citealt{reid2009}). This dominates the uncertainties in both distance and radius measurements and in the worst cases can be as large as 50\,per\,cent, however, for most clumps this will be 25\,per\,cent or less. 

	The dust to gas ratio ($R$), dust absorption coefficient ($\kappa_\mu$), kinetic temperature and distance all contribute to the uncertainties in the clump mass and column density measurements. Both the dust absorption coefficient and dust-to-gas ratio are poorly constrained, contributing to uncertainties  of a few, with the mass and column density and an error in the kinetic temperature of $\pm$5\,K having a similar impact. The combination of these parameters is probably only accurate to a factor of a few at best; however, it is worth keeping in mind that although the uncertainties in the absolute values are large they are likely to affect the sample uniformly and therefore the statistical results will not be affected.
	
	The uncertainty in the virial mass is dominated by the error in the radius and this is likely to be of order 20\,per\,cent. Given the uncertainties in distance, flux values (20-40\,percent) and the scaled fluxes, the luminosities are probably reliable to within a factor of 2-3.

\section{Discussion}
\label{sect:discussion}

 \begin{figure}
\begin{center}
\includegraphics[width=0.45\textwidth, trim= 0 0 0 0]{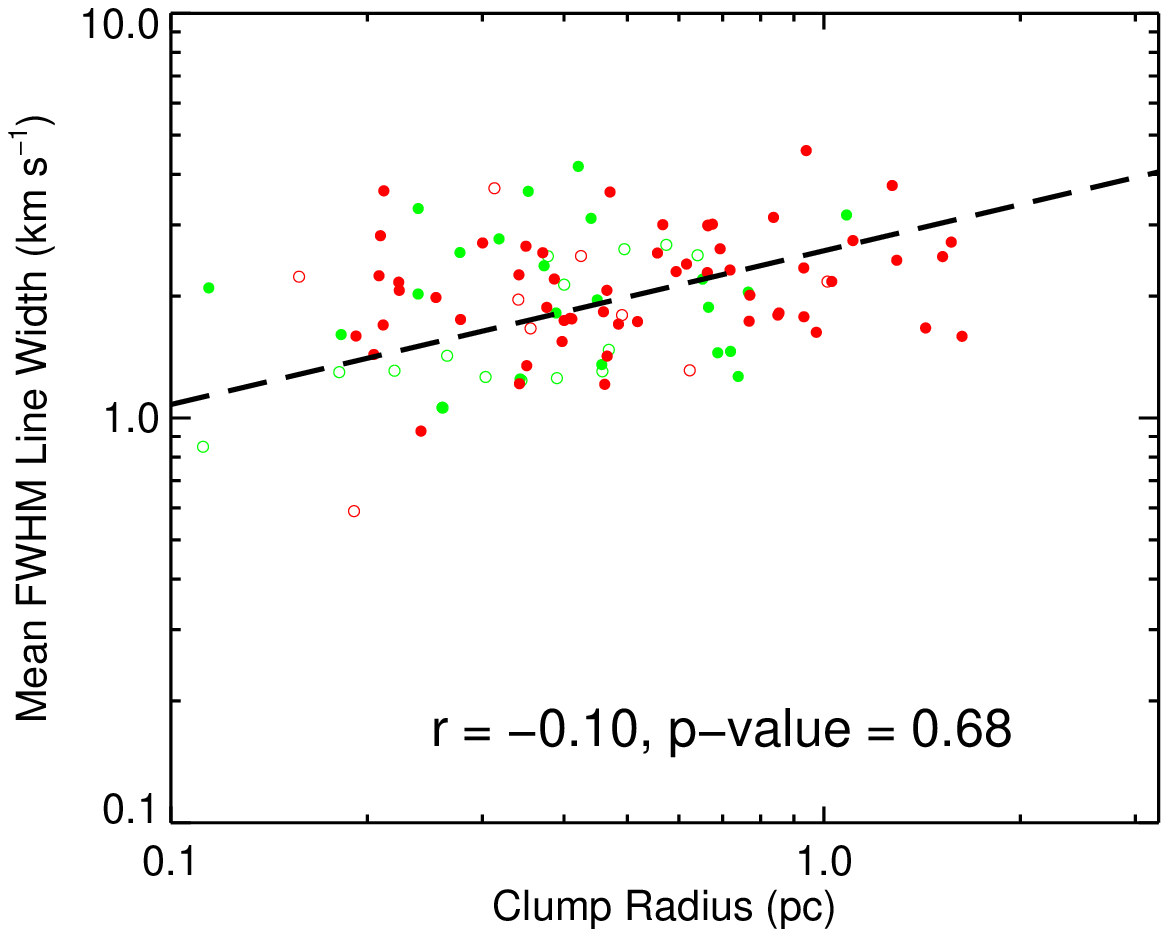}
\includegraphics[width=0.45\textwidth, trim= 0 0 0 0]{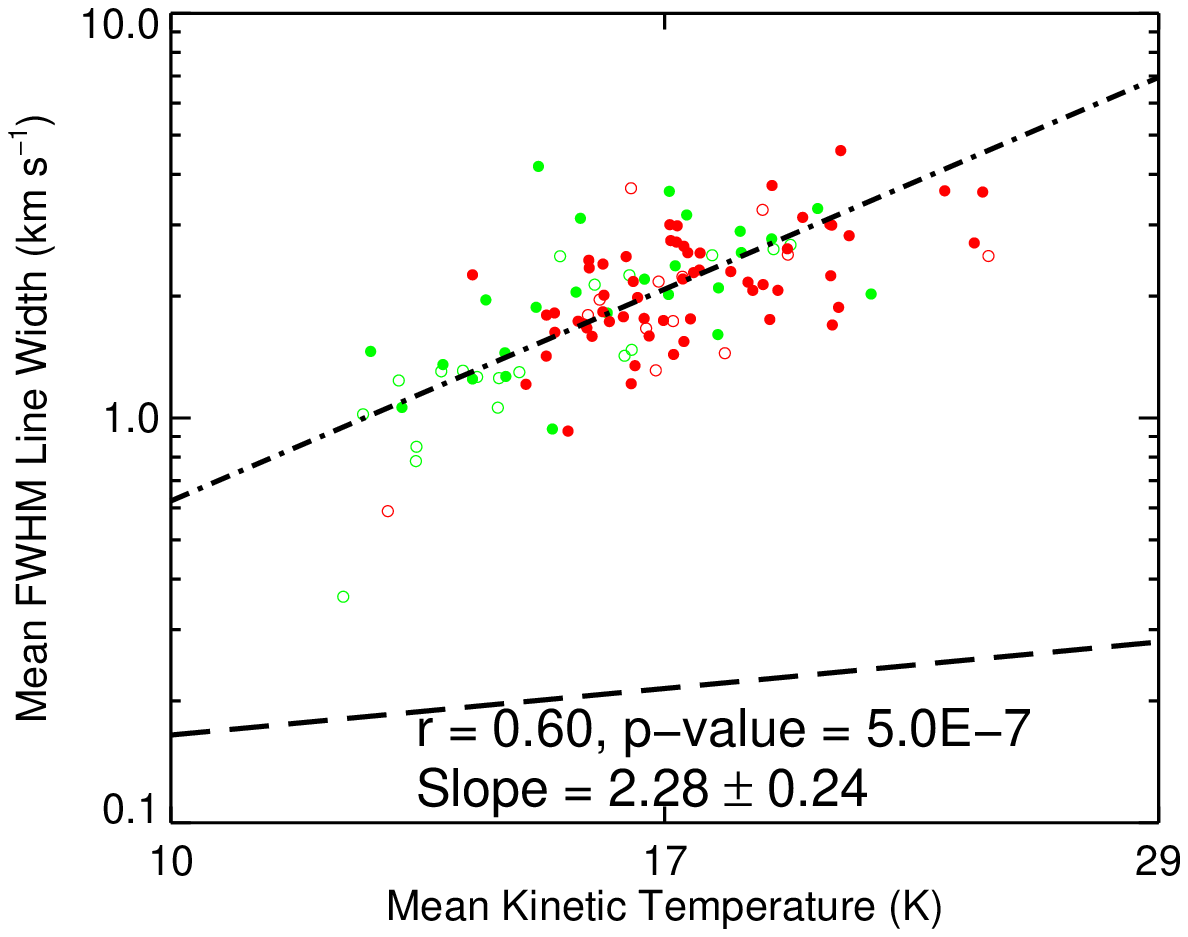}
\includegraphics[width=0.45\textwidth, trim= 0 0 0 0]{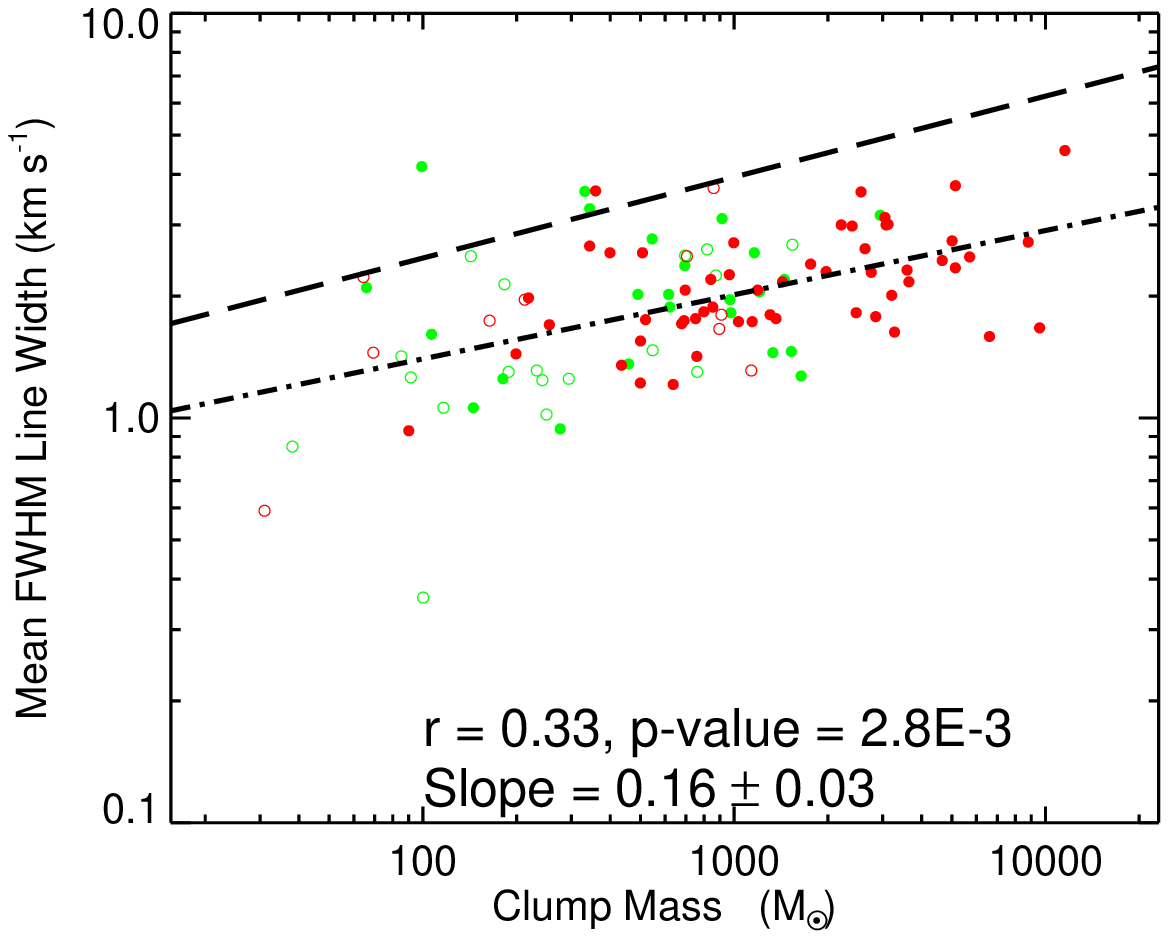}

\caption{FWHM line width relations: In the upper, middle and lower panels we present the intrinsic line widths as a function of radius, kinetic temperature of the gas and clump mass, respectively. The green and red colours distinguish between the quiescent and MSF clumps with the filled and open circles identify clumps above and below 10$\sigma$, respectively. The dashed-dotted lines show the results of a power-law fit to the data. The long-dashed lines shown on the upper and lower panels shows the expected size-line width and mass-line width relationships derived by \citealt{larson1981}; note that these have been scaled by 2.355 to obtain comparable FWHM line width values. The long-dashed line shown in the middle panel indicates the line width expected from purely thermal motions. Towards the bottom right corner of each plot we give the partial-Spearman correlation coefficient and corresponding $p$-values,  and slope and associated error obtained from the power-law fit.} 
\label{fig:fwhm_relations}

\end{center}
\end{figure}

\subsection{Correlation between physical parameters}
\label{sect:corrlations}

For all of the following analyses we calculate the partial-Spearman correlation coefficient ($r_{\rm{AB,C}}$; \citealt{yates1986,collins1998,urquhart2013a}) to remove any dependence of the parameter of interest on distance and thus quantify the effect of the \Mal\ bias. This has the form:

\begin{equation}
r_{\rm AB,C} \; = \; \frac {r_{\rm AB} -  r_{\rm AC} r_{\rm BC}}
{\left[(1-r^2_{\rm AC})(1-r^2_{\rm BC})\right]^{1/2}},
\end{equation}
\noindent where A, B are the parameters for which the correlation is being sought, and C is the variable on which they depend, and  $r_{\rm AB}$, $r_{\rm AC}$ and $r_{\rm BC}$ are the Spearman rank correlation coefficients for each pair of parameters. The significance of the partial rank correlation coefficients is estimated using  \mbox{$r_{\rm AB,C}[(N-3)/(1-r_{\rm AB,C}^2)]^{1/2}$} assuming it is distributed as Student's t statistic (see \citealt{collins1998} for more details; this is referred to as the \KSvalue). Furthermore, while including all of the available data in the correlation plots presented in this section, we make a distinction between the high and low SNR clumps (i.e., above and below 10$\sigma$) and restrict the correlation tests and power-law fits to the former.

\subsubsection{Line width relations}

Fig.\,\ref{fig:fwhm_relations} shows the correlation between line widths and  size, average kinetic temperatures and masses for all of the clumps. We find no significant correlation between the line width and the size of the clumps, which is rather surprising as we might have expected the clumps to follow the Larson size-line width relation (i.e., $\sigma_v$ (\kms) = $1.1\times R_{\rm{clump}}^{0.38}$\,(pc); \citealt{larson1981}). This is perhaps due to the fact that the ammonia (1,1) transition is not probing the full extent of the gas but rather is measuring the velocity dispersion of many smaller cores that are clustered towards the centre of the clumps. This was also seen in NH$_3$ data compiled from the literature by \citet{kauffmann2013} when looking at a similar range of clump sizes, although the Larson size-line width relation was recovered when the full size range was fitted (obtained a slope of 0.32 sizes from 0.01 to several tens of pc). The lack of correlation in our data may be the result of a lack of dynamic range in the clump sizes (cf {\color{red}Maud et al. 2015}).

There is a strong correlation between the temperature of the gas and the line width for all of the clumps.  Quiescent clumps appear to be a little cooler and have slightly narrower line widths than the MSF-associated clumps. All of the clump line widths are significantly broader than expected from thermal motions, indicating that non-thermal motions dominate the line widths. The strong correlation between these two parameters would be consistent with increased feedback from an embedded source for the MSF clumps. The quiescent subsample is very similarly distributed but dominates the lower temperature and line-width end of the distribution. As previously mentioned, many of the quiescent clumps also harbour embedded sources that are still in an early stage of their evolution, and others may also harbour even younger protostellar objects that do not yet have a mid-infrared counterpart.

There is a moderate correlation between the masses and line widths of the clumps, and although the line widths are a factor of a few lower than predicted by \citet{larson1981}, the slope of the distribution is consistent with that study (i.e., \mbox{$\sigma v$ (\kms) = $0.42\times M_{\rm{clump}}^{0.2}$\,\msun}). Again, we do not find any significant difference between the distributions of the MSF and quiescent clumps.

The fact that the line-width, radius and clump mass are broadly consistent with the Larson relation combined with the lack of difference between the MSF and quiescent samples would suggest that feedback from the embedded massive stars not having a significant impact on the {\it{dynamics}} of the clumps. However, there is significant evidence for localised increases in the line-widths and temperatures towards the centres of the MSF clumps, which are coincident with the positions of the embedded objects and so there is good reason to suspect that feedback is playing an important role but that this is not having a significant impact on the overall properties of their host clumps.

\subsubsection{Luminosity, mass and size relations}

 \begin{figure}
\begin{center}

\includegraphics[width=0.45\textwidth, trim= 0 0 0 0]{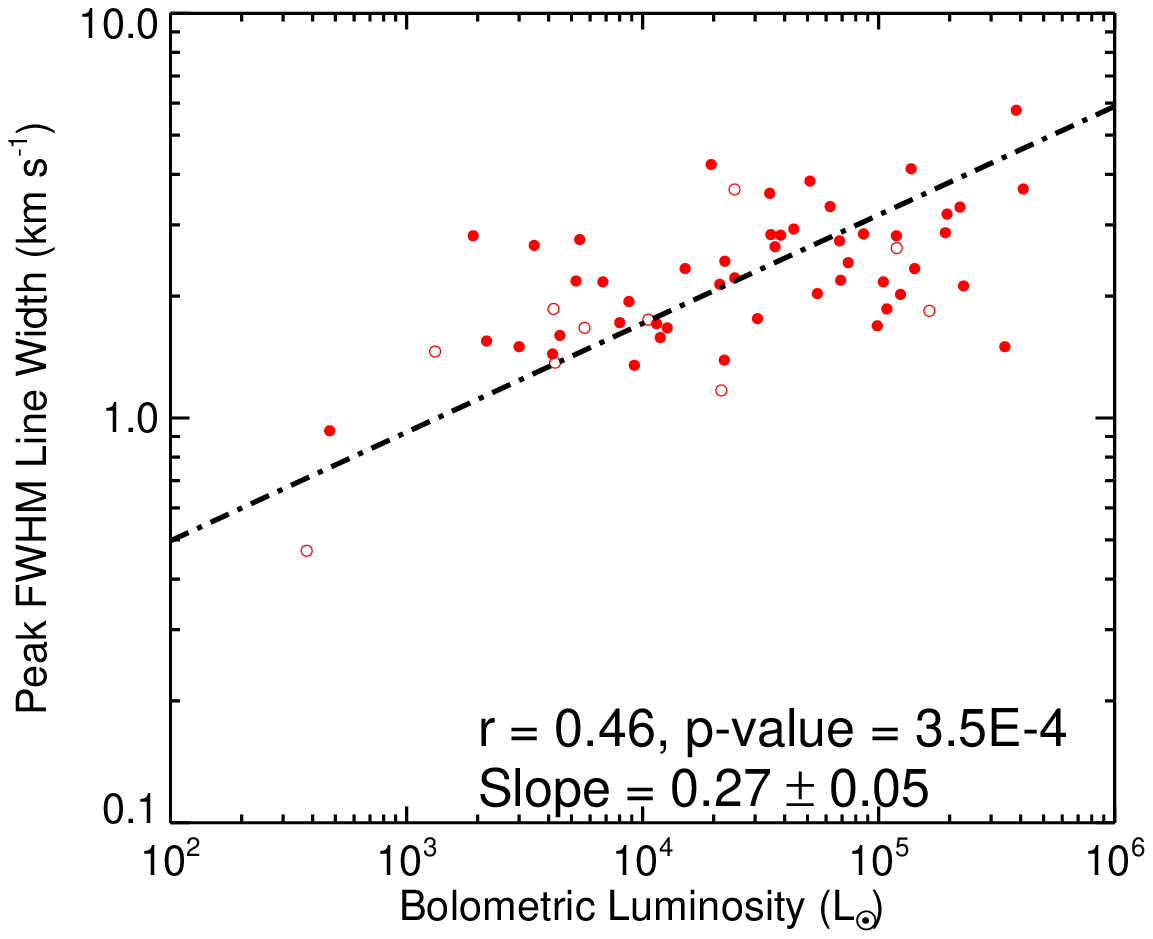}
\includegraphics[width=0.45\textwidth, trim= 0 0 0 0]{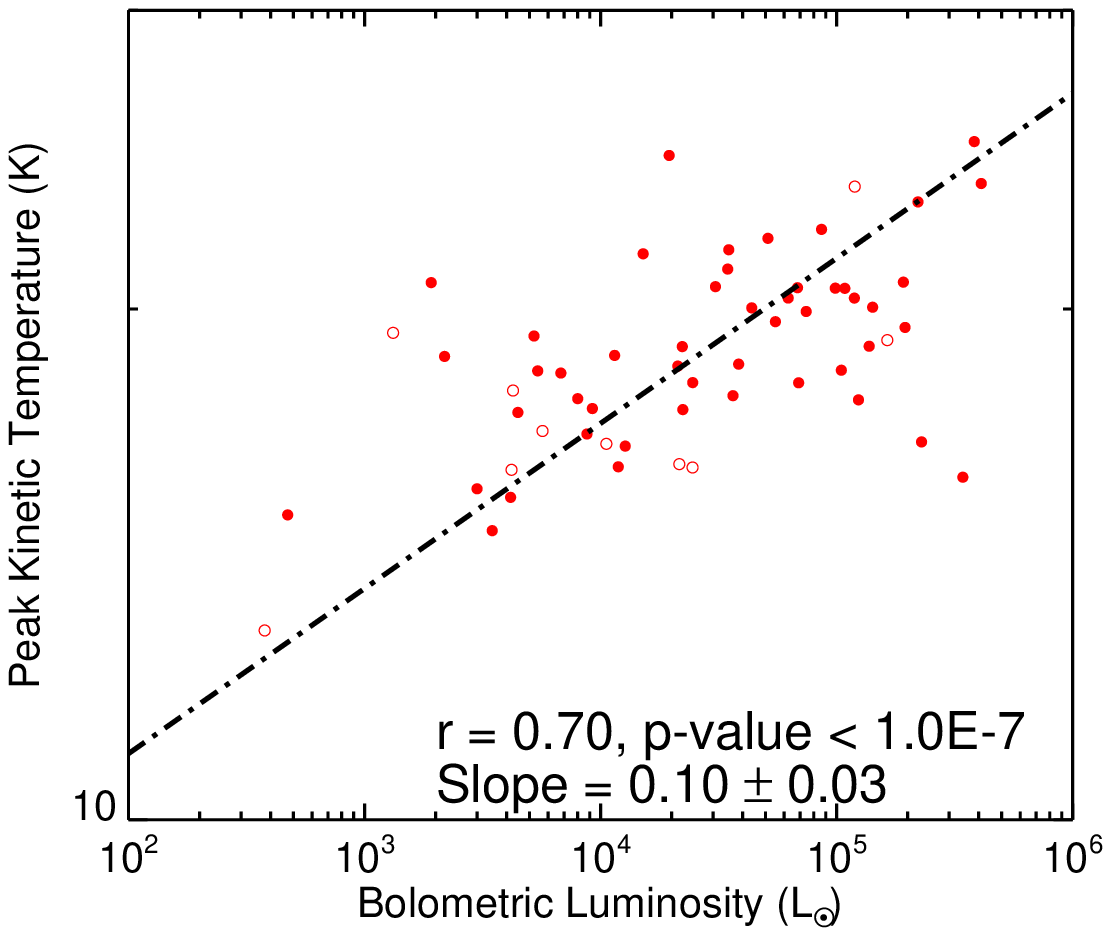}
\includegraphics[width=0.45\textwidth, trim= 0 0 0 0]{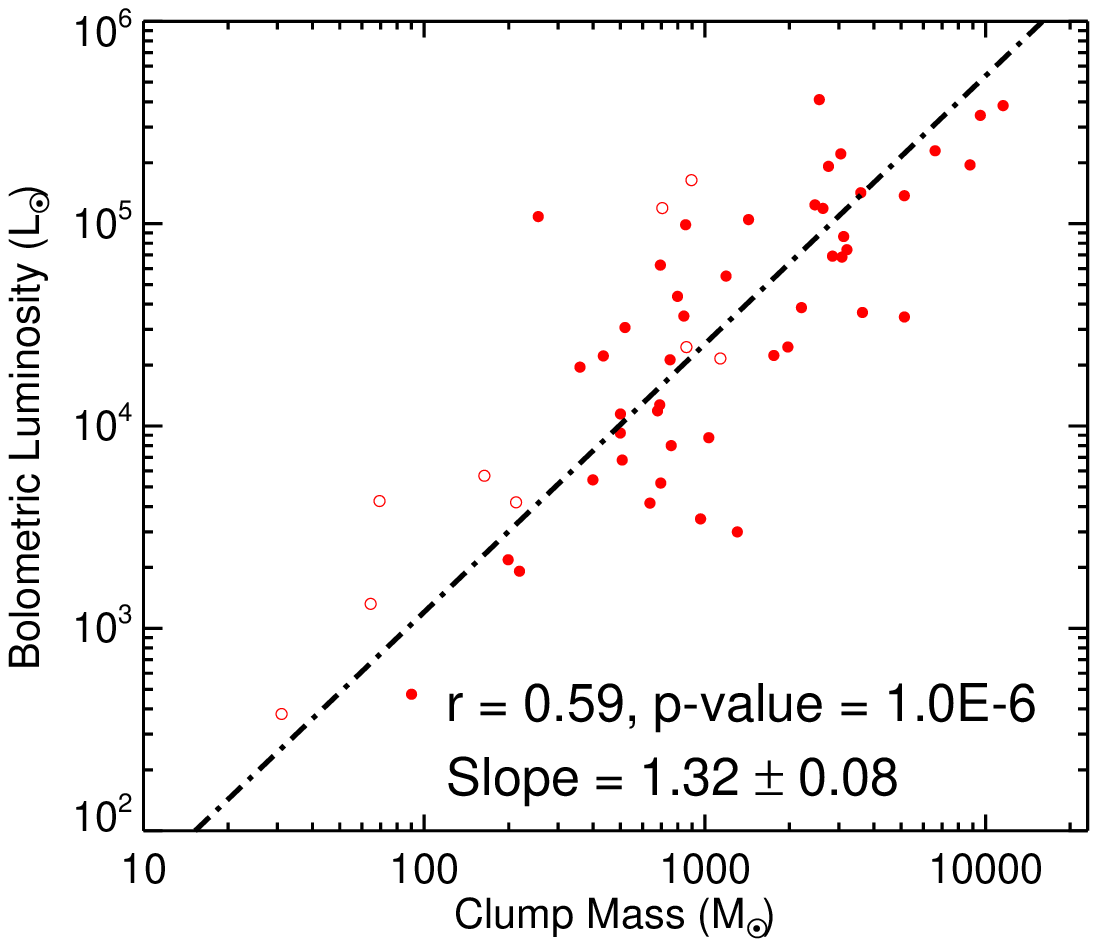}

\caption{Plots showing the relationships between the bolometric luminosities and line-width, kinetic temperatures and clump mass are presented in the upper, middle and lower panels, respectively. The filled and open circles identify clumps above and below 10$\sigma$, respectively. The dashed-dotted lines show the results of a power-law fit to the data. Note that luminosities are only available for the MSF clumps. Towards the bottom right corner of each plot we give the partial-Spearman correlation coefficient and corresponding $p$-values,  and slope and associated error obtained from the power-law fit.} 
\label{fig:lum_relations}

\end{center}
\end{figure}

In Fig.\,\ref{fig:lum_relations} we present a mixture of plots comparing the bolometric luminosity with clump mass, temperature, and line width. Luminosities are only available for the RMS sources, and so the quiescent clumps are not included in these plots. It is clear that there is a strong correlation between luminosity and all of these parameters. There are some slight differences with the correlation values presented in Paper\,I; however, improvements in distances and SED fitting and the smaller sample size are likely to account for these. Furthermore, we are using a partial-Spearman correlation test to remove the dependence on distance, which was not used in the previous analysis.  

In previous sections we have reported that the centres of clumps associated with embedded RMS sources are significantly warmer than the quiescent clumps and show a temperature gradient between the inner and outer envelopes. The strong correlation between the luminosity and temperature extends this relationship, and links the temperature directly to radiative feedback from the embedded object. It is tempting to link the moderate correlation between the line width and luminosity to feedback from the embedded object as well, but there is also a strong correlation between the clump mass and luminosity (as described below), and in the previous section we found that the line widths are correlated to the clump mass and follow the expected line width-mass relationship. It is therefore not clear to what extent feedback is responsible for the non-thermal line width, and what is inherent from the pre-stellar mass. This ambiguity seems contrary to the structured enhancements in the line width we see in the distribution maps, but perhaps this is a result of emission peaking farther from the clump centres and having a more local impact, which is diminished in the global analysis being presented here.

In the lower panel of Fig.\,\ref{fig:lum_relations} we show the clump mass-luminosity relation. There is a strong correlation between these parameters, and the correlation coefficient is very similar to the value reported from previous RMS studies (\citealt{urquhart2013b,urquhart2014b}). The correlation between clump mass and the bolometric luminosity of the embedded cluster is likely to be the result of a fairly uniform initial mass function (\citealt{salpeter1955}) and a limited range of star formation efficiencies (10-30\,per\,cent; \citealt{johnston2009, lada2003b}). The value  of the slope obtained from the power-law fit to the mass-luminosity data is in excellent agreement with the value obtained by \citet{urquhart2014b}.

\begin{figure}
\begin{center}

\includegraphics[width=0.45\textwidth, trim= 0 0 0 0]{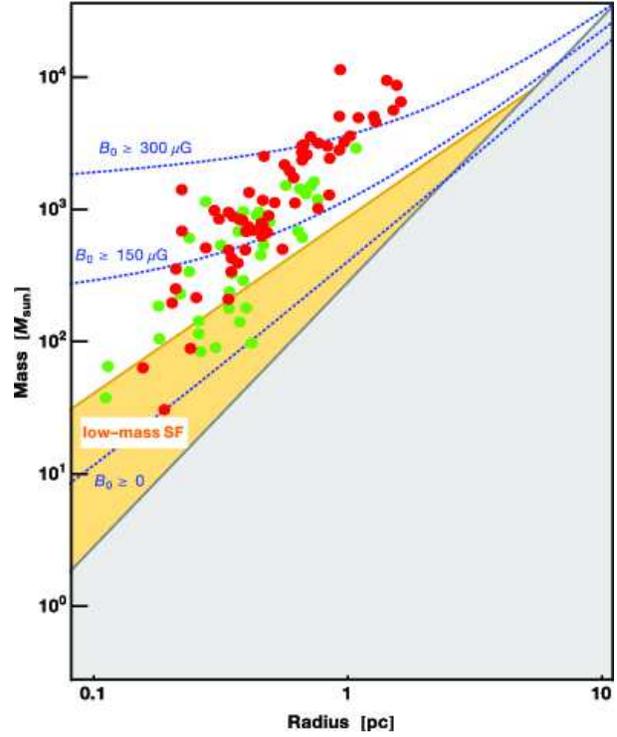}

\caption{Mass-size diagram showing the relationships of these parameters for the MSF and quiescent samples; these are shown as red and green circles, respectively. This diagram has been adapted from the a similar plot presented by \citet{pillai2015} to study the ability of magnetic fields to support clumps, i.e., equilibrium models with magnetic support; these are indicated by the dotted lines. The grey shaded area indicates the region of parameter space where column densities are too low for molecular hydrogen to form, while the yellow shaded and unshaded areas cover the regions where low-mass and high-mass star is thought to dominate, respectively. The partial-Spearman correlation coefficient is 0.67 ($p$-values $\sim$ 10$^{-7}$) and slope obtained from the power-law fit is 1.90$\pm$0.12.} 
\label{fig:mass_size_relations}

\end{center}
\end{figure}

In Fig.\,\ref{fig:mass_size_relations} we present the mass-size distribution of the whole sample of clumps. The value obtained for the slope of the mass-radius relation is a little steeper than that found by \citet{urquhart2014b} but is consistent within the uncertainty. The previous study covered larger ranges of clump masses and radii, particularly including larger and more massive objects.  Inspection of their Fig.\,25 suggests evidence of a slight turnover in the power-law for clumps larger than 1\,pc. The difference in the derived slopes may indicate that the mass-size relationship is stronger for clumps than for small clouds, where 0.15\,pc $< R_{\rm{clump}} <$ 1.25\,pc and 1.25\,pc $< R_{\rm{cloud}}$ (\citealt{bergin2007}). 

\subsubsection{Gravitational stability}

In Section\,\ref{sect:clump_mass} and Section\,\ref{sect:virial_mass} we derived both the clump masses and their corresponding virial masses. The ratio of these two masses can be used to determine the stability of these clumps, and we hereby define the virial parameter as:

 \begin{equation}
\alpha \, =  \frac{M_{\rm{vir}}}{M_{\rm{clump}}}
\end{equation}

\begin{figure}
\begin{center}
\includegraphics[width=0.49\textwidth, trim= 0 0 0 0]{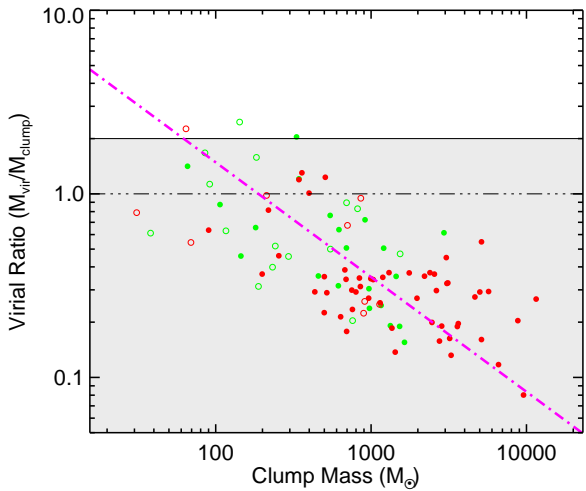}

\caption{\label{fig:virial_mass} Virial ratio ($\alpha$) as a function of clump mass ($M_{\rm{clump}}$) is shown for the two subsamples. The solid and dash-dotted lines indicate the critical values of $\alpha$ for an isothermal sphere in hydrostatic equilibrium with ($\alpha = 2$) and without magnetic support ($\alpha = 1$), respectively. The light grey shading indicates the parameter space where clumps are unstable and likely to be collapsing without additional support from a strong magnetic field. The green and red colours distinguish between the quiescent and MSF clumps with the filled and open circles identify clumps above and below 10$\sigma$, respectively. The dashed-dot magenta line shows the results of power law fits to all sources above 10$\sigma$.}

\end{center}
\end{figure}

The critical virial parameter ($\alpha_{\rm{cr}}$) for an isothermal sphere that is in hydrostatic equilibrium (i.e., a Bonnor-Ebert sphere; \citealt{ebert1955,bonnor1956}) that is not supported by a magnetic field is $\alpha_{\rm{cr}}$ = 2 (\citealt{kauffmann2013}). Clumps with values of alpha above $\alpha_{\rm{cr}}$ are subcritical, and will expand unless pressure-confined by their local environment. Conversely, clumps with values below $\alpha_{\rm{cr}}$ are supercritical: they are gravitationally unstable and likely to be in a state of global collapse unless supported by a strong magnetic field. The virial ratio can therefore provide a useful estimate of a particular clump's overall stability, and can provide a good indication of the stability of the population as a whole.

In Fig.\,\ref{fig:virial_mass} we plot the virial ratio versus the clump mass for 100 sources for which we are able to determine a distance and estimate the total clump mass. The distribution of the whole sample reveals a strong trend for decreasing values of $\alpha$ with increasing clump masses. This suggests that the most massive clumps are also the least gravitationally stable. There is a higher density of the RMS-associated clumps found towards the higher clump mass, although both quiescent and RMS-associated subsamples appear to form a continuous distribution. There is a strong correlation ($r_{\rm{AB,C}}=0.58$, $p$-value $\ll$ 0.0001) between clump mass and the virial parameter. A power law fit to the whole sample has a slope of $-0.63\pm0.06$, and although a slightly steeper slope was found for the RMS-associated subsample and a slightly steeper slope was found for the quiescent subsample, these slopes are not found to be significantly different. The slope of the whole distribution is similar to values determined from other massive star formation studies in the literature (e.g., \citealt{sridharan2002,wienen2012,kauffmann2013}). This trend for decreasing virial ratios with increasing clump mass may explain the apparent lack of any very massive pre-stellar clumps outside the Galactic centre (\citealt{ginsburg2012,tackenberg2012}).

It is clear from this plot that the vast majority of both the quiescent and star forming clumps have virial parameters below the critical value and are likely to be unstable against gravity. However, a recent study of two massive IRDCs (G11.11$-$0.12 and G0.253+0.016; \citealt{pillai2015}) has found that the magnetic fields are strong enough to support these clouds against collapse.\footnote{G11.11$-$0.12 is part of the same IRDC filament as the source we identify in this paper as G010.990$-$00.083.} They measured a magnetic field strength of 267$\pm$26\,$\mu$G for G11.11$-$0.12, and if this was proved to be typical for massive clumps, then it would be sufficient to  support them against gravitational collapse. This is nicely illustrated in Fig.\,\ref{fig:mass_size_relations} where a magnetic field of $\sim$300\,$\mu$G is sufficient to support the majority of clumps in our sample. This is based on a very small sample and the magnetic fields need to be measured for many more clumps before we can determine whether field strengths of hundreds of $\mu$G are indeed typical. However, even if the magnetic fields were able to globally support the clouds we know that smaller regions must be collapsing as most clumps are actively forming massive stars.  

It is interesting to note that we find that no massive clumps towards the upper right region of Fig.\,\ref{fig:virial_mass}. This is where we might expect to find massive quiescent clumps that will go on to form the next generation of massive clusters. This could indicate that the clumps form very rapidly and start forming stars very quickly after their formation. This is consistent with the results reported by \citet{kauffmann2013} from a comprehensive analysis of studies available in the literature. However, this may also be due to a sensitivity issue that we are not able to detect the massive clumps until they have condensed down and taken their place on the observed distribution.

\subsubsection{MSF vs quiescent clumps}

In the previous sections we have compared the properties of the MSF and quiescent clumps, and have found few significant differences between the two samples. Although not explicitly mentioned, we looked for differences between the aspect ratios, orientation with respect to the Galactic mid-plane and $Y$-factor (see Sect.\,\ref{sect:clump_structure}) but failed to find anything significant. This is in contrast to the differences in the aspect ratio and $Y$-factor reported by \citet{urquhart2014b} from a comparison of MSF and quiescent clumps identified by the ATLASGAL survey, which found that MSF clumps tended to be significantly more centrally condensed and spherical in shape. However, that comparison was based on the whole inner Galaxy population of dense clumps while the clumps included in the sample presented here are biased towards active massive star forming regions.

We have also found no significant difference between the clump masses, surface density and column densities of the two subsamples, although we note that these quantities are generally lower for the quiescent clumps. Furthermore, there are no differences in the correlations or relationships between different parameters we have examined in the previous sections. In fact, the only significant differences we have found between the two samples is that the MSF clumps are warmer and have larger line widths than the quiescent clumps. Both of these parameters peak towards the centre of the MSF clumps and are approximately coincident with the position of the embedded massive star, and are consistent with stellar feedback.

The similarities between these two samples suggests that they are likely to be part of the same parent population. In cases where both quiescent and MSF clumps are present in the same field it seems likely that they would have formed at a similar time and have similar initial conditions. We also note that in these fields it is always the most massive and unstable clump that is host to the embedded source. The similarity of surface densities between both samples implies that they also have similar volume densities and therefore similar free-fall collapse time scales. Although most clumps are unstable, the quiescent clumps have noticeably higher virial parameters and so may be more resistant to collapse in the presence of additional support mechanisms such as magnetic fields (e.g., \citealt{pillai2015}). 

The properties of the quiescent sample of clumps and their similarities to the MSF clumps suggests they may be in a pre-stellar state and an ideal sample of pristine clumps with which to study the initial conditions for massive star formation (cf the radio-loud and radio-quiet clumps in \citealt{thompson2006}).

\section{Summary and conclusions}
\label{sect:summary_conclusions}

We have used the K-band Focal Plane Array (KFPA) on the GBT to map the ammonia (1,1) and (2,2) inversion transition emission towards 66 massive star forming regions. We have identified 115 distinct clumps: we classify 21 of these with aspect ratios $>$1.8 as filaments while the rest can be described being roughly spherical in morphology. The column density distributions are strongly peaked towards the centre of the clumps and decrease towards the edges. The beam filling factors are significantly lower than unity indicating the presence of a clumpy substructure that is similarly concentrated towards the centre of the clumps. The majority of clumps appear to be so centrally concentrated. We find that the semi-major axis of the entire sample of clumps are preferentially aligned parallel to the Galactic mid-plane.

We find that 71 of the clumps are associated with embedded MYSOs and compact \hii\ regions identified by the RMS survey. We refer to these as massive star forming clumps and the remaining 44 clumps as quiescent. We compare the properties of both subsamples as well and the properties of the inner and outer envelopes. The MSF clumps are warmer, have marginally broader line widths and slightly lower pressure ratios than the quiescent clumps. The central regions of the MSF clumps also appear to be warmer and more turbulent than their outer envelopes, while the quiescent clumps show no evidence of a temperature or line width gradient across them. We also find the abundance ratio is fairly uniform for all clumps with no significant variations between the inner and outer envelopes.

Although the density structure of the MSF and quiescent clumps are similar, the MSF clumps are typically most massive and have significantly higher peak column densities. We find that all of the clumps are unstable against gravitational collapse; however, those already associated with massive star formation also appear to be the most unstable. Given that all of the clumps are located in massive star forming regions, where the initial conditions and environment are likely similar, we might expect the clumps that form in these regions to have broadly similar properties. This sample of quiescent clumps may therefore represent a potentially useful sample of massive clumps that are currently in a pre-stellar stage that can be used to determine the initial conditions of the gas and thus identify subtle evolutionary changes in the gas properties.

\section*{Acknowledgments}
 
We would like to thank the staff of the GBT telescope for their support during the observation and special thanks to Glen Langston for his assistance with the KFPA. We would also like to thanks the referee for their comment that have helped improve the clarity of this paper. This research has made use of the SIMBAD database operated at CDS, Strasbourg, France. We have used data from the ATLASGAL project, which is a collaboration between the Max-Planck-Gesellschaft, the European Southern Observatory (ESO) and the Universidad de Chile. This paper also made use of information from the Red MSX Source survey database at \url{http://rms.leeds.ac.uk/cgi-bin/public/RMS\_DATABASE.cgi} which was constructed with support from the Science and Technology Facilities Council of the UK. This work was partially funded by the ERC Advanced Investigator Grant GLOSTAR (247078) and was partially carried out within the Collaborative Research Council 956, sub-project A6, funded by the Deutsche Forschungsgemeinschaft (DFG). JSU would like to dedicate this work to the memory of his mother Marilyn Frances Urquhart.

\bibliography{rms}

\bibliographystyle{mn2e_new}



\end{document}